\documentclass[11pt,tightenlines,aps,prd,floatfix,onecolumn,superscriptaddress,nofootinbib]{revtex4-2}
\usepackage{amsmath, amsthm, amssymb,slashed,url,bbold,xcolor,upgreek}
\usepackage{graphicx}
\usepackage{epstopdf}
\usepackage{slashed}
\usepackage{physics}
\usepackage{appendix}

\usepackage[pdfpagelabels, pdfencoding=auto, psdextra]{hyperref}
\hypersetup{%
 pdfsubject=Paper,
 pdfkeywords={nuclear physics} 
 unicode = true,
 breaklinks = true,
 colorlinks = true,
 linkcolor = blue,
 menucolor = blue,
 citecolor = blue,
 urlcolor = blue
}

\usepackage{etoolbox}% http://ctan.org/pkg/etoolbox
    \makeatletter
    \patchcmd{\maketitle}{\@fpheader}{}{}{}
    \makeatother

\graphicspath{ {./Plots/} }

\def\XXint#1#2#3{{\setbox0=\hbox{$#1{#2#3}{\int}$}
     \vcenter{\hbox{$#2#3$}}\kern-.5\wd0}}

\def\d{\mathrm{d}}

\def\OMIT#1{{}}

\newcommand{\beq}{\begin{equation}}
\newcommand{\eeq}{\end{equation}}
\newcommand{\beqa}{\begin{eqnarray}}
\newcommand{\eeqa}{\end{eqnarray}}

\newcommand{\nsat}{n_{\rm sat}}

\newcommand{\ZN}{$\mathbb{Z}_{\mathcal{N}}$ }

\usepackage{xspace}
\newcommand{\ChiEFT}{ChiEFT\xspace}
\newcommand{\orcid}[1]{\href{https://orcid.org/#1}{\includegraphics[scale=0.055]{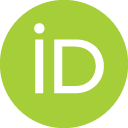}}}
\newcommand{\MeV}{\, \mathrm{MeV}}

\begin{document}

\title{Neutron Stars with Exceptionally Light QCD Axions} 

\author{Mia Kumamoto~\orcid{0009-0004-9515-9213}}
\email{mialk@uw.edu}
\affiliation{Department of Physics, University of Washington, Seattle, WA 98195-1560, USA}
\affiliation{Institute for Nuclear Theory, University of Washington, Seattle, WA 98195-1560, USA}

\author{Junwu Huang~\orcid{0000-0001-6007-7315}}
\email{jhuang@perimeterinstitute.ca}
\affiliation{Perimeter Institute for Theoretical Physics, Waterloo, Ontario N2L 2Y5, Canada}

\author{Christian~Drischler~\orcid{0000-0003-1534-6285}}
\email{drischler@ohio.edu}
\affiliation{Department of Physics and Astronomy, Ohio University, Athens, OH 45701, USA}
\affiliation{Facility for Rare Isotope Beams, Michigan State University, East Lansing, MI 48824, USA}

\author{Masha Baryakhtar~\orcid{0000-0002-7631-2604}}
\email{mbaryakh@uw.edu}
\affiliation{Department of Physics, University of Washington, Seattle, WA 98195-1560, USA}
% \affiliation{Institute for Nuclear Theory, University of Washington, Seattle, WA 98195-1560, USA}

\author{Sanjay Reddy~\orcid{0000-0003-3678-6933}}
\email{sareddy@uw.edu}
% \affiliation{Department of Physics, University of Washington, Seattle, WA 98195-1560, USA}
\affiliation{Institute for Nuclear Theory, University of Washington, Seattle, WA 98195-1560, USA}

\date{\today}

\begin{abstract}
{We present a comprehensive study of axion condensed neutron stars that arise in models of an exceptionally light axion that couples to quantum chromodynamics (QCD). These axions solve the strong-charge-parity (CP) problem, but have a mass-squared lighter than that due to QCD by a factor of $\varepsilon<1$.  Inside dense matter, the axion potential is altered, and much of the matter in neutron stars resides in the axion condensed phase where the strong-CP parameter $\theta =\pi$ and CP remains a good symmetry. In these regions, masses and interactions of nucleons are modified, in turn changing the equation of state (EOS), structure, and phenomenology of the neutron stars. We take the first steps toward the study of the EOS of neutron star matter at $\theta =\pi$ within chiral effective field theory and use relativistic mean field theory to deduce the resulting changes to nuclear matter and the neutron star low-density EOS. We derive constraints on the exceptionally light axion parameter space based on observations of the thermal relaxation of accreting neutron stars, isolated neutron star cooling, and pulsar glitches, excluding the region up to $5 \times 10^{-7} \lesssim \varepsilon \lesssim 0.2$ for $ m_a \gtrsim 2\times 10^{-9}\,{\rm eV} $. We comment on potential changes to the neutron star mass-radius relationship, and discuss the possibility of novel, nuclear-density compact objects with $\theta =\pi$ that are stabilized not by gravity but by the axion potential.
}
\end{abstract}

\maketitle

\section{Introduction}
The quantum chromodynamics (QCD) axion is a well-motivated and minimal solution to the Strong Charge-Parity (CP) problem~\cite{axion1,axion2,axion3,Peccei:1977ur}, the puzzle of the smallness of the neutron electric dipole moment (EDM). The neutron EDM is proportional to $\bar{\theta}$, a combination of Standard Model parameters,
\begin{equation}
    \bar{\theta} = \theta + \arg \det Y_u Y_d,
\end{equation}
where $\theta$ is the QCD $\theta$ angle, while $Y_u$ and $Y_d$ are the up- and down-type Yukawa matrices. The combination is experimentally bounded to be $\bar{\theta} \lesssim 10^{-10}$ based on the non-observation of the neutron electric dipole moment~\cite{Abel:2020pzs}. The smallness of $\bar{\theta}$ is particularly puzzling since the CP-violating angle in the Cabibbo-Kobayashi-Maskawa (CKM) matrix
has been measured to be $\mathcal{O}(1)$, in agreement with theoretical expectation. 

The axion is the pseudo-Nambu-Goldstone boson resulting from the spontaneously broken Peccei-Quinn symmetry and couples to gluons as
\begin{equation}
  \mathcal{L} \supset  \left(\frac{a}{f_a} -\bar{\theta} \right) \frac{\alpha_s}{8 \pi } G^{\mu\nu} \tilde{G}_{\mu\nu},
\end{equation}
where $a$ is the axion field, $f_a$ the axion decay constant, $\alpha_s$ the strong coupling constant, $G^{\mu\nu}$ the color field strength
tensor, and $\tilde{G}_{\mu\nu} \equiv \varepsilon_{\mu\nu\rho\sigma}G^{\rho\sigma}/2$ its dual. The entire dependence on $\bar{\theta}$ has been placed in the gluon term after a chiral rotation of the quarks.  At low energies, the axion obtains a periodic potential from the coupling to gluons, and the CP-violating angle $\bar{\theta}$ relaxes to zero during cosmic evolution of the axion field in its potential~\cite{Preskill:1982cy, Abbott:1982af,Dine:1982ah}. 

The QCD axion has inspired a diverse experimental campaign~\cite{ParticleDataGroup:2024cfk,ciaran_o_hare_2020_3932430,Hook:2018dlk,Adams:2022pbo}. Searches for the QCD axion and axion-like particles (ALPs) through their electron and photon couplings have seen tremendous progress in recent years. The axion-nucleon coupling---the coupling that defines the QCD axion and distinguishes it from ALPs that do not solve the strong CP problem---is much more difficult to probe in the laboratory~\cite{Graham:2013gfa,Budker:2013hfa,Arvanitaki:2014dfa,Arvanitaki:2021wjk,Berlin:2022mia}. The sensitivity of current ground-based experiments is rapidly improving~\cite{ARIADNE:2017tdd,Abel:2017rtm,Chang:2017ruk,JacksonKimball:2017elr,Roussy:2020ily,Aybas:2021nvn,JEDI:2022hxa,Schulthess:2022pbp,Zhang:2022ewz,Winter:2024cyp,Fan:2024pxs}, but is as yet orders of magnitude away from the ``QCD axion line'' on which the axion decay constant $f_a$ and the axion mass $m_a^{\rm (QCD)}$ are related by the QCD prediction~\cite{axion1},
\begin{equation}
m_a^{\rm (QCD)} =  \sqrt{\frac{m_u m_d}{(m_u+m_d)^2} }\frac{m_{\pi}f_{\pi}}{f_a}\,.
\label{eq:maQCD}
\end{equation} 
Furthermore, these laboratory searches are at present far weaker than astrophysical constraints from Big Bang Nucleosynthesis~\cite{Blum:2014vsa}, the evolution of supernova 1987A~\cite{Raffelt:1987yt,Chang:2018rso,Carenza:2019pxu,Lucente:2022vuo,Springmann:2024ret,Springmann:2024mjp}, stability of the Earth, the Sun and white dwarfs~\cite{Hook:2017psm,Balkin:2022qer},  LIGO-Virgo-Kagra (LVK) measurements of binary neutron star inspirals~\cite{Hook:2017psm,Huang:2018pbu,Zhang:2021mks} and isolated neutron star cooling~\cite{Gomez-Banon:2024oux}. See e.g.~\cite{OHare:2024nmr,Caputo:2024oqc}, for recent reviews.

Theoretically, there is renewed interest in the parameter space of axions which solve the strong CP problem but have an exceptionally light mass, $m_a < m_a^{\rm (QCD)}$,  in light of the discrete symmetry noted in Ref.~\cite{Hook:2018jle}. A discrete \ZN symmetry can suppress the axion mass below that predicted by Eq.~\eqref{eq:maQCD} in a technically natural way by a factor of
\begin{equation}\label{eq:defineep}
    \varepsilon^{1/2} \equiv \frac{m_a}{m_a^{\rm (QCD)}}  \approx \frac{1}{2^{{\mathcal{N}}/2 - \log \mathcal{N}}}
\end{equation}
in the large $\mathcal{N}$ limit~\cite{DiLuzio:2021pxd}. This motivates the study of lighter-than-expected QCD axions---hereafter referred to as exceptionally light QCD axions---that still solve the strong CP problem (for ${\mathcal{N}}$ odd). Even within this model, we note that as a small ${\theta}$ angle is itself technically natural, the very small $\varepsilon\ll 1$ parameter space, in particular $\varepsilon \lesssim 10^{-10}$, cannot be well-motivated: it would introduce a tuning at the level of the original tuning of the Strong CP problem. As we discuss below, due to their high density, neutron stars provide a unique environment to test exceptional axions close to the QCD line:  values of $\varepsilon$ close to unity, and \ZN models for modest values of $\mathcal{N}$. We note that the vacuum structure of the \ZN model is significantly different from the phenomenological $\varepsilon$ model and can result in different dynamics and constraints.

Observationally, the above-mentioned studies of dense astrophysical objects place strong constraints on the axion-gluon coupling parameter space, and hence $\varepsilon$, especially at small axion masses. These constraints stem from the effect first noted in Ref.~\cite{Hook:2017psm}. The axion potential is modified in the presence of matter, which suppresses the vacuum QCD contribution. For small $\varepsilon$, there is a critical density above which a large, dense object can contain a region with negative effective axion mass which, in turn, shifts the potential minimum from $\bar{\theta} = 0$ to $\bar{\theta} = \pi$ inside the object. For Earth and solar core densities, the $\bar{\theta} = \pi$  region emerges for $\varepsilon \lesssim 10^{-13}$, while in white dwarfs, $\bar{\theta} = \pi$ occurs at $\varepsilon \lesssim 10^{-7}$~\cite{Hook:2017psm}. Studies of the properties and stability of these objects have led to the most stringent constraint on the parameter space of $\varepsilon \lesssim 2\times 10^{-7}$ for $f_a \lesssim 10^{16} \, {\rm GeV}$~\cite{Balkin:2022qer}. 

In this study, we focus on axion condensation in neutron stars for $\varepsilon \gtrsim  5 \times 10^{-7}$. Earlier work has shown that for small axion masses ($m_a \lesssim 10^{-11} {\,\rm eV}$) the axion condensate extends far outside the radius of the star~\cite{Hook:2017psm}. In this scenario, the extended axion field mediates a force between neutron stars and is constrained by LVK observations of neutron star mergers~\cite{Hook:2017psm,Huang:2018pbu,Zhang:2021mks}. For a heavier axion with $m_a \gtrsim 10^{-8} {\,\rm eV}$, the axion field is largely confined inside the neutron star, and the effects of axion condensation on the neutron star interior need to be understood to predict the observational consequences.  Prior work has considered neutron stars in the axion condensed phase in the non-interacting Fermi gas approximation~\cite{Balkin:2023xtr} and in hybrid stars where axion interactions were included in the quark phase (but not the nuclear phase) via an effective 't Hooft determinant interaction in the NJL model~\cite{HybridNJLAxion}.

In this work, we provide a dedicated study of the equation of state (EOS) for $ 5 \times 10^{-7}\lesssim\varepsilon < 1 $  and its impact on neutron star structure. To calculate the EOS and phases present in the neutron star, we compare results from Chiral Effective Field Theory (\ChiEFT) and relativistic mean field theory (RMFT).
As noted in Ref.~\cite{Hook:2017psm}, the reduction in the effective mass of the axion in the presence of finite baryon density also leads to a reduction in the effective mass of nucleons. At fixed chemical potential, a smaller effective nucleon mass results in a larger baryon density. As a result, the transition from $\bar{\theta}=0$ to $\bar{\theta} =\pi$ introduces a discontinuity in the EOS. For different $\varepsilon$, this transition occurs at different locations in the neutron star, leading to a variety of observational effects. 
For $\varepsilon \gtrsim 0.1$, the neutron star will have an inner region with $\bar{\theta} = \pi$, separated by an axion field profile from the outside region where $\bar{\theta} = 0$, while for smaller $\varepsilon$ the entire star has $\bar{\theta} =\pi$ except for a thin region near the surface. 
When $\varepsilon \lesssim 0.1$, $\bar{\theta}=\pi$ becomes energetically favored at a baryon chemical potential lower than that of iron ($\mu_{\rm Fe}/A \simeq m_N - 8 \, \mbox{MeV}$) and the axion pressure itself becomes important at low density, producing a state with zero pressure at non-negligible energy density. This may result in a star that is almost entirely at a density above the neutron drip density or above the crust-core transition density.  In this case, the remaining crust phases will only appear in a thin region where $\theta$ changes from $\pi$ to $0$. 

Recent observations of neutron stars (NSs) have resulted in reliable measurements of NS masses and constraints on their size. Radio observations of pulsars have now provided compelling evidence for  neutron stars with masses greater than 2 M$_\odot$, while gravitational wave and x-ray observations suggest that typical NSs with mass $\simeq 1.4$ M$_\odot$ have radii in the range 11-13 km. Taken together, these observations and recent \ChiEFT-based calculations of the EOS of neutron-rich matter indicate that the pressure of dense matter inside neutron stars is relatively small up to $2\nsat$, where $\nsat=0.16$ nucleons per fm$^{3}$ is the nucleon density at nuclear saturation, and increases rapidly at higher densities encountered in the neutron star inner core to support massive neutron stars \cite{Watts:2016uzu,Drischler:2020fvz,Drischler:2021kxf}. A relatively soft EOS in the outer core and inner crust of a neutron star has important implications for neutron star structure. As we discuss, axion condensation at $n_B \lesssim \nsat$ could have dramatic effects on the structure and dynamics of the outer regions of the neutron star with observable consequences. 

The outline of this paper is as follows. In Sec.~\ref{sec:dilute}, we describe the most important features of axion condensation in a simple gas of non-interacting nucleons. In Sec.~\ref{sec:nuclear matter}, we discuss the effects of axion condensation including the potential effects of nuclear interactions, with a focus on \ChiEFT and RMFT models. Section~\ref{sec:neutron stars} describes the impact that axion condensation has on the structure and properties of neutron stars within our nuclear model. In Sec.~\ref{sec:piballs}, we describe novel nuclear density objects stabilized by an axion potential rather than gravity, first proposed in Refs.~\cite{Balkin:2022qer, Balkin:2023xtr} for a non-interacting Fermi gas. We present constraints in the $f_a$-$m_a$ parameter space for the phenomenological $\varepsilon$ model and comment on the \ZN model based on neutron star observables in Sec.~\ref{sec:constraints}. Section~\ref{sec:conclusion} concludes this manuscript with a summary and outlook.

\section{Axion condensation in a dilute gas of nucleons} 
\label{sec:dilute}

In this section, we summarize the main properties of axions discussed in Refs.~\cite{Hook:2017psm,Huang:2018pbu,Hook:2019pbh,Balkin:2020dsr,Balkin:2021wea,Zhang:2021mks,Balkin:2021zfd,Balkin:2022qer,Balkin:2023xtr}, which lead to the development of an axion profile around compact stellar objects with $\bar{\theta} = \pi$ inside these objects and $\bar{\theta} = 0$ outside. We focus here on the scenario of a dilute nucleon gas (see also~\cite{Balkin:2023xtr}) and leave the impact of interactions to Sec.~\ref{sec:nuclear matter}.  We also review the main observational consequences for the Earth, the Sun, white dwarfs and neutron stars. Henceforth, we take the quark mass matrix to be real and $\bar{\theta} = \theta$. Throughout the paper, we work with an axion with the following potential in vacuum
\begin{equation}\label{eq:axionpotential}
    V(\theta) =- \varepsilon m_{\pi}^2  f_{\pi}^2 ~\left[f(\theta) -1\right] \,,
\end{equation}
where~\cite{DiVecchia:1980yfw}
\begin{equation}\label{eq:foftheta}
    f(\theta)=   
     \sqrt{1-\frac{4 m_u m_d}{(m_u+m_d)^2}\sin^2\left(\frac{\theta}{2}\right)}\,,
\end{equation}
and $\theta =a/f_a$. The minimum of this potential is at $\theta=0$ and the mass of the axion is\footnote{The property of the potential in the \ZN model is similar near $\theta=0$ but can be quite different near $\theta = \pm \pi $. See sec.~\ref{ssec:domain_wall} and ~\ref{ssec:ZNmodel} for more details.}
\begin{equation}
m_a = \varepsilon^{1/2}m_a^{\rm (QCD)} = \varepsilon^{1/2}\sqrt{\frac{m_u m_d}{(m_u+m_d)^2} }\frac{m_{\pi}f_{\pi}}{f_a}.
\end{equation}

\subsection{Baryon and meson masses}
\label{sec:masses}
The spectrum of baryons and mesons and the vacuum energy are altered when $\theta \neq 0$ \cite{Vafa:1984xg,Ubaldi:2008nf}. The $\theta$ dependence of the pion mass, which is most important for our study, can be calculated by noting that the $\theta$ term in the QCD Lagrangian can be eliminated by a unitary rotation of the quark fields. When only up and down quarks are included, the transformation $u\rightarrow \exp{(i \gamma_5 \phi_u)} u $ and $d\rightarrow \exp{(i \gamma_5 \phi_d)} d$ where $\phi_u + \phi_d=\theta$ fully captures the effect of $\theta \neq 0$. From the relation $m_\pi^2 \propto m_q = m_u + m_d$, the $\theta$ dependence of the mass of the neutral pion is~\cite{Ubaldi:2008nf}
\begin{equation}\label{eq:pionmasstheta}
m_{\pi}^2 (\theta) = f(\theta )  m_{\pi}^2 (\theta = 0).
\end{equation}
At $\theta=\pi$, the net effect of this transformation is to change the sign of the up quark mass in the quark mass matrix and the neutral pion mass is $m_{\pi} (\theta = \pi) = (\sqrt{m_d-m_u}/\sqrt{m_d + m_u} ) m_{\pi}^{\rm phys}\approx  80 \,{\rm MeV}$ for $m_\pi^{\rm phys} = 138\, {\rm MeV}$, significantly lighter than at $\theta =0$.  The smaller pion mass implies a longer range of the force; implications for nuclear interactions will be discussed in Sec.~\ref{sec:nuclear matter}. 

We note that the $\theta$-dependence of the masses of the $\sigma$ and $\rho$ mesons have been calculated from analysis of pion-pion amplitudes in Refs. \cite{HanhartPelaezRios2008, Pelaez:2010sigmamass,Berengut:2013nh}. The amplitudes are calculated using chiral perturbation theory (ChiPT) but depend on the unitarization scheme which is not unique and introduces a systematic error that cannot be easily quantified. When $m_\pi$ is reduced to $80$ MeV, Refs. \cite{HanhartPelaezRios2008, Pelaez:2010sigmamass} find a 6\% reduction to the $\sigma$ mass. In Ref.~\cite{Berengut:2013nh} it was found that the logarithmic derivative $K_\sigma  = \dd \ln{m_\sigma}/\dd\ln{m_q} = 0.081\pm 0.007$, which implies $\approx 7$\% reduction to the $\sigma$ mass at $\theta=\pi$. The $\rho$ meson mass was also found to decrease with pion mass but to a lesser degree. In Sec.~\ref{sec:RMF}, we shall incorporate these observations to study the $\theta$ dependence of the nuclear EOS in a relativistic mean field model. 

Apart from the mass of the mesons, the masses of baryons, especially protons and neutrons, also depend on the $\theta$ angle. Such a dependence can be parameterized as
\begin{align}\label{eq:mnoftheta}
m_{n} (\theta) &= m_n (\theta=0) + \sigma_N \left( f(\theta) -1 \right) \left( 1 -\frac{\Delta \sigma }{\sigma_N f(\theta)}\right)\,,\nonumber\\
m_p (\theta)  &= m_p (\theta=0) + \sigma_N \left( f(\theta) -1 \right) \left( 1 +\frac{\Delta \sigma }{\sigma_N f(\theta)}\right)\, , 
\end{align}
where $\sigma_N$ is the isoscalar nucleon sigma term and $\Delta \sigma$ is the isovector sigma term. Despite much interest and effort the isoscalar nucleon sigma term is not precisely known and the best estimates from Lattice QCD and phenomenology indicate that $\sigma_N \simeq 50 \pm 10$ MeV \cite{Alarcon:2021dlz}. The value of $\Delta \sigma$ can be inferred from the mass splitting of baryons. Determining $\Delta \sigma$ from the mass splitting of the neutron and proton gives $\Delta \sigma = \left(m_n-m_p\right)_{\text{non}-\text{em}}/2 \simeq 1 \,{\rm MeV}$ \cite{Ubaldi2010deutBEtripAlph}. The mass splitting of the nucleons and $\Xi$ baryon gives a somewhat larger value of $3.1$ MeV \cite{CREWTHER1979123}. In what follows, we shall adopt  $\sigma_N = 50$ MeV and $\Delta \sigma =1$ MeV. For this choice, at $\theta = \pi$, the proton and neutron masses decrease by about 32 MeV and the neutron-proton mass difference increases by about 3.5 MeV. 

\subsection{Effective axion potential at finite density}

The dependence of the axion potential on the temperature and density of the medium is essential to  understanding the behavior of the axion field in the hot, early universe and in dense astrophysical objects. The temperature dependence of the axion mass in the early universe is well-known to be key to obtaining a precise prediction of the range of QCD axion masses that can be the dark matter in the Universe~\cite{GrillidiCortona:2015jxo}. The density dependence of the axion mass, on the other hand, is much less explored. In the limit where the medium can be treated as a dilute gas of nucleons with density $n_B$, the axion potential has been shown to become shallower as the nucleon density increases, following the equation~\cite{Hook:2017psm}

\begin{equation}\label{eq:madensity}
    m_{a, \rm eff}^2 (n_B, n_I) 
    %\simeq m_a^2 - \frac{\sigma_p n_p}{f_a^2} - \frac{\sigma_n n_n}{f_a^2} 
    =m_a^2\left(1 - \frac{1}{\varepsilon}\frac{\sigma_N  n_B}{f_\pi^2m_\pi^2} \left( 1-\frac{n_I}{n_B}
\frac{\Delta \sigma }{\sigma_N }\right) + \mathcal{O}\left[\left(\frac{\sigma_N n_B}{m_{\pi}^2 f_{\pi}^2} \right)^2\right]\right) \,,
\end{equation} 
where baryon density $n_B = n_p+n_n$ and  $n_I = n_p-n_n$ is the isospin density, with proton and neutron density $n_p$ and $n_n$, respectively. The critical density for a phase transition to axion condensed matter is given by
\begin{equation}
\label{eq:crit density}
n^c_B= \frac{\varepsilon m_{\pi}^2 f_{\pi}^2}{\sigma_N}\simeq 2.65 \nsat \varepsilon \left(\frac{50~\rm MeV}{\sigma_N}\right) \, . 
\end{equation} 
For $\varepsilon =1$, the critical density determined from Eq.~\eqref{eq:crit density} is sufficiently large that the expansion in $\sigma_N n_B / m_\pi^2 f_\pi^2$ breaks down near the critical density, and it is conceivable that axion condensation does not occur at any density. We will return to this question in Sec.~\ref{sec:ChiPTMBPT}. However, for $\varepsilon < 0.4$, Eq.~\eqref{eq:madensity} predicts axion condensation at $n_B \lesssim \nsat$. At these moderate densities where matter can be described reliably using nuclear physics, one can conclude that the axion potential is minimized at $\theta = \pi$. If the nuclear force is significantly modified by axion condensation, $\theta = \pi$ may be favored below the saturation density even for $\varepsilon \geq 0.4$, which we will discuss in more detail in Sec.~\ref{sec:nuclear matter} and Sec.~\ref{ssec:domain_wall}.

An interesting consequence of Eq.~\eqref{eq:crit density} is that, for small enough $\varepsilon$, the ground state of nuclear matter is $\theta = \pi$ and ordinary nuclei are only stabilized by the fact that the axion field changes on a length scale of $1/m_a$, much larger than the size of a typical nucleus. At finite nucleon density, it is more convenient to analyze the transition to the axion condensed phase as a function of the baryon chemical potential $\mu_B$. The phase transition will occur at a critical chemical potential $\mu_B^c$ where the free energy $\Omega(\mu_B^c, \theta = 0) = \Omega(\mu_B^c, \theta = \pi)$. Neglecting interactions between nuclei, the free energy of dilute non-relativistic matter at zero temperature and finite $\mu_B$ is given by, apart from gradient terms,
\begin{equation}
\label{eq:OFFG}
    \Omega(\mu_B,\theta) = -P(\mu_B,\theta)=-\frac{k^5_F(\theta)}{10 \pi^2 m_n(\theta)} + V(\theta)\,,
\end{equation}
where $P(\mu_B,\theta)$ is the pressure, $k_F(\theta) = \sqrt{\mu_B^2-m^2_n(\theta)}$ and the neutron mass $m_n(\theta)$ is given in Eq.~\eqref{eq:mnoftheta}. From Eq.~\eqref{eq:OFFG} we see that axion contribution to the pressure is negative, and for $\theta = \pi$, $P_{\rm axion}=-V(\theta=\pi)=\varepsilon f_\pi^2 [m_\pi^2(\theta=\pi) - m_\pi^2(\theta = 0)]$. The importance of this negative pressure was first pointed out in Ref.~\cite{Csaki:2018fls,Balkin:2022qer} where it was noted that 
for white dwarfs, which are stabilized by electron degeneracy pressure, this negative pressure would destabilize a large number of observed white dwarfs in the universe. This consideration leads to the strongest constraint on the axion parameter space for $\varepsilon< 2\times 10^{-7}$~\cite{Balkin:2022qer}. A neutron star, on the other hand, is much denser and balanced by the much larger neutron degeneracy pressure. It is unlikely that this negative pressure can lead to the collapse of a neutron star into a black hole, but it might significantly change the EOS of a neutron star~\cite{Balkin:2023xtr}.

\begin{figure}[tb]
    \centering
    \includegraphics[width=0.44\linewidth]{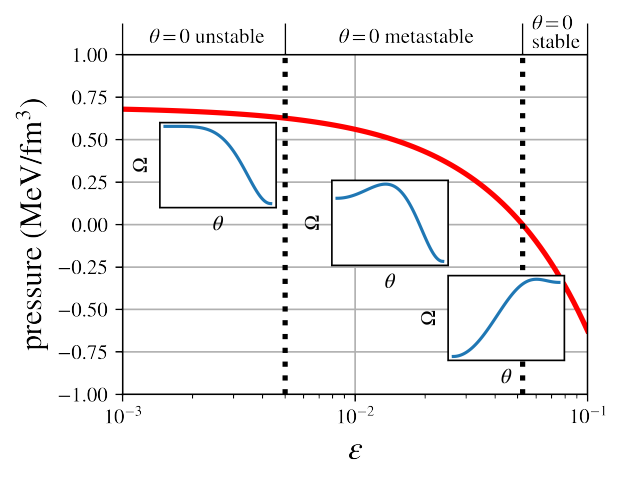}
    \includegraphics[width=0.42\linewidth]{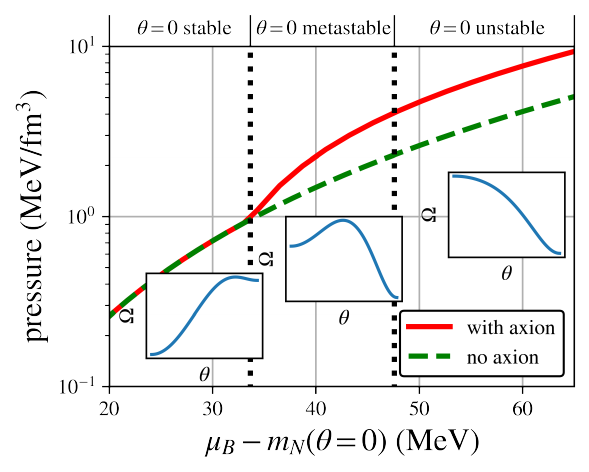}
    \caption{Left Panel: Pressure of axion condensed PNM at $\mu_B = m_n (\theta = 0)$ in the absence of nuclear interactions. Insets show free energy as a function of $\theta$ in each region. Right Panel: Phase diagram of a free neutron gas at $\varepsilon =0.3$. }
    \label{fig:nmstable}
\end{figure}
We begin by analyzing the effects of axions on a dilute gas of neutrons. The left panel of Fig.~\ref{fig:nmstable} shows the pressure of a free Fermi gas of neutrons at $\theta = \pi$ and $\mu_B = m_n (\theta = 0)$. The insets show the $\theta$-dependence of the free energy in each of the regions depicted. When the pressure is positive, axion condensation will be favored for dilute neutron matter. In the absence of nuclear interactions, dilute neutron matter with $\theta=0$ is metastable for $0.005 \lesssim \varepsilon \lesssim 0.065 $ and unstable when $\varepsilon \lesssim  0.005$. For $\varepsilon \gtrsim 0.065 $, the first-order transition to the $\theta=\pi$ phase occurs at $\mu_B > m_n$ and low-density neutron matter with $\theta=0$ is stable. In this case, the phase transition to axion condensed matter will occur at finite density. On the other hand, if $\varepsilon \lesssim 0.065$, the phase transition will occur at $\mu_B < m_n$, and bulk matter in its ground state will always have $\theta > \theta_c$, with $m_n(\theta_c) = \mu_B$. In the case where $2 \times 10^{-7} < \varepsilon \lesssim 0.005$, normal matter is unstable, and ordinary stars, planets, and white dwarfs are stabilized by Coulomb forces between atoms and nuclei. For $\varepsilon > 2 \times 10^{-7}$, a system of atoms or nuclei at fixed chemical potential must significantly increase its density to nucleate the $\theta = \pi$ phase, and an electrical barrier must be overcome to bring the nuclei closer together. For smaller $\varepsilon$, the required increase in density is more modest, and the constraint on axions from the stability of white dwarfs, stars, and planets applies.

% \begin{figure}[tb]
%     \centering
%     \includegraphics[width=0.45\linewidth]{p_of_mub_ffg.pdf}
%     \caption{Phase diagram of a free neutron gas at $\varepsilon =0.3$.}
%     \label{fig:PDFFG}
% \end{figure}
For a given $\varepsilon$, if $\mu_B$ inside the star exceeds the critical chemical potential and it is possible to nucleate the $\theta=\pi$ phase, axions would condense in the inner region, separated by a domain wall with thickness $\Delta r_{\rm DW} \approx 1/m_a $ from the  exterior $\theta=0$ phase. In the Fermi gas model, for  $\varepsilon \gtrsim 0.065$, the neutron star would consist of three regions: a normal outer region with $\theta=0$, a metastable region which could, in principle, exist in the $\theta=0$ phase if the nucleation timescale was long, and inner core characterized by $\theta=\pi$. In the right panel of Fig.~\ref{fig:nmstable}, we plot pressure as a function of $\mu_B$ for $\varepsilon =0.3$. It consists of the three regions just mentioned, and the insets show the $\theta$ dependence of the free energy, which delineates the stable, metastable, and unstable regions.

\subsection{Axion field profile interpolating between $\theta =\pi$ and $\theta =0$}\label{sec:axion dilute neutron}

The axion field profile across the domain wall, which minimizes the energy, is  obtained by solving 
\begin{equation}
f_a^2 \nabla^2\theta = \frac{\dd\Omega(\theta,\mu^c_B)}{\dd\theta}\,,
\end{equation}
where $\mu^c_B$ is the critical baryon chemical potential at which the transition occurs. This equation, though used here for a dilute gas of nucleons, remains exact when nuclear interaction is included in $\Omega(\theta,\mu^c_B)$.
Such an axion profile will develop as long as the object that contains this dilute gas of nucleons is both dense enough (see Eq.~\eqref{eq:crit density}) and its radius, $ r_{\rm NS}$, large enough such that the bulk of the star can be subsumed within a domain wall of thickness
\begin{equation}\label{eq:conradius}
    \Delta r_{\rm DW} \simeq \frac{1}{|m_{a,\rm eff} (n_B)|} \leq r_{\rm NS}. 
\end{equation}
We will focus on the condition on the density in Eq.~\eqref{eq:crit density} for now and return to the effects of the finite size of stellar objects in Sec.~\ref{sec:piballs}. For axions with $m_a < 10^{-11} \,{\rm eV}$, the axion profile can be treated as situated outside the neutron star. In binary systems, this field profile results in  axion-mediated forces and axion radiation, which lead to the most stringent constraint from LVK observations of neutron star binaries~\cite{Zhang:2021mks}. On the other hand, when $m_a \gg 10^{-11} \,{\rm eV}$, the axion field profile can reside inside the neutron star, separating regions with different densities and pressure. Given the prior constraints on the exceptionally light QCD axion parameter space, we are mostly interested in the region of parameter space where $|m_a^2 (n_B)| > m_a^2 > 1/r_{\rm NS}^2$. 

For these larger axion masses, the thickness of the region in which $\theta$ changes from $\pm \pi$ to 0 is much smaller than the radius of a neutron star. On the other hand,  constraints on QCD axions from SN cooling imply that the axion mass is below the tens of meV scale~\cite{Springmann:2024ret}, i.e. the domain wall thickness is much larger than the relevant microphysical scales: the size of a nucleon, typical separation between nucleons, Debye screening length, and mean free path of neutrons and protons in a neutron star. As a result, we can treat the axion field profile in the parameter space of interest as smooth when considering interactions between the nucleons, while at the same time, studying the properties of the gas of nucleons in domains where $\theta = \pm \pi$ and $\theta = 0$ separately, and glue them together with a boundary condition at the location of a thin axion ``domain wall.''

Unlike a normal neutron star where density and pressure are continuous, with an axion field profile, both the density and pressure are no longer continuous when adopting the approximation that the domain wall is infinitely thin. In addition to the negative pressure from the condensation of the axion field itself, the domain wall also exerts a force inward due to domain wall tension of order $m_a f_a^2/r_{\rm NS}$. For $m_a \gg 1/r_{\rm NS}$, this additional pressure is parametrically smaller in the thin wall limit in the model in Eq.~\eqref{eq:axionpotential}. We return to discuss the importance of domain wall pressure in the \ZN model in Sec.~\ref{ssec:domain_wall}.

\section{Neutron and nuclear matter at $\theta=\pi$}
\label{sec:nuclear matter}

In this section, we describe our efforts to model the properties of dense nuclear matter at $\theta = \pm \pi$, taking into account changes to the strength of nuclear interactions. Despite developments in Effective Field Theory (EFT) methods to describe nuclear interactions, the connection between fundamental parameters that appear in the QCD Lagrangian (the quark mass matrix ${\cal M}_q$ and $\theta$) and the parameters needed to describe interactions between nucleons remains elusive. This is because the quark mass dependence of the low-energy constants (LECs) in the EFT that describe the strength of interactions at short distances is poorly known. Nevertheless, earlier studies have provided useful insights to address the question of the quark mass dependence of the nucleon-nucleon interaction. 

We first briefly summarize earlier work in Refs.~\cite{Beane:2002vs,Epelbaum:2002gb,Berengut:2013nh} which are based on nuclear potentials derived from \ChiEFT as they are better suited to address how nuclear interactions are altered by changes to the quark mass matrix. To assess the quark mass dependence of the nuclear interactions at momentum scales of relevance to the large nuclei and neutron-rich matter encountered in neutron stars, we next compare results from \ChiEFT and the three scenarios in RMFT which we will use to extract phenomenological results, summarized in Table~\ref{tab:Ksigma}.

In \ChiEFT, nuclear forces are derived by explicitly including the contribution from pion-exchange as well as the short-distances components, which are included through a systematic expansion of operators of increasing dimension constructed from nucleon fields consistent with the symmetries of low-energy QCD. Such a low energy EFT description is expected to be applicable for $n_B \lesssim 2n_{\rm sat}$, and is described in Sec~\ref{sec:ChiPTMBPT}. We employ \ChiEFT potentials at order N$^2$LO and perform calculations of the energy per particle in pure neutron matter (PNM) using many-body perturbation theory (MBPT). In this case, the quark mass dependence is taken into account through its modification of the pion mass, nucleon masses, the axial coupling constant $g_A$, and the pion decay constant$f_\pi$.

RMFT is based on the assumption that scalar and vector interactions provide the dominant contributions to the interaction energy of baryons in large systems. In this phenomenological model, the interaction between nucleons is mediated by the exchange of scalar and vector mesons. The model and the quark mass dependence of the model parameters are described in Sec.~\ref{sec:RMF} (see table~\ref{tab:Ksigma} for a summary). While \ChiEFT is a microscopic theory in which potentials are constrained by nucleon-nucleon scattering data and properties of light nuclei, RMFT is a phenomenological model designed to provide a good description of large nuclei. RMFT also provides a convenient extrapolation to high density and permits efficient calculation of heterogeneous crust phases. In RMFT Scenario A, the only modification we make to nuclear physics is to modify the nucleon masses according to Eq.~\eqref{eq:mnoftheta} since the pion does not explicitly appear in RMFT. In Scenario B, we modify the meson masses with $\theta$ dependence given by calculations in Refs.~\cite{HanhartPelaezRios2008,Pelaez:2010sigmamass,Berengut:2013nh}. In Scenario C, we consider a more extreme modification to the RMFT based on the results around saturation density given by \ChiEFT.

\begin{table}[tb]
\renewcommand{\arraystretch}{1.2}
    \setlength{\tabcolsep}{12pt}
    \caption{Summary of the three  scenarios used in this work as benchmarks of nuclear force modification in the $\theta=\pi$ region. Parametrized by  the logarithmic derivative evaluated at the physical quark mass $K_\sigma = \dd \ln{m_\sigma}/\dd\ln{m_q}$. See Section~\ref{sec:RMF} for discussion.}
    \centering
    \begin{ruledtabular}
    \begin{tabular}{ccl}
    % \hline
         {\rm Scenario} & $K_\sigma$ & interpretation \\
         \hline 
        A & $0$  &  Nuclear forces are not modified\\
         % \hline
        B & $0.08$ & Corresponds to an $m_\sigma$ decrease of $\sim6\% $ as found in Refs.~\cite{HanhartPelaezRios2008, Pelaez:2010sigmamass, Berengut:2013nh}     \\
         % \hline 
        C & $0.16$ & Matches the largest binding of neutron matter found using \ChiEFT. See Figs.~\ref{fig:ChiEFT_Eofmpi}~\&~\ref{fig:rmf epb}. \\
         % \hline
    \end{tabular}
    \end{ruledtabular}
    \label{tab:Ksigma}
\end{table}

As noted earlier, in the low energy theory, finite $\theta$ manifests as a reduction of the pion mass as in Eq.~\eqref{eq:pionmasstheta}. In what follows, we shall assume that the effect of $\theta=\pi$ can be adequately characterized by the reduction in the pion mass and calculate all other parameters of relevance to the two-nucleon interaction using ChiPT. 

\subsection{Pion mass dependence of two-nucleon observables}

By explicitly accounting for the $m_\pi$ dependence of the parameters that describe the nucleon-nucleon (NN) interaction in \ChiEFT, one can calculate the quark mass dependence of low-energy NN observables such as the NN scattering phase shifts and the deuteron binding energy. In \ChiEFT, at leading order (LO) in the momentum expansion, the NN potential is characterized by long-range one-pion exchange (OPE) and short-range NN contact interactions. While the $m_\pi$ dependence of the parameters associated with the OPE potential is well understood, the $m_\pi$ dependence of the NN contact interaction, which is typically called $D_2$, is not well determined and the resulting uncertainty in the scattering length and scattering phase shifts have been studied in earlier work \cite{Beane:2002vs,Epelbaum:2002gb}. 

In Ref.~\cite{Epelbaum:2002gb}, the authors find that when the quark mass is reduced, the binding energy of the deuteron increases, and the spin-singlet and spin-triplet scattering lengths decrease. To quantify the quark mass dependence of an observable $\cal O$, the authors define the logarithmic derivative at the physical point 
\begin{equation}
K_{\cal O} = \frac{m_q}{\cal O}   ~\left(\frac{\dd \cal O }{\dd m_q}\right)_{\rm phys}\simeq \left(\frac{m^2_\pi}{\cal O}   ~\frac{\dd \cal O }{\dd m^2_\pi}\right)_{m_\pi=m^{\rm phys}_\pi}\,,
\label{eq:KO}
\end{equation}
since at leading order in ChiPT,  $m_\pi^2 \propto m_q $. The $m_q$ dependence of the $^1S_0$ and $^3S_1$ scattering lengths and the deuteron binding energies are captured by the values of $K_{a1S0}$, $K_{a3S1}$, and $K_{\rm deut}$, and their values obtained in earlier studies are shown in Table~\ref{tab:Kqs}. Results shown in the first row were obtained using potentials derived from \ChiEFT at next-to-leading order (NLO) in Weinberg power counting~\cite{Epelbaum:2002gb} and NLO results obtained in Ref.~\cite{Beane:2002vs} using Kaplan-Savage-Wise power counting~\cite{Kaplan:1998tg} are shown in the second row. In this case, the uncertainty is dominated by the poorly known LEC, $D_2$, which was mentioned earlier. Results obtained more recently in Ref.~\cite{Berengut:2013nh} are shown in the third row; here, the authors employ strategies to go beyond the NLO approach adopted in Refs.~\cite{Beane:2002vs,Epelbaum:2002gb} and use information from resonance saturation models to constrain the quark mass dependence of the short-range contact operators and incorporate N$^2$LO corrections to $g_A$, $f_\pi$, and the nucleon mass.   

\begin{table}[tb]
\renewcommand{\arraystretch}{1.2}
    \setlength{\tabcolsep}{15pt}
    \caption{Quark mass dependence of the scattering lengths and the deuteron binding energy $K_{\cal O}$, as defined in Eq.~\eqref{eq:KO}.}
    \centering
    \begin{ruledtabular}
    \begin{tabular}{lccc}
    % \hline
         {\rm Reference} & $K_{a1S0}$ &  $K_{a3S1}$ & $K_{\rm deut}$ \\
         \hline 
         Epelbaum et al.~\cite{Epelbaum:2002gb} & $5 \pm 5$  & $1.1 \pm 0.6$ & $-2.8 \pm 1.2$\\
         % \hline
         Beane et al.~\cite{Beane:2002vs} & $2.4 \pm 3.0$ & $3.0 \pm 3.5$  & $-7 \pm 6$  \\
         % \hline 
         Berengut et al.~\cite{Berengut:2013nh} & $2.3^{+1.6}_{-1.5}$ & $0.32^{+0.17}_{-0.18}$  &  $-0.86^{+0.45}_{-0.50}$\\
         % \hline
    \end{tabular}
    \end{ruledtabular}
    \label{tab:Kqs}
\end{table}

With decreasing quark mass, these earlier studies found the $^1S_0$ phase shifts were reduced and the $^3S_1$ phase shifts were enhanced. Thus, in dilute neutron matter where spin singlet interactions dominate, the interaction energy will be less attractive when s-wave interactions dominate. In dilute nuclear matter with an equal number of protons and neutrons (symmetric nuclear matter), where spin triplet interactions dominate, the s-wave interaction energy is larger and more attractive when the quark mass is reduced.  However, when the pion becomes lighter and of longer range, higher partial wave contributions are in general enhanced \cite{Bulgac:1997ji}. The attraction in the $^3P_0$ and  $^3P_2$ channels is enhanced and the repulsion in the $^1P_1$ channel is reduced \cite{Epelbaum:2002gb}. For these reasons, the analysis of the scattering lengths and the deuteron binding energy is insufficient to inform about the behavior of matter at densities of relevance to nuclei and neutron stars where contributions due to higher partial waves and higher-order quark mass-dependent short-distance operators can contribute. 

\subsection{Axion condensation in Chiral Effective Field Theory}
\label{sec:ChiPTMBPT}
To address how nuclear interactions influence axion condensation, we first make the simple observation that axion condensation is favored when the interaction energy per particle decreases with pion mass. From the discussion in Sec.~\ref{sec:dilute}, we infer that in this case, condensation would occur at $n_B \lesssim 2.6 \nsat$ for $\varepsilon =1$. In \ChiEFT, the $m_\pi$ dependence of the long- and intermediate-range parts of the nucleon-nucleon and three-nucleon (3N) potentials is explicit because pion-exchanges are systematically included in Weinberg power counting. The $m_\pi$ dependence of the nuclear interaction energy in PNM and symmetric nuclear matter (SNM) has been studied in the context of addressing how $\langle \bar q q \rangle$, the chiral condensate, evolves at finite density \cite{Cohen:1991nk, Kaiser:2007nv, Plohl:2007dp, Kruger:2013iza}. In these studies, the Feynman-Hellmann theorem was exploited to relate the change in   $\langle \bar q q \rangle$ to the derivative of the interaction energy with respect to $m_\pi^2$. Explicitly, one finds that   
\begin{equation}
\frac{n_B}{f_\pi^2} \left(\frac{\dd E_{\rm int}(n_B)}{\dd m_\pi^2} \right)_{m_\pi =m^{\rm phys}_\pi}  = R_0(n_B) - R(n_B) \,,
\end{equation} 
where $E_{\rm int}(n_B)$ is the interaction energy per nucleon in nuclear matter, $R(n_B)=\langle \bar q q \rangle_{n_B}/\langle \bar q q \rangle_0$ is the ratio of the chiral condensate at a finite density to that in vacuum, and $R_0(n_B)$ is the same ratio calculated by neglecting interactions. 

\ChiEFT-based calculations reported in Refs.~\cite{Kaiser:2007nv,Plohl:2007dp,Kruger:2013iza} find that   $R(n_B)/R_0(n_B) > 1$ over the entire range of densities where the \ChiEFT is expected to be useful. Thus, the interaction energy increases with decreasing pion mass in the vicinity of $m_\pi =m^{\rm phys}_\pi$ in SNM and PNM. However, the study in Ref.~\cite{Kaiser:2007nv}, which was restricted to the case of SNM, found that the ratio $R(n_B)/R_0(n_B) < 1$ for smaller $m_\pi$. Their calculations indicate that $R_0(\nsat) - R(\nsat)\simeq 0.1$ at $m_\pi=70$ MeV and $R_0(\nsat) - R(\nsat)\simeq 0.3$ in the chiral limit.  Together, these results imply that the interaction energy increases with decreasing $m_\pi$ for $m_\pi \simeq m^{\rm phys}_\pi$ and then decreases rapidly for smaller values of $m_\pi$. The microscopic calculations of the energy per particle of PNM as a function of $m_\pi$ that we shall describe in the following show a similar trend. 

%%%%%%%%%%%%%%%%%%%%%%%%%%%%%%%%%%%%%%%%%%%%%%%%
\begin{figure}[tb]
    \centering
\includegraphics[]{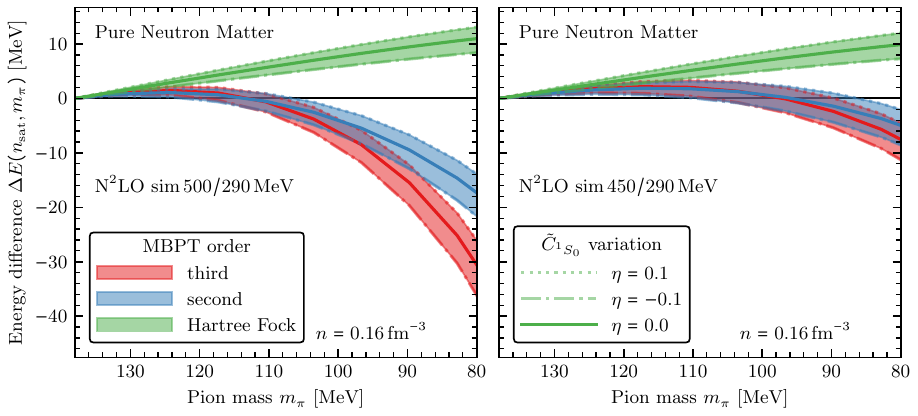}
    \caption{Difference in the energy per particle~\eqref{eq:energy_diff} for the N$^2$LO sim potentials~\cite{Carlsson:2015vda} with two momentum cutoffs $\Lambda$ at $n_B=\nsat$ as a function of $m_\pi$: $\Lambda = 500$~MeV (left panel) and $\Lambda = 450$~MeV (right panel), as annotated.
    In these potentials, we modify $m_\pi$, $f_\pi$, and the LO  LEC $\tilde{C}_{{}^1S_0}$ (different line styles), whereas $g_A$ is kept at its physical value. 
    The bands depict the variation of the LO LEC $\tilde{C}_{{}^1S_0}$ according to Eq.~\eqref{eq:C1S0_mpi}.
    The results were obtained at the Hartree-Fock level (green bands) and using second-order (blue bands) and third-order MBPT (red bands). 
    The energy per particle at the physical pion mass obtained from on our third-order calculations is $E(n_0) = 16.94 \MeV$  and $E(n_0) = 17.21 \MeV$ for $\Lambda = 450 \MeV$ and $\Lambda = 500 \MeV$, respectively.
    See the main text for details.}
    \label{fig:ChiEFT_Eofmpi}
\end{figure}
% Christian, please describe the calculation and the curves in Figure.  
%%%%%%%%%%%%%%%%%%%%%%%%%%%%%%%%%%%%%%%%%%%%%%%%
% \SR{I am wondering if it would suffice to only focus on PNM in this paper.}

Figure~\ref{fig:ChiEFT_Eofmpi} shows the difference in the PNM energies per particle,
\begin{equation} \label{eq:energy_diff}
    \Delta E(\nsat, m_\pi) = E(\nsat, m_\pi) - E(\nsat, m^{\rm phys}_\pi),
\end{equation}
for two chiral potentials at $n_B=\nsat$ as a function of $m_\pi$ (see annotations). %\SR{ Can we replace $E/A$ by $E$ and define $E$ to be the energy per particle?}
Here, we use the family of bare N$^2$LO potentials with chiral NN and 3N forces developed in Ref.~\cite{Carlsson:2015vda}.
These nonlocal potentials are called the sim potentials because they were simultaneously fit to scattering and bound-state observables in the pion-nucleon, NN, and few-nucleon sectors.
Specifically, we use the two N$^2$LO potentials with the NN and 3N momentum cutoffs $\Lambda = 500$ (left panel) and $450$~MeV (right panel); both constrained by NN scattering data at laboratory energies $T_\mathrm{lab} \leqslant 290$~MeV~\cite{Carlsson:2015vda}.
The analytic expressions that define these potentials and were used in this work are summarized in the Appendix of Ref.~\cite{Entem:2014msa} and Section~5.1.2 in Ref.~\cite{Machleidt:2011zz}.
To obtain the PNM energy per particle, we apply the Monte Carlo framework for MBPT nuclear matter calculations developed in Ref.~\cite{Drischler:2017wtt}. 
The green bands in Fig.~\ref{fig:ChiEFT_Eofmpi} depict the Hartree-Fock reference energy and the blue and red bands the MBPT results at second and third order, respectively.
All MBPT calculations are performed at the normal-ordered two-body level.
More details on MBPT for nuclear matter can be found in Ref.~\cite{Drischler:2021kxf}.
This MBPT framework uses the operatorial definition of the chiral potentials, allowing us to straightforwardly vary the values of physical constants, such as the pion mass, low-energy couplings, and more.
Not all chiral potentials are given in their operatorial form; e.g., when evolved to lower resolution scales using renormalization group (RG) methods, their operatorial definition is usually unknown.
For the solid lines in Fig.~\ref{fig:ChiEFT_Eofmpi}, we keep the values of the low-energy couplings at their best-fit values determined in Ref.~\cite{Carlsson:2015vda} and vary only the pion mass, nucleon mass, and the pion decay constant. The $m_\pi$ dependence of the nucleon mass was discussed in Sec.~\ref{sec:masses}, and the $m_\pi$ dependence of the pion decay constant is given by    
\begin{equation}
    f_\pi = f \left[ 1 + \frac{m_\pi^2}{(16 \pi^2 f^2)} \bar{l}_4  \right] \,,
\end{equation}
with its value in the chiral limit $f = 86.2$~MeV and the LEC $\bar{l}_4 = 4.3$~\cite{Berengut:2013nh}.
The pion-mass dependence of the axial coupling constant $g_A$ is relatively weak in the regime considered here, as can be seen in Figure~1 in Ref.~\cite{Berengut:2013nh}, and thus we keep $g_A = 1.29$ at the physical value.
We use these potentials to identify overall trends and emphasize the need to explore the pion-mass dependence of a wider range of chiral potentials in the future.

% %%%%%%%%%%%%%%%%%%%%%%%%%%%%%%%%%%%%%%%%%%%%%%%%
% \begin{figure}[tb]
%     \centering
% \includegraphics[]{ChiEFT_Eofmpi_450_eta.pdf}
%     \caption{Similar to Fig.~\ref{fig:ChiEFT_Eofmpi} but for the $\Lambda = 450$~MeV potential and probing the $\eta$-dependence.
%     See the main text for details.}
%     \label{fig:ChiEFT_Eofmpi_eta}
% \end{figure}
% % Christian, please describe the calculation and the curves in Figure.  
% %%%%%%%%%%%%%%%%%%%%%%%%%%%%%%%%%%%%%%%%%%%%%%%%

As noted earlier, the pion-mass dependence of the short-range components of the interaction that are incorporated through scale-dependent LECs is not well understood or constrained by data. 
To study the sensitivity of our results to variations in the short-range interaction, we have calculated the energy per particle by changing the LEC $\tilde{C}_{{}^1S_0}$ that governs the strength of the NN contact interaction at LO in the ${}^1S_0$ partial-wave channels. 
Specifically, we perform calculations for the energy per particle with a scaled value for this coupling constant,
\begin{equation} \label{eq:C1S0_mpi}
    \tilde{C}_{{}^1S_0}(\theta) = \tilde{C}_{{}^1S_0}^{\rm fit} \left( \frac{1 + \eta f(\theta)}{1+ \eta}\right) \,,
\end{equation}
where $f(\theta)=\left(m_\pi(\theta)/m^{\rm phys}_\pi\right)^2$, $\eta = D_2 (m^{\rm phys}_\pi)^2/\tilde{C}_{{}^1S_0}^{\rm fit} $ and $\tilde{C}_{{}^1S_0}^{\rm fit}$ is the best-fit value in each isospin channel obtained in Ref.~\cite{Carlsson:2015vda}. Here, $D_2$ is the LEC discussed in Ref.~\cite{Kaplan:1998tg,Beane:2002vs} that captures the pion-mass dependence at NLO. 
We choose the two values $\eta = \pm 0.1$ to set the limits of this variation and adopt $m^{\rm phys}_\pi = 138.04$~MeV for the physical pion mass.
The dotted and dash-dotted lines in Fig.~\ref{fig:ChiEFT_Eofmpi} depict the results for the PNM energy per particle with the modified LO coupling constant~\eqref{eq:C1S0_mpi}; the solid lines correspond to $\eta = 0$ and thus $ \tilde{C}_{{}^1S_0}(\theta) \equiv \tilde{C}_{{}^1S_0}$.
Increasing the strength of $\tilde{C}_{{}^1S_0}(\theta)$, corresponding to $\eta = -0.1$, reduces $\Delta E(\nsat, m_\pi)$, whereas decreasing the strength, corresponding to $\eta = +0.1$, increases $\Delta E(\nsat, m_\pi)$.

% \begin{figure}[ht]
% \begin{center}
%     \includegraphics[width=0.67\textwidth]{xeft_epb_by_order.pdf}
%     \caption{Kinetic and nuclear interaction energy per baryon for SNM and PNM in \ChiEFT order by order in MBPT for $\theta = \pi$.}
%     \label{fig:xeft epb}
% \end{center}
% \end{figure}

Several important insights can be drawn from the results in Fig.~\ref{fig:ChiEFT_Eofmpi}. 
MBPT appears to converge more slowly with decreasing pion mass, which could be further investigated using a Weinberg eigenvalue analysis~\cite{Hoppe:2017lok}. Although the leading Hartree-Fock energies indicate a systematic increase in interaction energy with $m_\pi$ across the range shown in Fig.~\ref{fig:ChiEFT_Eofmpi}, the second-order correction reverses this trend. The third-order correction is modest and of the same sign as the second-order correction. For relatively small reductions of the pion mass ($m_\pi \gtrsim 120$~MeV for $\Lambda = 500$~MeV and $m_\pi \gtrsim 110$~MeV for $\Lambda = 450$~MeV), the energy difference~\eqref{eq:energy_diff} is positive.
The sensitivity of the results at the second and third order to the cutoff variation is larger than that of the LEC variation.
These general trends are not altered by varying the momentum cutoff over the range shown in Fig.~\ref{fig:ChiEFT_Eofmpi}. 
However, the rather large dependence on the momentum cutoff is disconcerting and warrants a careful inclusion of the $m_\pi$ dependence of the short-range components of the force. As discussed earlier, we have studied the effect of the $m_\pi$ dependence of the leading LEC associated with scattering in the $^1S_0$ channel shows that the qualitative behavior is not changed.  Based on these trends, we conclude that the interaction energy at $\theta=\pi$, where $m_\pi \approx 80$ MeV, can be smaller than that at $\theta=0$, where $m_\pi \approx 138$ MeV. However, the energy difference $\Delta E$ remains quite sensitive to the regularization scheme.   

Constructing a neutron star EOS with $\theta=\pi$ using \ChiEFT poses challenges as more work is needed to include beta-equilibrium, charge neutrality, and heterogeneity in the crust. We defer this to future work and instead use the \ChiEFT results in PNM to inform the simpler RMFT model, which we shall describe in the next section. In particular, we utilize the energy per particle of PNM at $\eta = 0$ and momentum cutoff $\Lambda = 500 \, {\rm MeV}$ to fix a key parameter to account for changes to the nuclear interaction at $\theta=\pi$.

\subsection{Axion condensation in Relativistic Mean Field Theory}
\label{sec:RMF}

To derive concrete results for neutron stars, we use a model within RMFT (see Ref.~\cite{Dutra:2014rmfreview} for a review).  The Lagrangian of this theory is given by the following, omitting terms that will be discarded after taking the meson fields to have their mean field values,
\begin{equation}
    \begin{split}
        \mathcal{L}_{\rm RMF} &= \sum_{i=n, \,p} \overline{\psi}_i \left[ i \slashed{\partial} - g_\omega \omega \gamma^0  - \frac{1}{2} g_\rho \rho \gamma^0 \tau_3 - (m_i - g_\sigma \sigma)\right] \psi_i \\
        &+ \sum_{\ell = e, \, \mu} \overline{\psi}_\ell (i\slashed{\partial} - m_\ell) \psi_\ell + \frac{1}{2}(- m_\sigma^2 \sigma^2 + m_\omega^2 \omega^2 + m_\rho^2 \rho^2)+ \mathcal{L}_{\sigma \omega \rho}\,,
    \end{split} 
\end{equation}
where the nucleon-meson couplings are denoted by $g_\sigma, g_\omega$ and $g_\rho$, the meson masses by $m_\sigma,m_\omega$ and $m_\rho$, and $m_i$ are the nucleon masses. The term $\mathcal{L}_{\sigma \omega \rho}$ contains all meson-meson couplings that are fit, along with the nucleon-meson couplings to known properties of nuclear matter for the canonical choice of meson masses $m_\sigma=550$ MeV and $m_\omega=m_\rho=770$ MeV. $\tau_3$ is the third Pauli matrix in isospin space. We use the IUFSU$^*$ parameter set from Ref.~\cite{Agrawal:2012rmfmodel} because it is optimized for the calculation of finite nuclei and our study will focus on the crust.  The mean field values of the mesons and Fermi momenta of the nucleons and leptons are calculated by solving the Euler-Lagrange equations for each field and enforcing charge neutrality and $\beta$ equilibrium. 

Since mean field models do not explicitly include pion fields, incorporating the effects of a reduced pion mass on nuclear interactions presents a challenge. To address this, we model the increased attraction predicted by ChiEFT at lower pion masses by reducing the mass of the $\sigma$ meson. This approach is motivated by the fact that the two-pion exchange potential—which provides significant attraction in ChiEFT—can be effectively represented by scalar ($\sigma$) meson exchange in mean field models.

We consider three scenarios in our analysis. In Scenario A, we assume no change in the nuclear interaction, meaning the $\sigma$-meson mass is independent of $\theta$. In Scenarios B and C, we introduce a modest reduction in the $\sigma$-meson mass to mimic the decrease in energy per particle predicted by ChiEFT at $m_\pi = 82$ MeV. The specific values of $m_\sigma$ used in these scenarios are informed by studies of the $\theta$ dependence of meson masses reported in Refs.\cite{HanhartPelaezRios2008, Pelaez:2010sigmamass}.

\begin{figure}[tb] %[ht]
    \includegraphics[width=\textwidth]{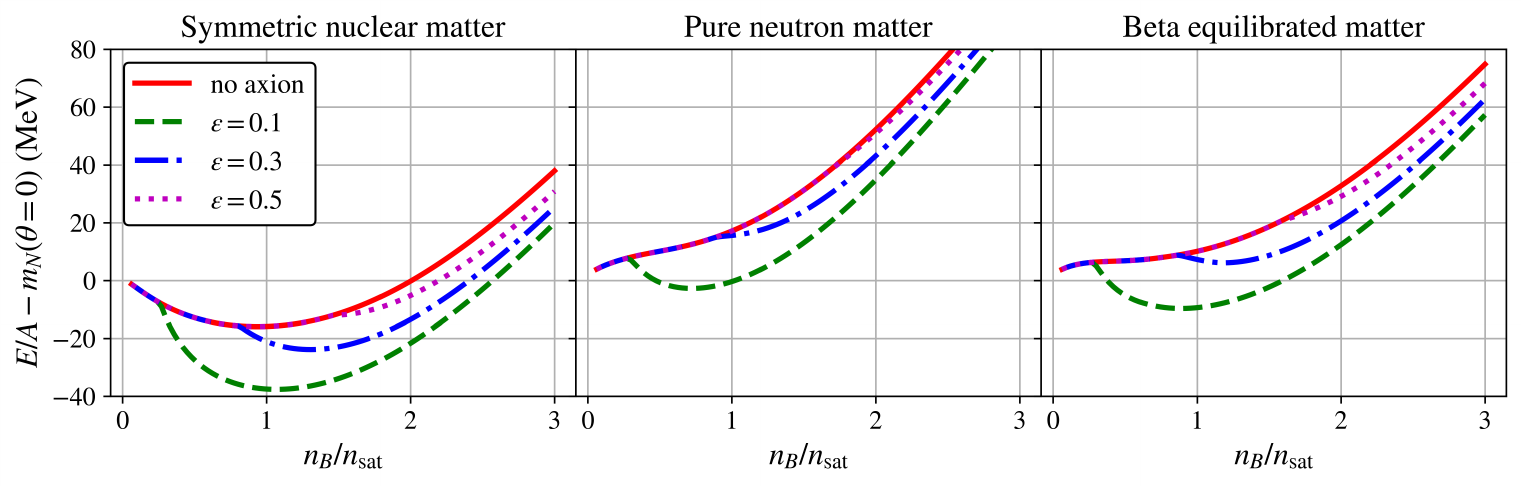}
    \caption{Energy per baryon for SNM, PNM, and $\beta$ equilibrated matter with RMFT EOS with IUFSU$^*$ parameter set including axion energy in Scenario~A.}
    \label{fig:rmf epb no d}
\end{figure}

As noted earlier,  Scenario~A is the simple case where interactions are unmodified and only the nucleon masses are modified by finite $\theta$. For this case, Fig.~\ref{fig:rmf epb no d} shows the energy per baryon, including axion energy and the reduction of the mass of the nucleons, of SNM, PNM, and beta equilibrated matter in this model. % Note that the results shown in PNM at $n / \nsat = 1$ correspond to those shown in the left panel in Fig.~\ref{fig:ChiEFT_Eofmpi}.
At fixed baryon density, the energy per baryon is minimized, leading to axion condensation due to the reduced nucleon mass. For small $\varepsilon$, the energy cost of having non-zero $\theta$ is small and axion condensation is favored at lower densities.  Since the axions themselves contribute to the energy density, the energy per baryon of axion condensed matter will always diverge in the limit of zero density and a phase without axions will be favored. This has the effect of producing a minimum in the energy per baryon of axion condensed matter when $\varepsilon$ is sufficiently small and allows a stable zero pressure phase of nuclear matter at lower energy than normal nuclei. We will return to discuss objects made out of such matter in Sec.~\ref{sec:piballs}.

The density where the energy per baryon of uniform nuclear matter has a minimum should be differentiated from a state with zero pressure \textit{not} including the axion contribution to the energy per baryon that will be found in the crust. In the crust, the length scale over which the axion field profile changes is much longer than the length scale of the lattice of nuclei. When there are no dripped neutrons, these nuclei must be in a state with zero pressure after including the surface energy of nuclear matter since the lepton and axion fields take on the same value inside and outside the nuclei. If axion condensation causes the nuclear force to become more attractive, nuclear matter can be at zero pressure at a larger isospin asymmetry, resulting in more neutron-rich nuclei.

In a neutron star, the more natural parameter to fix is the baryon chemical potential $\mu_B$. Due to the decreased nucleon mass, the value of $\mu_B$ corresponding to a system of nuclei in a lattice with $\theta=0$ corresponds to bulk matter at $\theta = \pi$. As a result, the axion domain wall never separates two heterogeneous phases. While there can be a transition from homogeneous matter with $\theta = \pi$ to a normal crust, there will be no matter outside the domain wall in any star where an axion-condensed crust phase appears. 

Of particular relevance to observations of neutron stars is the pressure at which neutron drip occurs. Neutron drip refers to the density at which it becomes energetically favorable for neutrons to come out of nuclei and form a background of free neutrons. As such, the neutron drip phase transition will occur when the baryon chemical potential is equal to the mass of the neutron (possibly modified by finite $\theta$) in the case where PNM is not self-bound. If the axion-condensed phase persists to the surface of the neutron star, but the negative axion pressure is larger than the pressure of electrons at neutron drip density (nuclei make a small contribution to the pressure), the neutron drip phase transition will not be found in the neutron star and the surface of the star will have a thin domain wall covering either a phase of nuclei in a dripped neutron background or bulk nuclear matter. 

Not included thus far is the pion mass dependence of the nuclear force itself.  As noted earlier, we  use a simple ansatz for modifications to the nuclear force, we modify the $\sigma$ mass with a tunable parameter $d_\sigma$ as follows:
\begin{equation}\label{eq:defineKsigma}
\begin{split}
    m_\sigma^2(\theta) &= m_\sigma^2(0) \frac{1 + d_\sigma f(\theta)}{1+d_\sigma} \,.
\end{split}
\end{equation}
The $\theta$-dependence of the masses of the $\sigma$ and $\rho$ mesons have been calculated in Refs.~\cite{HanhartPelaezRios2008, Pelaez:2010sigmamass}. They find that, comparing $\theta = 0$ and $\theta = \pi$, the $\sigma$ mass decreases by about 6\% and the $\rho$ mass decreases by about 2\%.  At low density, the RMFT EOS is dominated by the $\sigma$ meson and its modification is much larger than that of the vector mesons so we focus on the effect of modifying the scalar interaction.
Note that the logarithmic derivative evaluated at the physical quark mass $K_\sigma = \dd \ln{m_\sigma}/\dd\ln{m_q} = d_\sigma / (2 + 2 d_\sigma)$. Scenario A, in which the nuclear forces are not modified, corresponds to $K_\sigma=0$. For positive $K_\sigma$, the $\sigma$ mass is lower, leading to a larger mean field value of the $\sigma$ field and a more attractive nuclear force. The limit $K_\sigma = 1 / 2$ corresponds to the $\sigma$ mass having the same dependence as the $\pi$ mass. The 6\% reduction to the $\sigma$ mass found in Refs.~\cite{HanhartPelaezRios2008, Pelaez:2010sigmamass} is reached for $K_\sigma = 0.08$. This is the value we adopt in Scenario B. We also note that Ref.~\cite{Berengut:2013nh} found that $K_\sigma = 0.081\pm 0.007$. For Scenario C, we consider a more extreme modification to the nuclear force with $K_\sigma = 0.16$, which achieves the same binding for neutron matter found in our most attractive estimate from \ChiEFT\ found for cutoff $\Lambda=500 \MeV$. These three scenarios together span the parameter space supported by our \ChiEFT calculation.  We will refer to our three scenarios based on the value of $K_\sigma$ evaluated at the physical point, but the actual $\theta$-dependence is given by Eq.~\eqref{eq:defineKsigma} and $K_\sigma$ is not a constant when $\theta$ is varied. Scenarios B and C are tuned to the properties of axion condensed matter at $\theta = \pi$; $K_\sigma$ at the physical point is merely a convenient benchmark. The leading modification to the effective axion mass from this modification to the nuclear force is given by
\begin{equation}
    m^2_{a, \rm eff} (n_B, n_I) = m_a^2 \left[ 1 - \frac{1}{\varepsilon} \frac{\sigma_N n_B}{f_\pi^2 m_\pi^2} \left(1 + \frac{K_\sigma g_\sigma^2 n_B}{\sigma_N \chi m_\sigma^2} - \frac{n_I}{n_B} \frac{\Delta \sigma}{\sigma_N} \right) \right]
\end{equation}
where $g_\sigma$ is a parameter of the RMFT model and $\chi$ is the volume fraction of nuclei when in a heterogeneous phase and unity in a homogeneous phase. At saturation density, $g_\sigma^2 n_{\rm sat}/\sigma_N m_\sigma^2 \simeq 11$ for the RMFT model we use so for Scenario B, axion condensation should be expected below $n_{\rm sat}$ for $\varepsilon \lesssim 0.8$ and in Scenario C, axion condensation should be expected below $n_{\rm sat}$, even for $\varepsilon = 1$! While this is a striking prediction, it should be noted that RMFT is a model for the nuclear interaction that does not have explicit pions and we are parameterizing the expected effects of a reduced pion mass via the meson masses and where Scenario C is chosen to recreate the most extreme prediction of \ChiEFT. While the $\sigma$ meson is typically understood as a two pion resonance, this comparison is certainly incomplete. More detailed study is required to determine the fate of neutron matter at saturation density. The predictive power of this model is additionally limited by the fact that we only consider isoscalar effects on the nuclear force. Axion condensation may also have an important impact on the symmetry energy of nuclear matter beyond the effect of $\Delta \sigma$. We expect this to have a more pronounced effect at high density where our model of the nucleon-axion interaction breaks down.

\begin{figure}[tb] %[ht]
    \includegraphics[width=\textwidth]{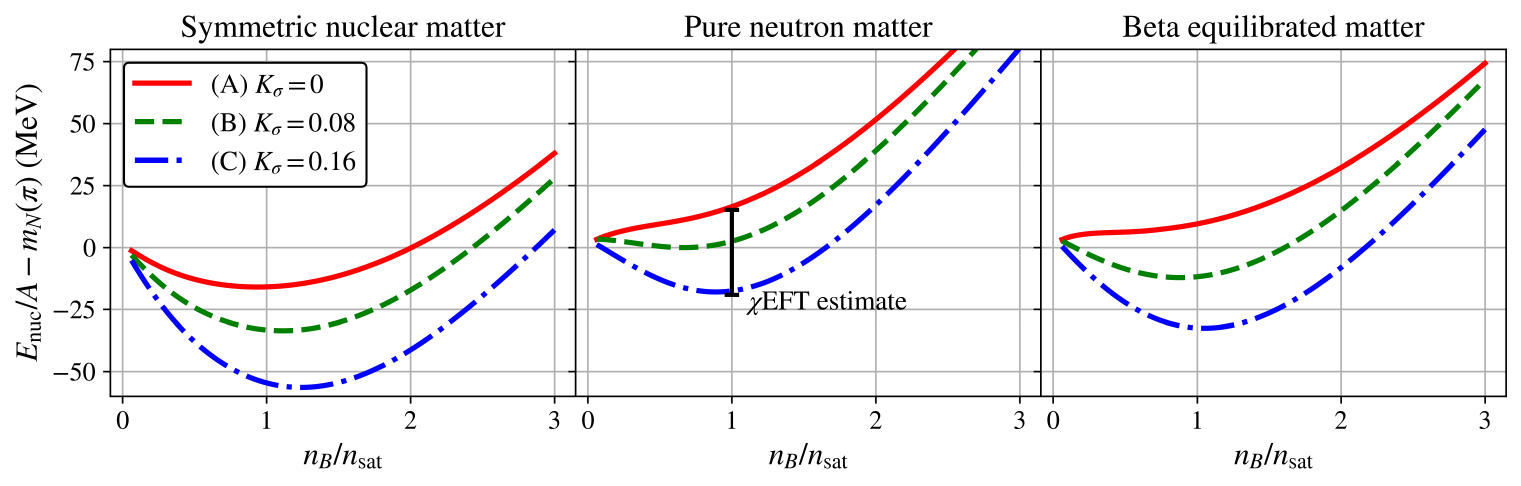}
    \caption{Kinetic and nuclear interaction energy per baryon for SNM, PNM, and $\beta$ equilibrated matter at $\theta = \pi$ with RMFT EOS with IUFSU$^*$ parameter set and our three choices of $K_\sigma$. The estimate for the binding of PNM from ChiEFT is shown with the approximate error bar including cutoff- and $\eta$-dependence.}
    \label{fig:rmf epb}
\end{figure}

Figure~\ref{fig:rmf epb} shows the kinetic and interaction energy per baryon for Scenarios A, B, and C at $\theta = \pi$. Since this only includes the energy from the kinetic energy of baryons and nuclear forces, this quantity is independent of $\varepsilon$. Note that this differs from Fig.~\ref{fig:rmf epb no d} in that the reduction of the nucleon masses is not included in this value. 

When $\varepsilon$ is small enough that axion condensed matter is favored down to zero pressure, there can be striking effects on the crust of the star. We construct an axion-condensed inner crust of nuclei in a background of dripped neutrons, including Coulomb and surface energies. Since the constraints we derive on axion parameter space are tied to the properties of the crust, we choose a surface tension $\sigma = 0.75 \,\mbox{MeV}/\mbox{fm}^2$ tuned to recreate neutron drip pressure at $\theta = 0$ calculated with more detailed microphysics (see, e.g., Ref. \cite{BPS1971}). To be self-consistent between the two phases, the phase transition will be calculated between the RMFT model in all crust phases for all values of $\theta$ rather than using a tabulated crust EOS. Where relevant, we compare quantitative results to the calculation of crust properties calculation in the BPS and NV EOSs~\cite{BPS1971, NegeleVautherin1973}. 

For sufficiently large $K_\sigma$, PNM saturates even for $\varepsilon = 1$, giving zero pressure solutions at finite density. For $0.035 \gtrsim K_\sigma \gtrsim 0.025$, there are two positive pressure branches to the PNM EOS, producing a jump in density in the inner crust. For $K_\sigma \gtrsim 0.035$, the neutron drip phase transition would become a first-order phase transition to a phase of saturated neutrons if it appeared in the neutron star, but as we will show later, within this model the neutron dripped phase is never favored for such large values of $K_\sigma$.

\begin{figure}[tb] %[ht]
\begin{center}
\includegraphics[width=\textwidth]{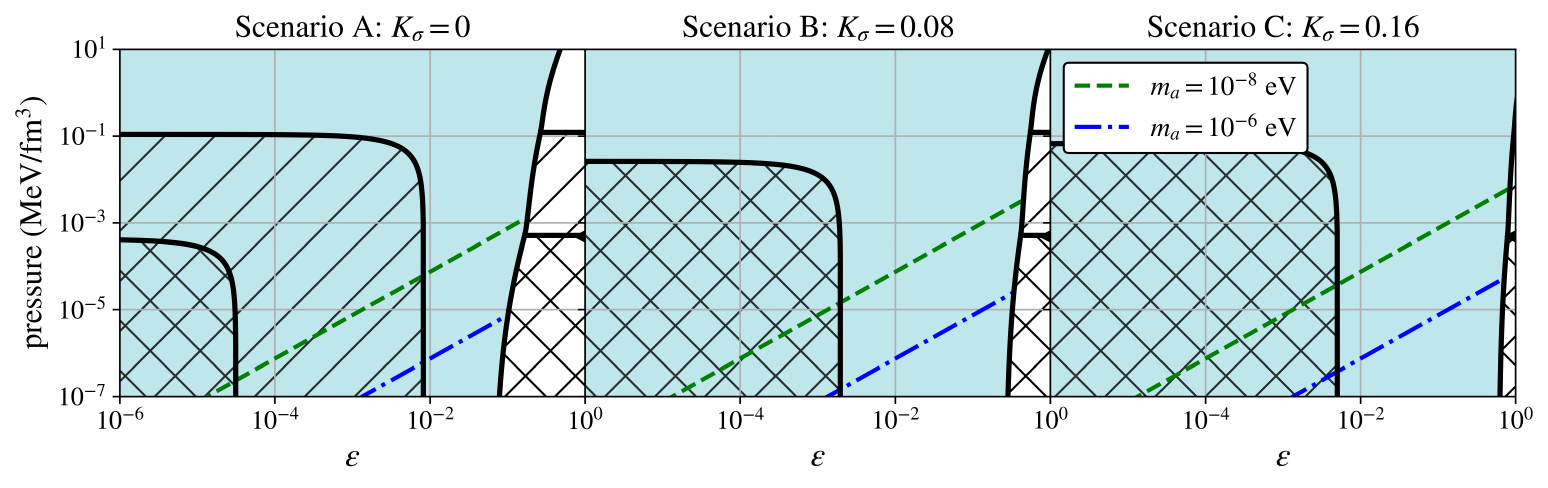}
\caption{Phase diagram for RMFT model with IUFSU$^*$ parameter set and our three choices of $K_\sigma$. The colored diagonal lines show the domain wall pressure for a fiducial neutron star radius $R= 12$ km and two example axion masses. The blue shading (no shading) indicates the pressures for which $\theta = \pi$ ($\theta = 0$) is the favored phase. The hatching indicates the regions of increasing density within the neutron star: outer crust (cross-hatched), inner crust (diagonal hatched) and core (no hatch).}
\label{fig:rmf phase diagram}
\end{center}
\end{figure}

Figure~\ref{fig:rmf phase diagram} shows a phase diagram for the RMFT model as a function of pressure and $\varepsilon$. In the blue-shaded region, $\theta = \pi$ is favored while in the unshaded region, $\theta = 0$ is favored.  The cross-hatched regions contain the outer crust of nuclei and electrons and the diagonal-hatched regions contain the inner crust of nuclei, dripped neutrons, and electrons. Regions without hatching are composed of homogeneous nuclear matter. The diagonal colored lines show the domain wall pressure for an axion of a given mass with the specified $\varepsilon$ if the radius of the domain wall is 12 km. When the boundary region is treated as infinitely thin, this is the pressure difference across the domain wall. For small $\varepsilon$ where the axion condensed phase is favored for all densities, this acts as a cutoff pressure at the surface of the star; lower pressures can only exist in the finite size of the domain wall itself. Effects of the domain wall will be discussed in more detail in Sec. \ref{ssec:domain_wall}. 

Note that points in the axion condensed phase with the same pressure but different $\varepsilon$ have different energy densities. Equal pressure but larger $\varepsilon$ corresponds to larger energy density both from the energy of the axions themselves, but also because pressure equality results in a larger baryon density to compensate for the negative axion pressure. In the normal phase, this effect is not present and all points of a given pressure have the same density. For $\varepsilon \rightarrow 0$, if the nuclear force is not modified the phases present in the neutron star look very similar to what is found in a star without axions, but for intermediate values of $\varepsilon$, some phases may be missing because the negative axion pressure cuts off the edge of the neutron star before the density for that phase is reached. In this case, the missing phases will only be found in the domain wall region.

As $K_\sigma$ is increased from $0$, the pressure of the PNM surrounding the nuclei in the inner crust decreases significantly and, as a result, the chemical potential corresponding to the crust-core transition (the boundary between the diagonally hatched and unhatched regions) decreases. At the same time, a more attractive nuclear interaction causes nuclei to become more neutron-rich, increasing the electron chemical potential. An increased electron density corresponds to a larger pressure at neutron drip (the boundary between the diagonal hatched and cross-hatched regions). For sufficiently large $K_\sigma$, if $\varepsilon$ is small enough for an axion condensed crust to be present, there will be a phase transition directly from bulk nuclear matter to a phase with nuclei and free electrons, with no dripped neutrons. This occurs because bulk matter is favored when the chemical potential is less than $m_n (\theta = \pi) + \Gamma_n (\theta = \pi)$ where $\Gamma_n (\theta)$ is the binding of PNM in the case where neutron matter is self-bound (Scenarios B and C) and $0$ in the case where it is not. 

\subsection{EOS of axion condensed matter}
\label{sec:EOS}

\begin{figure}[tb] %[ht]
\begin{center}
\includegraphics[width=\textwidth]{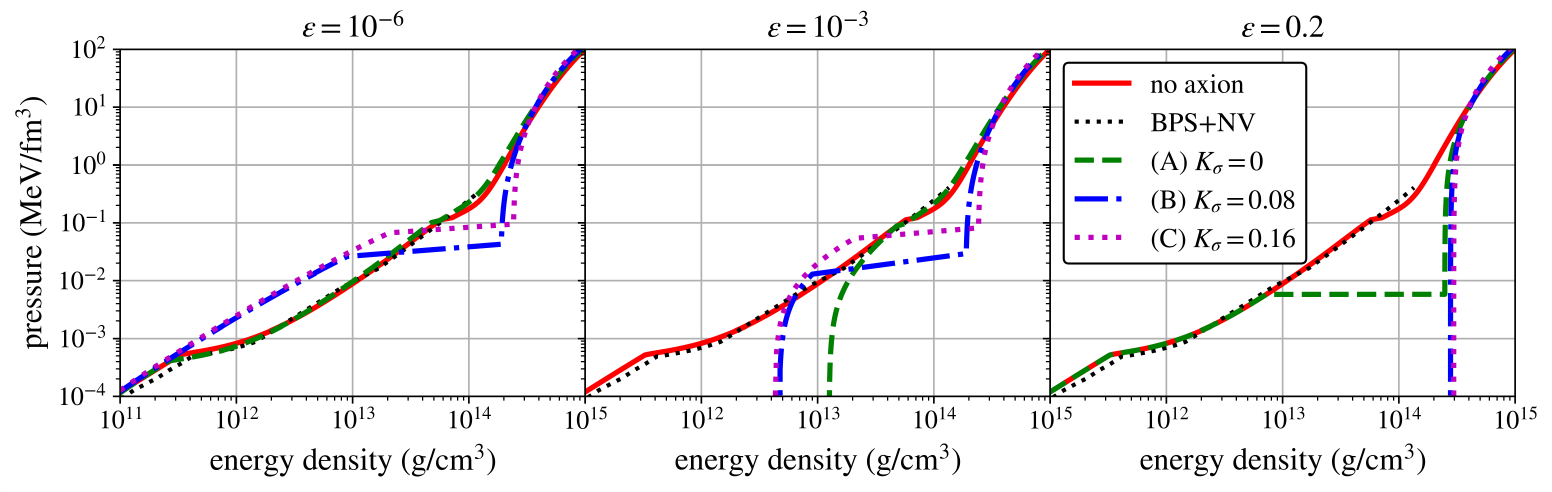}
\caption{The RMFT EoS for three example values of $\varepsilon$ and the three $K_\sigma$ scenarios using the IUFSU$^*$ parameter set. The standard no axion RMFT EoS  and the BPS+NV EoS for the crust are shown for comparison \cite{BPS1971,NegeleVautherin1973}. See section~\ref{sec:EOS} for details. }
\label{fig:iufsu eos}
\end{center}
\end{figure}

Figure~\ref{fig:iufsu eos} shows the RMFT EOS for charge neutral, $\beta$-equilibrated matter for several choices of $\varepsilon$ in the three scenarios discussed. We focus on the crust and outer regions of the core as those are the densities at which the effects of axion condensation are well understood. The solid red curve corresponds to the normal EOS without axions. For comparison, the black dotted curve shows the BPS+NV crust EOS \cite{BPS1971, NegeleVautherin1973}.

We first focus on the left panel of Fig.~\ref{fig:iufsu eos}. For $\varepsilon = 10^{-6}$, the axion field itself contributes very little to the energy density and pressure at the densities pictured, and Scenario A looks very similar to the case with no axions. At pressures of order $\varepsilon m_\pi^2 f_\pi^2 \simeq 10 \times \varepsilon \, {\rm MeV/fm}^3$ the curve for Scenario A will diverge from the EOS without axions and go to zero pressure at finite energy density. For $\varepsilon = 10^{-6}$ this corresponds to $p \simeq 10^{-5} \rm{MeV/fm}^3$, outside the plot range shown. This occurs because zero pressure in the axion condensed phase occurs when the pressure of nuclear matter exactly cancels the negative axion pressure, corresponding to a much larger baryon density than normally found in nuclear matter at zero pressure. The point where the curves for Scenario B and C in the left panel diverge from Scenario A and the EOS without axions corresponds to the neutron drip phase transition for Scenario A and the EOS without axions. When $K_\sigma$ is large enough that the inner crust is absent, the pressure is larger (the EOS is ``stiffer") at energy densities where ordinarily dripped neutrons would be present because there are fewer degrees of freedom in the system, resulting in the lepton density being larger. At the crust-core transition, however, there is a large jump in energy density from the mixed phase to bulk nuclear matter, resulting in a lower pressure (the EOS is ``softer") in the outermost part of the core.

The center and right panels of Fig.~\ref{fig:iufsu eos} can be understood in terms of the basic trends in the left panel. For $\varepsilon = 10^{-3}$, the outermost regions of the star are still axion condensed, and the only difference between the center and left panels is the much larger energy density and negative pressure from the axion field itself. This results in the equation of state going to zero pressure at large energy density because the nuclear matter at zero net pressure must contribute enough pressure to cancel the much larger axion pressure. In the right panel, Scenario B and C remain axion condensed down to zero pressure but the much larger value of $\varepsilon$ means that this occurs before the crust-core phase transition and the jump in energy density seen at lower $\varepsilon$ is not present. In Scenario A in the right panel, axion condensation is not favored at the lowest densities and there is a phase transition from axion condensed matter to normal matter, producing a discontinuity in the energy density due to the much lower baryon density in the normal phase.

\section{Neutron stars with $\theta=\pi$ matter}
\label{sec:neutron stars}

In this section, we describe the modification to the neutron star EOS taking into account the effects we discussed in Secs.~\ref{sec:dilute} and~\ref{sec:nuclear matter}, including the negative pressure at $\theta=\pi$ and modifications to the nucleon masses and interactions, as well as the axion field profile and domain wall pressure. Neutron star structure can be derived by combining gravity and these new effects that depend on the axion coupling and masses, parameterized by $\varepsilon$ and $m_a$. Qualitatively, for larger $m_a$ ($m_a\gg 1/R_{\rm NS}$), the axion field profile can be treated as a thin wall over which the gravitational potential does not change, and the gravity and axion effects can be accounted for separately. We will first discuss neutron star structure in the simpler case where this approximation is valid, before commenting on the criteria for the validity of the approximation and working out the more general cases.  

\subsection{Neutron star structure in axion condensed stars}
\label{ssec:NSstructure}

\begin{figure}[tb] %[ht]
\begin{center}
\includegraphics[width=0.4\textwidth]{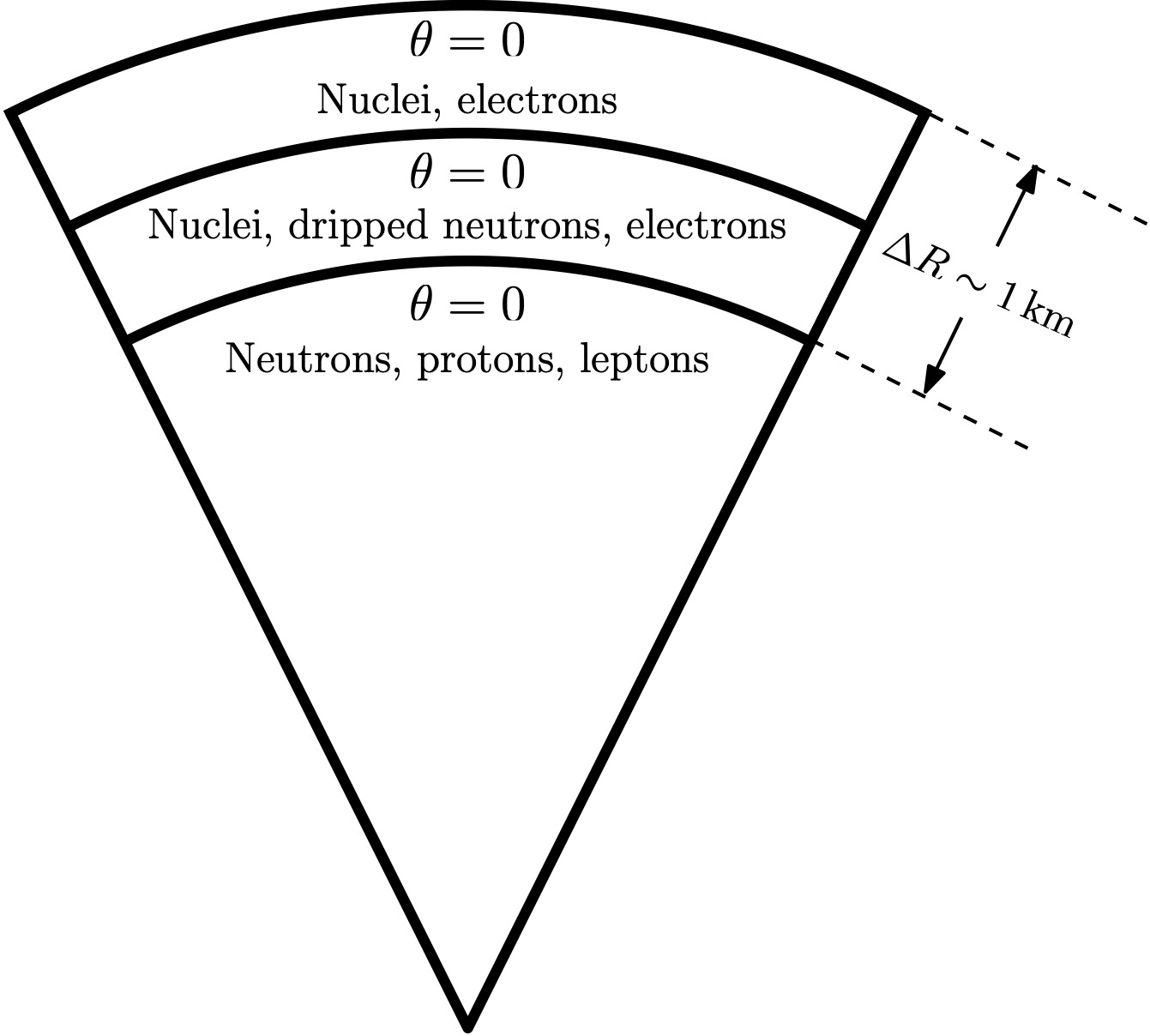}
\includegraphics[width=0.4\textwidth]{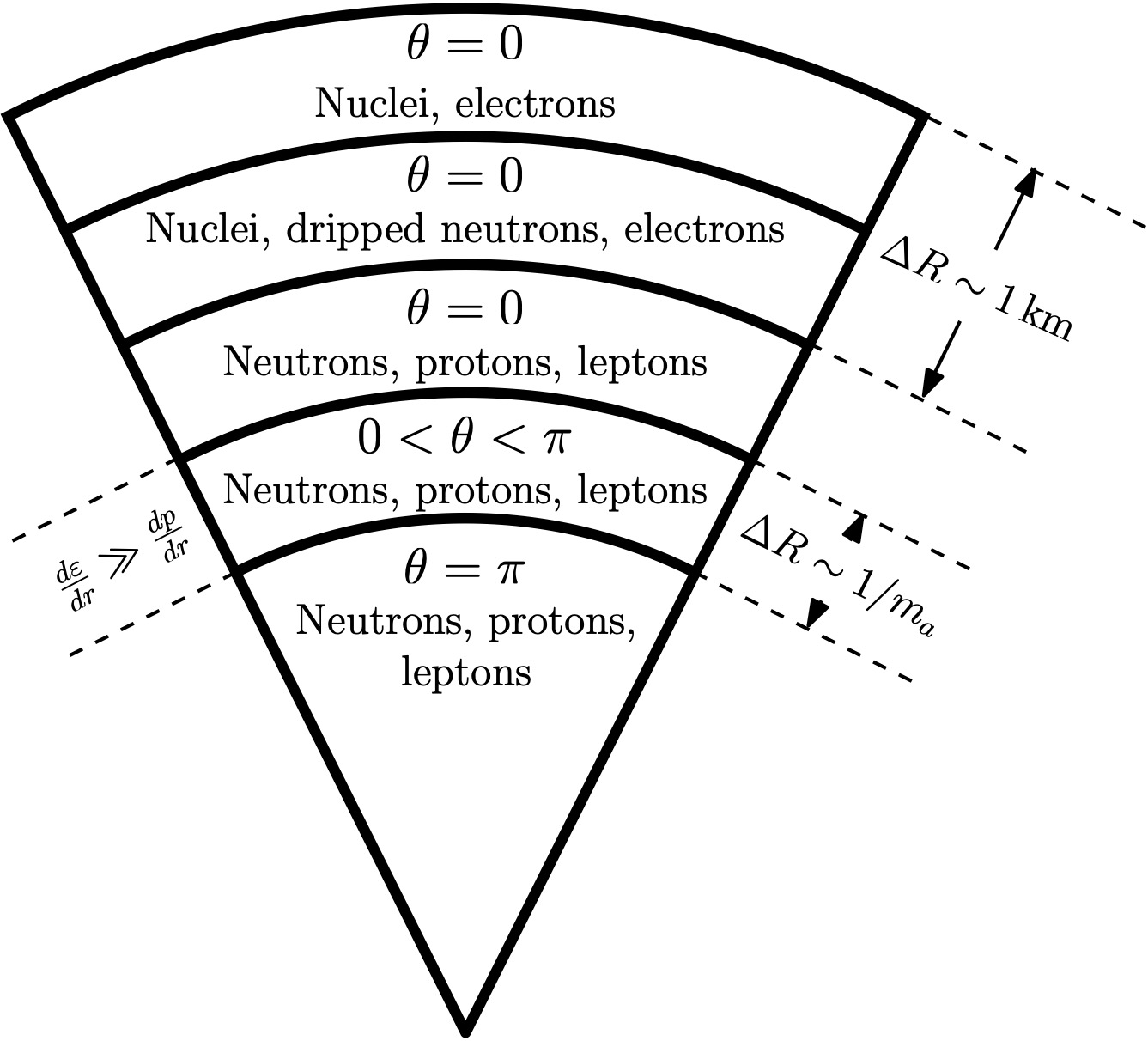}
\caption{Layers of a normal neutron star (left) and a neutron star with an axion condensed inner core (right). The thickness of layers is not to scale. See section~\ref{ssec:NSstructure} for discussion.}
\label{fig:no ax and norm crust}
\end{center}
\end{figure}

An ordinary neutron star has an outer crust composed of nuclei in a background of free electrons and an inner crust composed of nuclei in a background of free electrons and dripped neutrons, together having a thickness of around a kilometer (see the left panel of Fig.~\ref{fig:no ax and norm crust} for a sketch). The possibilities for the outer structure of a neutron star become more varied in the presence of axion condensation, depending on the value of $\varepsilon$ and any modifications to the nuclear force.  In the simplest case, for $\varepsilon \gtrsim 0.1$, the center of the star is axion condensed with a transition to normal matter at some intermediate radius. That radius may be large enough that portions of the normal crust are missing with, e.g., a transition directly from bulk matter at $\theta = \pi$ to an outer crust at $\theta = 0$ with possibly a thin inner crust in the domain wall region. For larger $\varepsilon$, the entire normal crust is present and only a portion of the bulk matter in the center of the star is axion condensed, with a domain wall region of thickness $\Delta r_{\rm DW} \sim 1 / m_{a, \rm eff}$ in which the energy density changes rapidly but the pressure is nearly constant. This case is sketched in the right panel of Fig.~\ref{fig:no ax and norm crust}.

\begin{figure}[tb] %[ht]
\begin{center}
\includegraphics[width=0.412\textwidth]{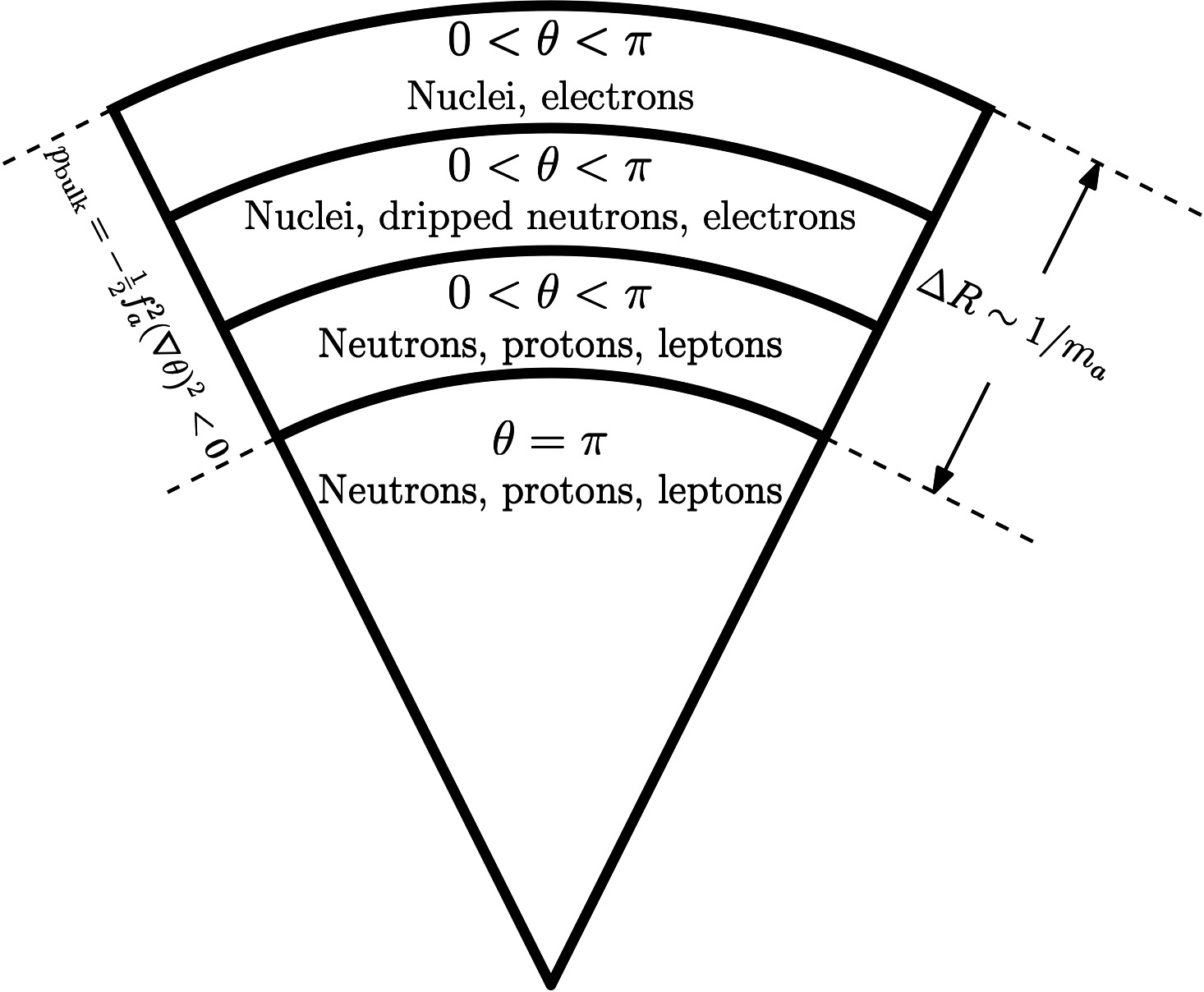}
\includegraphics[width=0.45\textwidth]{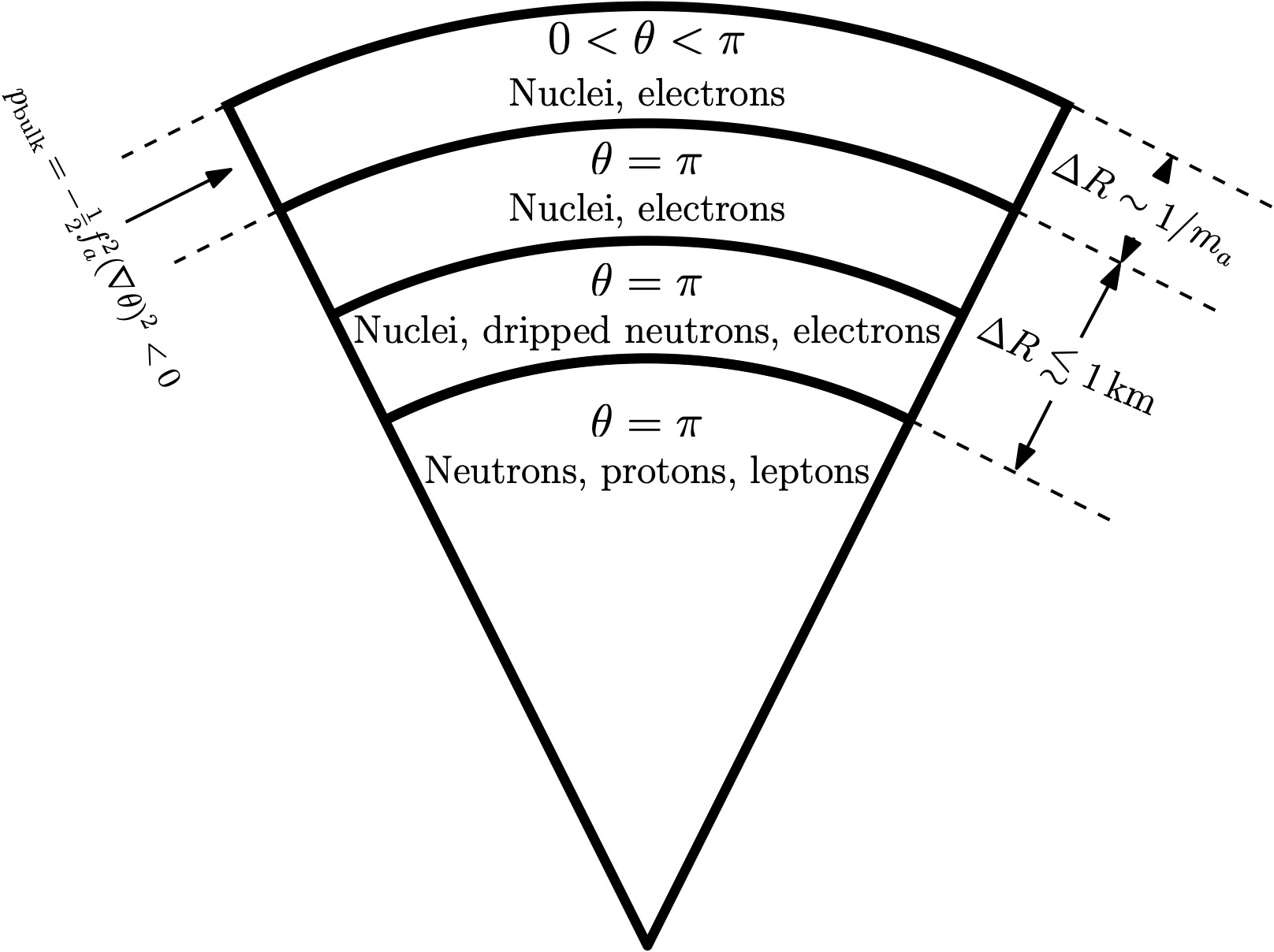}
\caption{Layers of a neutron star with a crust only in the domain wall region (left) and one with an axion condensed crust (right) in Scenario A. Scenario B and C are the same, but with no dripped neutron phase. The thickness of layers is not to scale.  See section~\ref{ssec:NSstructure} for discussion.}
\label{fig:axcrusts}
\end{center}
\end{figure}

For smaller $\varepsilon$, different structures for the outermost layers of the star are possible.  For $0.01 \lesssim \varepsilon \lesssim 0.1$, there is no positive pressure where a heterogeneous phase is favored in all three scenarios and bulk matter at $\theta = \pi$ persists almost to the edge of the star. A crust in this case is only present in the thin domain wall region where the pressure of bulk matter is negative and hydrostatic equilibrium is maintained by the pressure produced by the axion gradient. In this case, $p_{\rm bulk} \approx - p_{\rm grad} = - f_a^2 (\nabla \theta)^2 / 2$. This case is sketched in the left panel of Fig.~\ref{fig:axcrusts} for Scenario A. Scenarios B and C look the same, but without a phase of dripped neutrons. If $\varepsilon$ is smaller, it is possible to have an axion-condensed crust at positive pressure. In this case (sketched in the right panel of Fig.~\ref{fig:axcrusts} for Scenario A), if the nuclear force is not significantly modified, some or all of the normal layers of the crust will be present in the $\theta = \pi$ region. The domain wall region still has a positive gradient pressure in this case and will support more crust at a negative bulk pressure. 

\begin{figure}[tb] %[ht]
\begin{center}
\includegraphics[width=\textwidth]{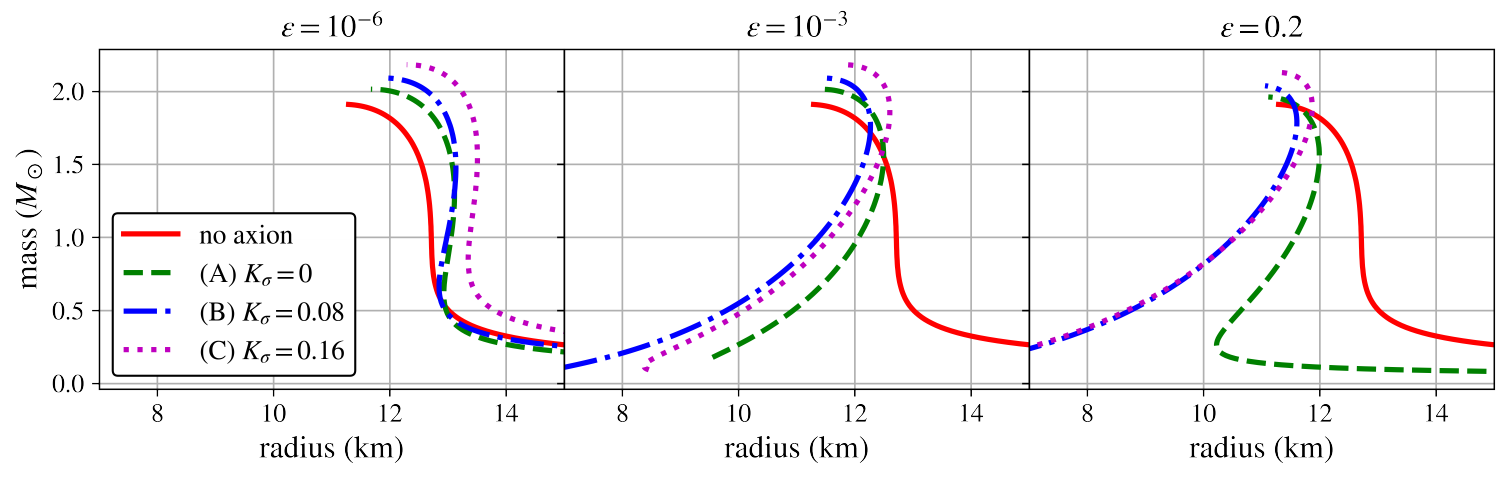}
\caption{Mass-radius relationship for neutron stars with three fiducial $\varepsilon$ values and $K_\sigma$ in RMFT. We consider only the leading interaction between nucleons and axions (which leads to unknown uncertainties for $M \gtrsim 1.4 \, M_\odot$) and neglect the thickness of the domain wall.}
\label{fig:mr}
\end{center}
\end{figure}

Although our treatment of axion condensed matter does not apply at arbitrarily high density, we can still speculate about the effects of axion condensation on neutron star masses and radii. A similar calculation was performed in Ref.~\cite{Balkin:2023xtr} for a non-interacting gas of neutrons, see also Ref.~\cite{Gao:2021fyk} for scalar effects on the crust thickness and EoS. As can be seen in Fig.~\ref{fig:iufsu eos}, at low density, the EOS is typically softer for $\theta = \pi$ and $\varepsilon$ not too small because of the contribution of the axion field.  A naive extrapolation of nuclear matter to high density with only the axion-nucleon interaction we have considered thus far will eventually yield a stiffer EOS because of the lighter nucleon mass. This results in heavy neutron stars having larger radii and larger maximum mass. Figure~\ref{fig:mr} shows the relationship between mass and radius for neutron stars with various $\varepsilon$ within this simplistic picture of the high density EOS and treating the domain wall as thin ($1/m_a \ll r_{\rm NS})$. 

Counter-intuitively, larger $K_\sigma$ yields larger NS radii for very small $\varepsilon$, pointing to a stiffer EOS.  At high density, larger $K_\sigma$ results in a stiffer EOS because of greater reduction to the effective nucleon mass. At low density, larger $K_\sigma$ corresponds to nuclei with larger isospin asymmetry increasing the electron pressure once the neutron dripped phase is missing, again giving a stiffer EOS.  Only in a range of energy densities just above the crust-core phase transition do Scenario B and C give a softer EOS than Scenario A with Scenario C having such a small region of softening that it has a minimal effect on the mass-radius curve and radii are larger for all masses.

For $\varepsilon$ where the lowest density parts of the crust are missing, low mass stars may have very small radii, with a mass-radius relation akin to strange quark stars. Unlike a strange quark star, however, neutron stars with a condensed axion field lack a confining potential at their surface, and as a result have a thin crust of matter at intermediate $\theta$ in the domain wall region. Since the density and chemical potential smoothly go to zero on a length scale determined by the axion field profile, unusual surface properties of strange quark stars (see, e.g., Ref.~\cite{StrangeStars} for a detailed discussion) are not present in axion condensed stars. For the same reason, the radius of an axion condensed star does not go to zero for vanishing mass for any $\varepsilon$ but will always be limited by the finite size of the domain wall. The minimum stable mass and radius of a compact axion condensed object will be addressed in greater detail in Sec.~\ref{sec:piballs}.

Since our treatment of the interaction between axions and nucleons is only strictly accurate at densities around or below nuclear density, predictions from our calculation of the mass-radius relationship should be considered speculative for $M \gtrsim 1.4 \, M_\odot$. As such, the constraints we draw will all be based on the structure of the crust in axion condensed stars. We will use masses and radii of $1.4 \, M_\odot$ stars calculated with the uncertain high density equation of state as a starting point to calculate the crust and domain wall, but these will only have an effect by altering the gravitational potential and explicit dependence of gradients on the radius in spherical coordinates. Since the inclusion of axion condensation changes the radius of a $1.4 \, M_\odot$ star by up to approximately a kilometer, and the uncertainties from nuclear physics are $\pm 1$-$3 \, {\rm km}$ (cf. Fig.~3 in Ref.~\cite{PRC_limitingMR}), this is a sub-leading source of uncertainty.

\begin{figure}[tb] %[ht]
\begin{center}
\includegraphics[width=\textwidth]{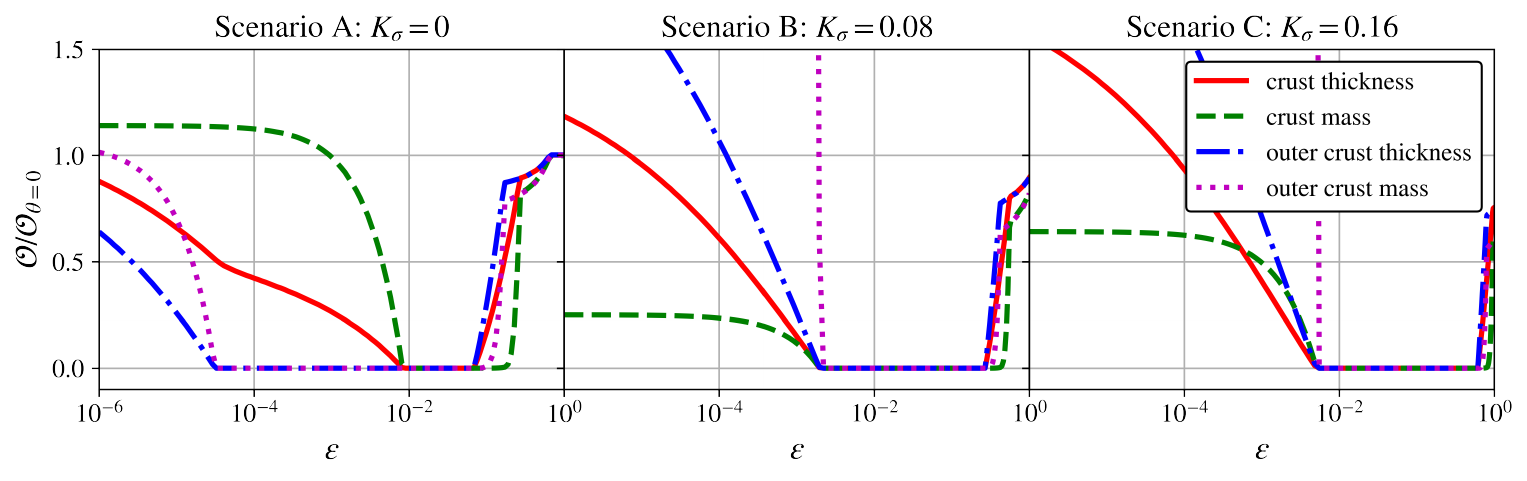}
\caption{Crust and outer cruss mass and thickness of a $1.4 \, M_\odot$ neutron star as a function of $\varepsilon$ for RMFT EOS. The three panels correspond to the three interaction scenarios (Table~\ref{tab:Ksigma}). The curves are normalized to $1$ for a star without axions.}
\label{fig:crust properties}
\end{center}
\end{figure}

Focusing on the crust, Fig.~\ref{fig:crust properties} shows the total thickness and mass of the crust and thickness and mass of the outer crust alone of a $1.4\, M_\odot$ neutron star as a function of $\varepsilon$, normalized to the value for a star without axions. In Scenario A, the stars at very small $\varepsilon$ have nearly the entire crust and look very similar to a star without axions, though the envelope (the region of the crust with $\rho < 10^{10} \, {\rm g/cm}^3$) is still diminished, providing a constraint from the cooling of isolated neutron stars, as will be discussed in Sec.~\ref{ssec:isolatedcooling}. In Scenarios B and C, there is less overall mass in the crust because of the lack of dripped neutrons. This will reduce the moment of inertia in the crust and can be constrained with pulsar glitches, as will be discussed in Sec.~\ref{ssec:glitches}. In all three scenarios, when $10^{-4} \lesssim \varepsilon \lesssim 0.1$ the overall thickness of the crust is diminished. Observations of the thermal relaxation timescale of neutron stars in x-ray binaries put a lower limit on the necessary thickness of the crust and will be discussed in Sec.~\ref{ssec:thermalrelaxation}.

\subsection{Axion domain wall}
\label{ssec:domain_wall}

An axion domain wall separates regions with densities that are larger and smaller than $n_B \sim \varepsilon m_\pi^2 f_\pi^2 / \sigma_N $. This profile, in general, should be solved together with the density and pressure profile of the neutron star. In the limit where $m_a^{\rm QCD} > m_a > 1/r_{\rm NS}$, the axion profile is a thin domain wall, located at a radius $r_a$ from the center of the neutron star. The tension $\sigma_a $ of this field profile wrapping the nuclear matter inside the neutron star exerts a pressure of $\sigma_a/2 r_{a}$, the minimal pressure of the matter inside the domain wall.

We solve the axion field profile in two regimes, extending the formalism described in section~\ref{sec:axion dilute neutron} to include nuclear interactions. First, consider the limit where an axion domain wall is in a smooth background of nuclear matter which is changing slowly over length scales much longer than $1/m_a$. In this case, we can solve the axion field profile in PNM as
\begin{equation}
\begin{split}
    \nabla^2 a &= \left(m_a^2 - \frac{\sigma_N (1 + \delta^{\rm (loc)}) n_n(r)}{f_a^2}\right) f_a\, \sin [a/f_a] \\
    &\simeq \frac{\sigma_N (1 + \delta^{\rm (loc)}) }{f_a}\left.\frac{\d n_n(r)}{\d r} \right|_{r= r_a}(r-r_a) \,\sin [a/f_a],
\end{split}
\end{equation}
where $\delta^{\rm (loc)} \propto K_\sigma n_B / \sigma_N m_\sigma^2$ is an order-one quantity that encodes the $m_\sigma$ dependence of the effective axion mass. Details can be found in App.~\ref{app:deltaloc}.

For axion profiles with thickness much smaller than $r_a$, at linear order in $a/f_a$, the solution of this equation is an Airy function, where the field profile changes over a length scale of $(m_a^2/r_{a})^{-1/3}$. The domain wall tension of this axion profile is $\sigma_a \leq m_a f_a^2$ and the resulting pressure difference due to this domain wall profile between inside and outside is smaller than that due to the vacuum energy difference of $m_a^2 f_a^2$.

The smooth density profile would be a good approximation only for the largest $\varepsilon$, where the axion domain wall is situated close to nuclear density. As we showed in Sec.~\ref{sec:nuclear matter}, for most of the $\varepsilon$ we consider, the domain wall is between regions with drastically different densities, and the density (and pressure) of the matter outside the domain wall is effectively zero. In this second case, inside the domain wall, the density $n_{\rm in}$ is such that the degeneracy pressure is
\begin{equation}\label{eq:pressurenin}
    \frac{(3\pi^2)^{2/3}}{5 m_n} n_{\rm in}^{5/3} \simeq  m_a^2 f_a^2\,,
\end{equation}
and the axion mass in the region inside the domain wall is much larger than the axion mass in vacuum at small $\varepsilon$, by a factor of
\begin{equation}\label{eq:massratio}
    \frac{m_a(n_{\rm in})}{m_a} = \left(\frac{\sigma_N n_{\rm in}}{m_a^2 f_a^2}\right)^{1/2} = \frac{\sigma_N^{1/2} (5 m_n)^{3/10}}{(3\pi^2)^{1/5} (m_a^2 f_a^2)^{1/5}} \approx \frac{1.5}{\varepsilon^{1/5}}.
\end{equation}
In this limit, the field profile is changing mainly inside the region of low density, and the field profile has an extent of $1/m_a$. This exerts a pressure of $\sim m_a f_a^2/r_{\rm NS}$ on the matter inside, again smaller than the pressure difference due to the changing vacuum energy\footnote{In the \ZN model, the vacuum energy difference is $m_a^2 f_a^2/\mathcal{N}^2$, while the domain wall pressure becomes comparable to this vacuum pressure when $r_{\rm NS}\simeq\mathcal{N}^2/m_a$, that is $m_a \simeq 10^{-8} \,{\rm eV}$ for $\varepsilon \simeq 2\times 10^{-7}$. The density $n_{\rm in}$ for which Eq.~\eqref{eq:pressurenin} is satisfied is still much larger than the critical density for which Eq.~\eqref{eq:crit density} saturates, since the equation Eq.~\eqref{eq:massratio} is only modified by logarithmic corrections.}.

For axion masses that are not too small, the chemical potential can be treated as approximately constant in the domain wall crust. Across a domain wall crust, the changing nucleon mass modulates the Fermi momenta in the same way that a changing baryon chemical potential does in a normal star, but on a much shorter distance scale. This approximation is good as long as the condition $\dd\mu_B /\dd r \ll \dd m_N(\theta)/\dd\theta \times \dd\theta/\dd r$. Since $\theta$ changes on a distance scale $1/m_a$ and the local chemical potential can be expressed in terms of the local spacetime metric $e^{\nu(r)} = g_{00} (r)$ and redshifted chemical potential $\tilde{\mu}_B$ according to $\mu_B(r) = \tilde{\mu}_B e^{-\nu(r)}$, this condition can be translated into a condition on the axion mass and the gravitational potential near the surface of the star,
\begin{equation}
\label{eq:surf ax limit}
    m_a \gg \frac{GM_{\rm NS}}{r_{\rm NS}^2 (1 - 2GM_{\rm NS}/r_{\rm NS})} \frac{m_N}{\sigma_N} \simeq 9 \times 10^{-11} \, \mbox{eV}\,,
\end{equation}
with the final value given for a $1.4\, M_\odot$ star with a radius of $12$ km. In what follows, we will consider axion masses $m_a > 10^{-10} \, \mbox{eV}$ when calculating the domain wall crust and will treat the chemical potential as approximately constant in this regime. 

The equivalent condition in the case where the domain wall is inside the star is
\begin{equation}
    |m_{a, \rm eff} (n_B)| \gg \frac{GM_a}{r_a^2 (1 - 2GM_a /r_a)} \frac{\mu_c}{\sigma_N} \left(1 + \frac{4\pi r_a^3 p_c}{M_a} \right)\,,
\end{equation}
where $r_a$ is the radius of the domain wall, $M_a$ is the mass contained within $r < r_a$, and $p_c$ and $\mu_c$ are the critical pressure and chemical potential where the phase transition occurs. In practice, $p_c \lesssim 0.1 \,\mbox{MeV}/\mbox{fm}^3$ for $\varepsilon$ where the crust is strongly affected and the GR correction is at most a few percent and we can instead use the approximate form
\begin{equation}
\label{eq:inner ax limit}
    |m_{a, \rm eff} (n_B)| \gg \frac{GM_a}{r_a^2 (1 - 2GM_a/r_a)} \frac{\mu_c}{\sigma_N}.
\end{equation}
This approximation fails in the case where the domain wall is very near the center of the star and $M_a$ is small. In practice, this is a challenging constraint to implement since $m_{a, \rm eff}$ changes sign in the vicinity of the domain wall.  In the core where the baryon density changes more slowly, this leads to a very extended domain wall even for axion masses that are larger than $1/R_{\rm NS}$, with the extent of the domain wall becoming comparable to the gravitational scale for $m_a \lesssim 10^{-9} \, \mbox{eV}$. For the sake of calculating the contribution of the domain wall to the crust, we limit ourselves to axion masses $m_a \gtrsim 10^{-9}\, \mbox{eV}$ when the domain wall is inside the star. This does not affect the final constraint we determine. 

Apart from the two special cases, more generally, in order to numerically solve the axion field profile, we will work in the limit where the baryon chemical potential changes on a length scale that is long compared to the length scale for $\theta$ to change, set by Eq.~\eqref{eq:surf ax limit} for the case where the domain wall is at the surface of the star and Eq.~\eqref{eq:inner ax limit} for the case where the domain wall is in the interior of the star. In this case, we can solve the axion field profile at fixed chemical potential according to
\begin{equation}
\label{eq:theta de}
\begin{split}
    \nabla^2 \theta &= \frac{1}{f_a^2} \frac{d\Omega}{d\theta} \\ 
    &= \frac{1}{f_a^2} \bigg\{ \sigma_N (1 + \delta^{\rm (loc)}) [n^{(s)}_p (\theta) + n^{(s)}_n (\theta)] + \frac{\Delta \sigma}{f^2(\theta)} [n^{(s)}_p (\theta) - n^{(s)}_n (\theta)] - \varepsilon m_\pi^2 f_\pi^2 \bigg\} f'(\theta)\,,
\end{split}
\end{equation}
where $n^{(s)}_i = \partial E / \partial m_i$ is the scalar density of baryon species $i$ which at leading order in bulk matter is given by $n^{(s)}_i = n_i - k_{Fi}^5/[10\pi^2 m_i^2 (\theta)]$, the spatial gradients include corrections from gravity, and the scalar densities are found at fixed baryon chemical potential. The quantity $\delta^{\rm (loc)}$ comes from the explicit $m_\sigma$-dependence of the RMFT Lagrangian and is approximately given by
\begin{equation}
    \delta^{\rm (loc)} \approx \frac{K_\sigma}{\chi \sigma_N} \left( \frac{g_\sigma}{m_\sigma} \right)^2 \left(n_p^{(s)} + n_n^{(s)}\right),
\end{equation}
where $g_\sigma$ is the coupling of the $\sigma$ meson to nucleons in the RMFT model and $\chi$ is equal to the volume fraction of nuclei in the outer crust and unity in the core. Exact expressions including higher-order RMFT corrections and the more complicated expression for the inner crust can be found in App.~\ref{app:deltaloc}. This correction is order one for both Scenarios B and C in the outer crust and core and should not be neglected, but is suppressed in the inner crust. Note that this correction is due only to explicit $m_\sigma$ dependence of the RMFT Lagrangian; implicit dependence via the effect of $m_\sigma$ on the Fermi momenta of the nucleons and the mean-field value of the $\sigma$ field vanishes since the mean-field approximation will always give Fermi momenta and a mean-field value for the $\sigma$ field that minimize the free energy. 

The fixed baryon chemical potential approximation leads to overestimating the amount of matter in the crust irrespective of the location of the crust. In the case where the domain wall is at the surface of the star, the small decrease to the baryon chemical potential in the domain wall will cause the baryon density to decrease more rapidly, resulting in a thinner crust. In the case where the domain wall is inside the star, a domain wall that is too broad would result in overestimating the size of the crust since, in reality, part of the domain wall would extend into the crust rather than being a separate layer that gets added to a normal crust. Since our constraint is derived from layers of the crust being missing or diminished, this approximation is conservative.

It is worth noting that the equations of hydrostatic equilibrium do not need to be solved separately in the domain wall because they are automatically satisfied by solving the Euler-Lagrange equation for the axion field given by Eq.~\eqref{eq:theta de}. This can be seen by observing that
\begin{equation}
    \frac{\dd p_{\rm bulk}}{\dd r} = - \frac{\dd \Omega}{\dd \theta} \frac{\dd\theta}{dr} = - f_a^2 \nabla^2 \theta \frac{\dd \theta}{\dd r} = - \frac{1}{2} \frac{\dd}{\dd r} [f_a^2 (\nabla \theta)^2] = - \frac{\dd p_{\rm grad}}{\dd r}.
\end{equation}
The differential equation given by Eq.~\eqref{eq:theta de} can be solved by a shooting procedure, varying the initial conditions at the inner edge of the domain wall until a solution is found that smoothly goes from $\theta = \pi$ to zero. In practice, solving this equation is greatly simplified by making a change of variables to confine $\theta$ to be between zero and $\pi$, $\theta (y) \equiv \pi / (1 + e^y)$ and solving the resulting equation for $y$. Numerical results from this calculation will be presented in Sec.~\ref{sec:constraints}.

Since we perform calculations in the domain wall region only in RMFT, the entire $\theta$-dependence in the domain wall is contained in the nucleon and meson masses and the energy of the axion field itself. There are additional terms in the nuclear Lagrangian that contribute only when $\theta$ is at some intermediate value that violates parity and a detailed calculation of the domain wall from first principles should include these terms that we neglect. For further discussion, see App.~\ref{app:pv_dw}.

\section{Finite size objects at zero pressure}
\label{sec:piballs}

As was first noted in Refs.~\cite{Balkin:2022qer, Balkin:2023xtr}, an object comprised of nuclear matter at $\theta = \pi$ can be stabilized by the negative pressure $m_a^2 f_a^2$, rather than gravity. These objects, which we call $\pi$-balls, are held together by the attractive axion potential as long as the baryon chemical potential is too small to allow nucleons to escape (see also ~\cite{Witten:1984rs,Zhitnitsky:2002qa, Gao:2021fyk}). Given the constraint from white dwarf stability of $\varepsilon > 2 \times 10^{-7}$, a zero-pressure object balanced by electron degeneracy pressure cannot exist. If a neutron star has a central density large enough to be unstable against axion condensation, it is possible that when such a neutron star is disrupted in a binary neutron star merger, $\pi$-balls may be part of the ejecta. Having an axion field surrounding them, these objects may produce a unique electromagnetic signature. Such questions of production and observational signatures of $\pi$-balls will be explored in the future; in this work, we merely comment on their basic properties.

\begin{figure}[tb] %[ht]
\begin{center}
\includegraphics[width=0.8\textwidth]{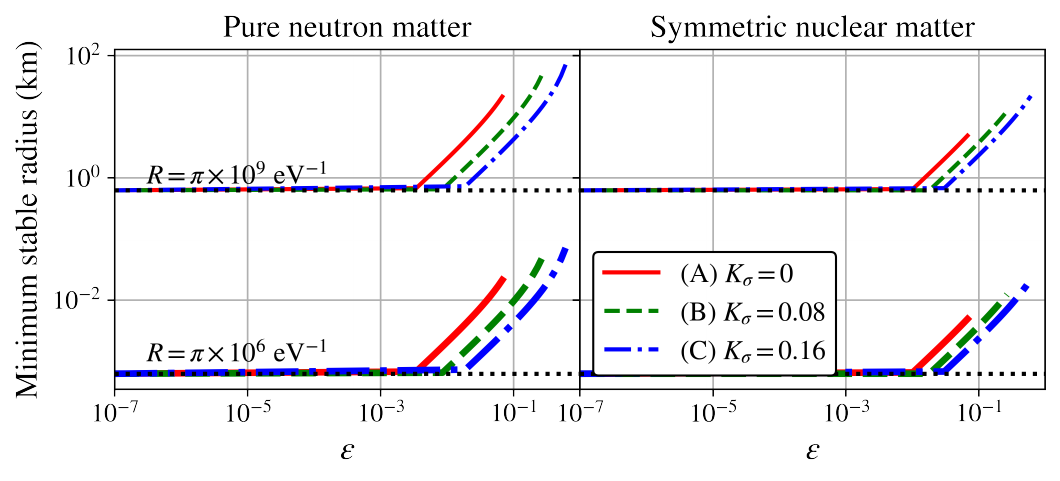}
\caption{Minimum stable radius of a $\pi$-ball made of SNM and PNM as a function of $\varepsilon$. Thin lines: $m_a = 10^{-9} \,\mbox{eV}$, thick lines: $m_a = 10^{-6} \, \mbox{eV}$. The dotted black lines show lines of $R=\pi/m_a$ for the two axion masses.}
\label{fig:minr pi ball}
\end{center}
\end{figure}

\begin{figure}[tb] %[ht]
\begin{center}
\includegraphics[width=0.8\textwidth]{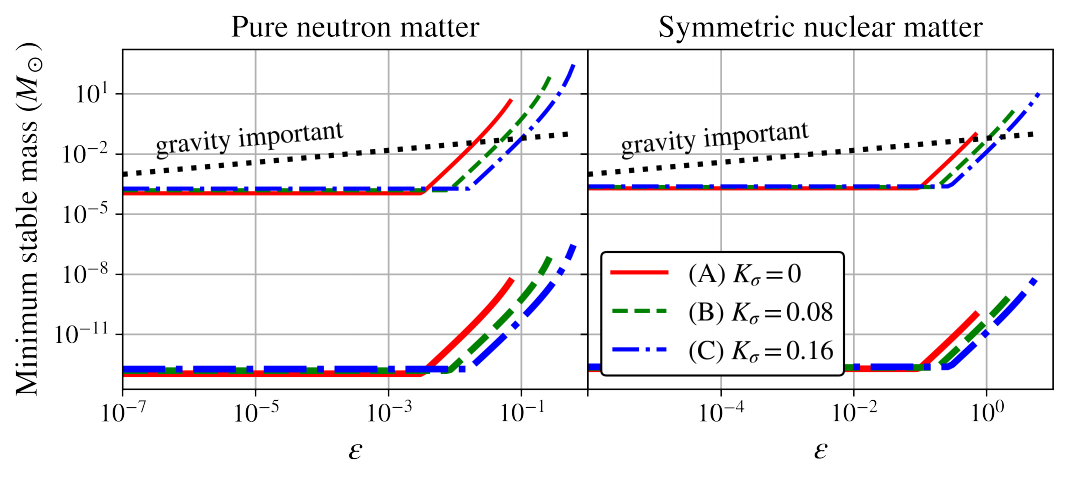}
\caption{The corresponding minimum stable mass of a $\pi$-ball made of SNM and PNM. Thin lines: $m_a = 10^{-9} \,\mbox{eV}$, thick lines: $m_a = 10^{-6} \, \mbox{eV}$. The black dotted line is the maximum mass before gravity becomes important from Eq.~\eqref{eq:piball_mmax}.}
\label{fig:minm pi ball}
\end{center}
\end{figure}

An object with size $R \gg 1/m_a$ made of PNM is stable at almost zero pressure when
\begin{equation}\label{eq:pressurecon}
    p_{\rm DW} = \frac{m_a f_a^2}{2R} = - \varepsilon m_\pi^2 f_\pi^2 [1 - f(\pi)] + \frac{(3\pi^2)^{2/3}}{5 m_n} n_n^{5/3}\,,
\end{equation}
and the Fermi energy of the neutrons is smaller than the kinetic energy required to escape the $\theta = \pi$ region
\begin{equation}\label{eq:escapecon}
    \frac{(3\pi^2)^{2/3}}{2 m_n} n_n^{2/3} <\sigma_N [1 - f(\pi)],
\end{equation}
which, when combined with Eq.~\eqref{eq:pressurecon}, is identical to the condition for the appearance of the $\theta = \pi$ region ($\sigma_N n_n > \varepsilon f_\pi^2 m_\pi^2$) if the domain wall pressure is small. The mass-radius relation of this object can be found to be
\begin{equation}
\begin{split}
    M &\simeq \frac{4\pi R^3}{3}\frac{m_n (5 m_n \varepsilon m_\pi^2 f_\pi^2)^{3/5}}{(3 \pi^2)^{2/5}}\left[ 1 - f(\pi) + \frac{m_u m_d}{2Rm_a (m_u + m_d)^2} \right]^{3/5} \,, \\
    &\simeq 2 \times 10^{18} \varepsilon^{3/5} \left( \frac{R}{1 \,\rm m} \right)^3 \,{\rm kg} \,,
\end{split}
\end{equation}
where the last equality holds if $m_a R \gg 1$. The maximal radius of this object where gravity can be ignored can be found by equating the gravitational potential energy with the Fermi energy of the neutrons on the surface of this object, which yields $R\simeq 5 \varepsilon^{-1/10}\,{\rm km}$ and the resulting maximal mass is
\begin{equation}
\label{eq:piball_mmax}
    M_{\rm max} \simeq 0.12 \, \varepsilon^{3/10} M_{\odot}\,, 
\end{equation}
beyond which the gravitational binding energy/pressure cannot be ignored. The minimal mass of a $\pi$-ball can either be found at $R= \pi/m_a$, when the thin wall approximation fails, or when the domain wall pressure becomes large enough that the baryon chemical potential exceeds the bare neutron mass. When the baryon chemical potential becomes larger than the bare neutron mass, it is energetically favorable for neutrons to leave the $\theta = \pi$ region. In the dilute approximation where nuclear interactions are ignored, the condition set by Eq.~\eqref{eq:escapecon} is saturated when $ R \simeq 1.2 \varepsilon^{1/2}/m_a$, which is smaller than $1/m_a$ for any $\varepsilon < 0.7$. This leads to the final minimal mass of
\begin{equation}
    M_{\rm min} \simeq  8\times 10^{9} \varepsilon^{3/2} \,{\rm MeV}^4/m_a^3 = 2 \times 10^{-7} M_{\odot} \left(\frac{\varepsilon}{10^{-1}}\right)^{21/10}\left(\frac{10^{-9}\,{\rm eV}}{m_a}\right)^{3}.
\end{equation}

In practice, to calculate the minimum mass of a stable object, nuclear interactions must be included. In the case of PNM, the minimum stable size is determined by the requirement that the baryon chemical potential not exceed the bare neutron mass.  In the case of SNM, the minimum stable size is determined by the requirement that the baryon chemical potential be less than the energy per baryon of the most bound symmetric nucleus, $\mu_{\rm Fe}/A = m_N - 8$~MeV.

Figure~\ref{fig:minr pi ball} shows the minimum stable radius for $\pi$-balls made of SNM and PNM and Fig.~\ref{fig:minm pi ball} shows the minimum stable mass, including nuclear interactions from the RMFT model and stabilized by axions. The horizontal dotted lines in Fig.~\ref{fig:minr pi ball} show when $m_a R = \pi$ and the domain wall approximation fails. The black dotted lines in Fig.~\ref{fig:minm pi ball} show the mass at which gravity cannot be neglected from Eq.~\eqref{eq:piball_mmax}. Note that when the constraints on axion parameter space from neutron star measurements we will derive in section~\ref{sec:constraints} are taken into account, these $\pi$-balls should typically have masses larger than the Earth. They will have masses comparable to heavy planets or brown dwarfs, and sizes as large as a few kilometers. Observation of one of these objects would be quite remarkable.

\section{Observable implications and constraints}
\label{sec:constraints}

In this section, we estimate constraints on the exceptionally light axion parameter space from neutron star observations.   We focus on three observables based on modifications to the crust of an axion-condensed neutron star: thermal relaxation times of the crust in accreting systems (Sec.~\ref{ssec:thermalrelaxation}), cooling rates of isolated neutron stars based on a diminished envelope, first considered in Ref.~\cite{Gomez-Banon:2024oux}, (Sec.~\ref{ssec:isolatedcooling}), and fractional moment inertia of the crust as inferred from pulsar glitches (Sec.~\ref{ssec:glitches}). All of these rely only on the understanding of neutron star matter and dynamics below saturation density. In Sec.~\ref{ssec:ZNmodel}, we summarize the constraints in the context of the \ZN model of exceptionally light axions.

We note also that the mass-radius relation of axion condensed stars has some unusual features (see Sec.~\ref{ssec:NSstructure}) which could lead to constraints on axion models. Current measurements of the masses and radii of neutron stars are not yet precise enough to draw conclusions. Next-generation experiments may significantly reduce the error bars and yield a useful constraint if the EOS for a range of densities, and for both $\theta=0$ and $\theta=\pi$, is better understood. 

To understand the constraint from observation, we  apply the methods discussed in Sec.~\ref{sec:neutron stars} and specifically the understanding of the axion domain wall established in Sec.~\ref{ssec:domain_wall}.
There are three distinct regions for this calculation. First, when $\theta = \pi$ is the favored phase all the way to zero pressure, the domain wall sits at the surface and forms a thin shell surrounding the rest of the star at negative bulk pressure, resulting in a thinner crust if $\varepsilon$ is not too small (cf. Fig.~\ref{fig:crust properties}). Second, when the transition from $\theta = \pi$ to $\theta = 0$ occurs in the crust of the normal phase, the domain wall connects a homogeneous axion condensed phase to a heterogeneous normal phase and the size of the crust is diminished. Third, when the phase transition to $\theta = \pi$ occurs at a density above the crust-core phase transition, the entire normal crust is present and we draw no constraint. 

If the domain wall is inside the star, the entire normal envelope is present and constraints can only be drawn from thermal relaxation and glitches, although the constraint from thermal relaxation is much weaker. When the domain wall is at the surface, all three constraints are relevant. Thus far we have focused on results for $\varepsilon\gtrsim 10^{-6}$; here, for the sake of calculating a constraint from the lack of an envelope, we consider $\varepsilon$ as low as $5 \times 10^{-7}$ as this is the $\varepsilon$ at which the envelope appears for positive pressure at $\theta = \pi$ (i.e. not only in the domain wall). Table~\ref{tab:crust ranges} shows the ranges of $\varepsilon$ relevant for each region.

\begin{table}[tb]
\renewcommand{\arraystretch}{1.2}
    \setlength{\tabcolsep}{12pt}
         \caption{Range of $\varepsilon$ corresponding to different locations of the domain wall in the thin-wall approximation ($m_a\gtrsim10^{-8}$~eV).}
    \centering
        \begin{ruledtabular}
    \begin{tabular}{p{4.2cm}cccc}
    % \hline
        Region & (A) $K_\sigma = 0$ & (B) $K_\sigma = 0.08$ & 
        (C) $K_\sigma = 0.16$ \\
        \hline 
        DW at surface & $\varepsilon < 0.068$ & $\varepsilon < 0.27$ & $\varepsilon < 0.61 $\\
         % \hline
        DW in crust & $0.068 < \varepsilon < 0.27$ & $0.268 < \varepsilon < 0.56$  & $0.61 < \varepsilon < 0.96$  \\
         % \hline 
        DW in core at $n_B<n_{\rm sat}$ & $0.27 < \varepsilon < 0.38$ & $0.56 < \varepsilon< 0.75$ & $0.96 < \varepsilon < 1$  \\
        % \hline
    \end{tabular}
        \end{ruledtabular}
    \label{tab:crust ranges}
\end{table}

\subsection{Constraint from crust thermal relaxation in x-ray binaries}
\label{ssec:thermalrelaxation}
Neutron stars in transiently accreting x-ray binaries go through periods of outburst and quiescence, reaching a quasi-steady state on astronomical timescales \cite{Brown_1998}. During outburst, matter is rapidly accreted onto the neutron star surface from a companion and pycnonuclear reactions deep in the crust heat the star. On timescales of tens to hundreds of days (see, e.g., Ref.~\cite{Cackett_2008}) the crust cools and comes back into thermal equilibrium with the core. Properties of the crust of neutron stars in these x-ray binaries can be inferred by observing the timescale for the crust to thermally relax following an outburst. The thermal timescale is roughly given by $\tau_{\rm th} \simeq C_V (\Delta r)^2 / \kappa$ where $C_V$ is the specific heat of the crust, $\kappa$ is the thermal conductivity, and $\Delta r$ is the thickness of the crust \cite{PRL_pagereddy}. 

The specific heat capacity, $C_V$, in the crust of a neutron star is influenced by electrons, ions, and neutrons, depending on the ambient temperature. However, the thermal conductivity, $\kappa$, is primarily determined by electrons across all relevant temperatures \cite{Chamel:2008ca}. For typical crustal temperatures in the range of $10^7$-$10^8$ K, as found in accreting neutron stars, the electron contribution provides a lower bound for $C_V $, while the neutron contribution—typically suppressed due to superfluidity—sets an upper bound.  

In Scenario A, since nuclear interactions remain unchanged, the crust's composition is also relatively unaltered for pressures much larger than the axion pressure. In Scenarios B and C we expect an increase in electron density above the neutron drip pressure of normal nuclear matter with a maximum increase by a factor of $3$-$4$ near the crust-core boundary in Scenario C. In all scenarios, electron conduction is primarily limited by electron-phonon scattering. To estimate uncertainties in electron scattering rates due to electron-phonon interactions in Scenarios B and C, we modified the phonon spectrum for solids with varying nuclear charge $Z$, mass number $A$, and the corresponding plasma frequency, $ \omega_p \sim (4\pi e^2 Z^2 n_I / A m_n)^{1/2}$ \cite{Chamel:2013}. Axion condensation softens the spectrum of phonons, enhancing electron-phonon scattering and suppressing the contribution of phonons to $C_V$ (for a review of thermal transport in neutron star crusts, see Ref.~\cite{page2012thermal}).

At equal nuclear pressure (i.e. not including the negative axion pressure) the lack of dripped neutrons generically enhances $\kappa$ in Scenarios B and C due to a significantly larger electron density when nuclei are treated in mean field theory. In order to outweigh the effects of increased electron density, $Z / A$ would need to decrease by about an order of magnitude. At these densities, $A > 100$ and including shell effects is unlikely to produce such a massive change. Nonetheless, lacking an \textit{ab initio} calculation of neutron-rich nuclei in the axion condensed phase, we place a generous uncertainty on this prediction. At low temperatures, the effect on $C_V$ from axion condensation is dominated by a decreased contribution from phonons, which as with $\kappa$ is susceptible to modifications from shell effects. For $T \gtrsim 5 \times 10^7 \, \mathrm{K}$ the increased electron density dominates the change to the heat capacity and shell effects are less important for $C_V$. When shell effects are neglected, the increase to $C_V$ at high temperature is always smaller than the increase to $\kappa$, leading to a net decrease in $C_V / \kappa$.

We do not include any free neutrons in the heat capacity in our comparisons as the neutrons form a superfluid at most densities they appear; including them would only strengthen this claim by introducing an additional reservoir of heat. Furthermore, $C_V / \kappa$ decreases as a function of nuclear pressure at these densities, so at equal pressure after including the axion pressure, $C_V / \kappa$ would be expected to increase in the axion condensed phase at equal total pressure. At the lowest densities, shell effects in nuclei are expected to dominate and it is challenging to predict the exact properties of these nuclei. However, in the range of $\varepsilon$ that we constrain with thermal relaxation, the part of the crust with positive pressure is all above the normal neutron drip pressure and these shell effects are only dominant in the thin domain wall. 

 To establish a bound on axion condensation from crust relaxation—without performing detailed self-consistent calculations of $C_V$ and $\kappa$—we exploit the quadratic dependence of the thermal relaxation timescale on crust thickness. As a conservative constraint, we require that the total crust thickness be at least 0.2 km, which is approximately one-fifth of its value in a normal neutron star. This assumption effectively limits the increase of $C_V / \kappa$ to no more than a factor of 25 when $\theta = \pi$, larger than the maximum increase in $C_V / \kappa$ of a factor of 10 at $\theta=\pi$ suggested by the preceding arguments.

% The calculation of $C_V$ and $\kappa$ is not straightforward, depending on pasta phases, superfluidity, magnetic field configuration, and the EOS~\cite{PRL_pastabfieldcrustcooling2015, ChaikinKaminkerYakovlev2018, PotekhinChabrier2021} and is likely to be significantly modified in Scenarios B and C where the crust has more electrons and lacks free neutrons.

% Without making this detailed calculation, we rely on the quadratic dependence of the thermal relaxation timescale on the thickness of the crust and place the conservative constraint that the total crust thickness should be at least 0.2 km, roughly one-fifth of its value in a normal neutron star. 

\begin{figure}[tb] %[ht]
\begin{center}
\includegraphics[width=\textwidth]{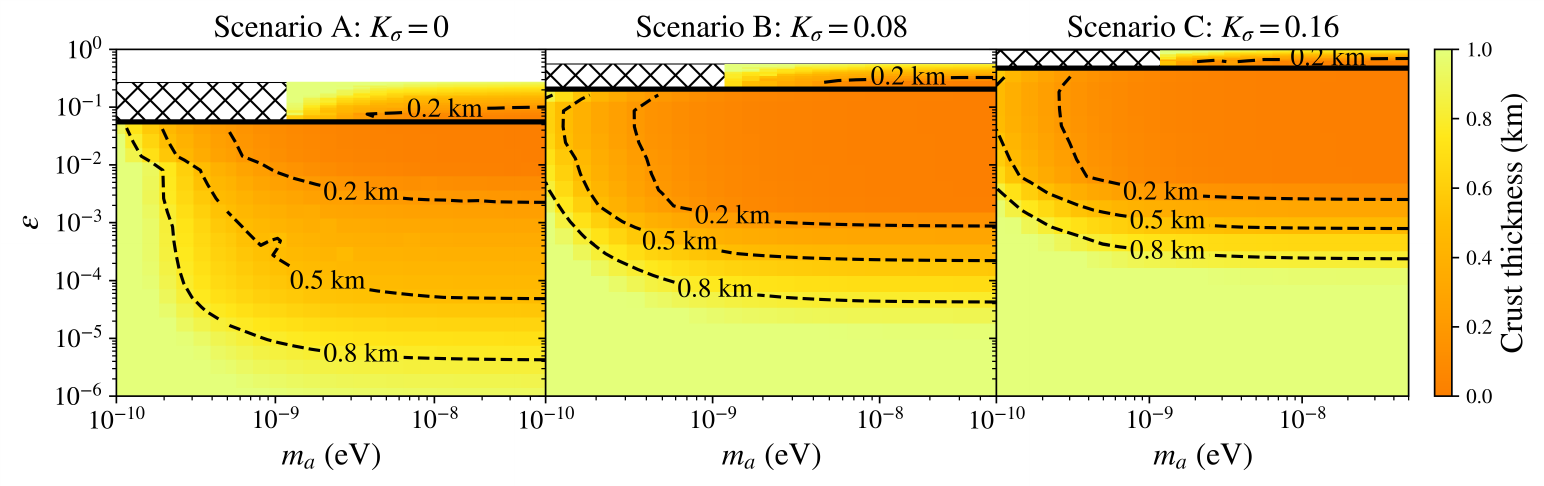}
\caption{Total crust thickness of a $1.4 \, M_\odot$ star as a function of $m_a$ and $\varepsilon$ in the RMFT model for three values of $K_\sigma$. In the hatched region, the domain wall approximation fails. In the white band, the entire normal crust is present.}
\label{fig:crust_thickness_eps_ma}
\end{center}
\end{figure}

\begin{figure}[tb] %[ht]
\begin{center}
    \includegraphics[width=0.7\textwidth]{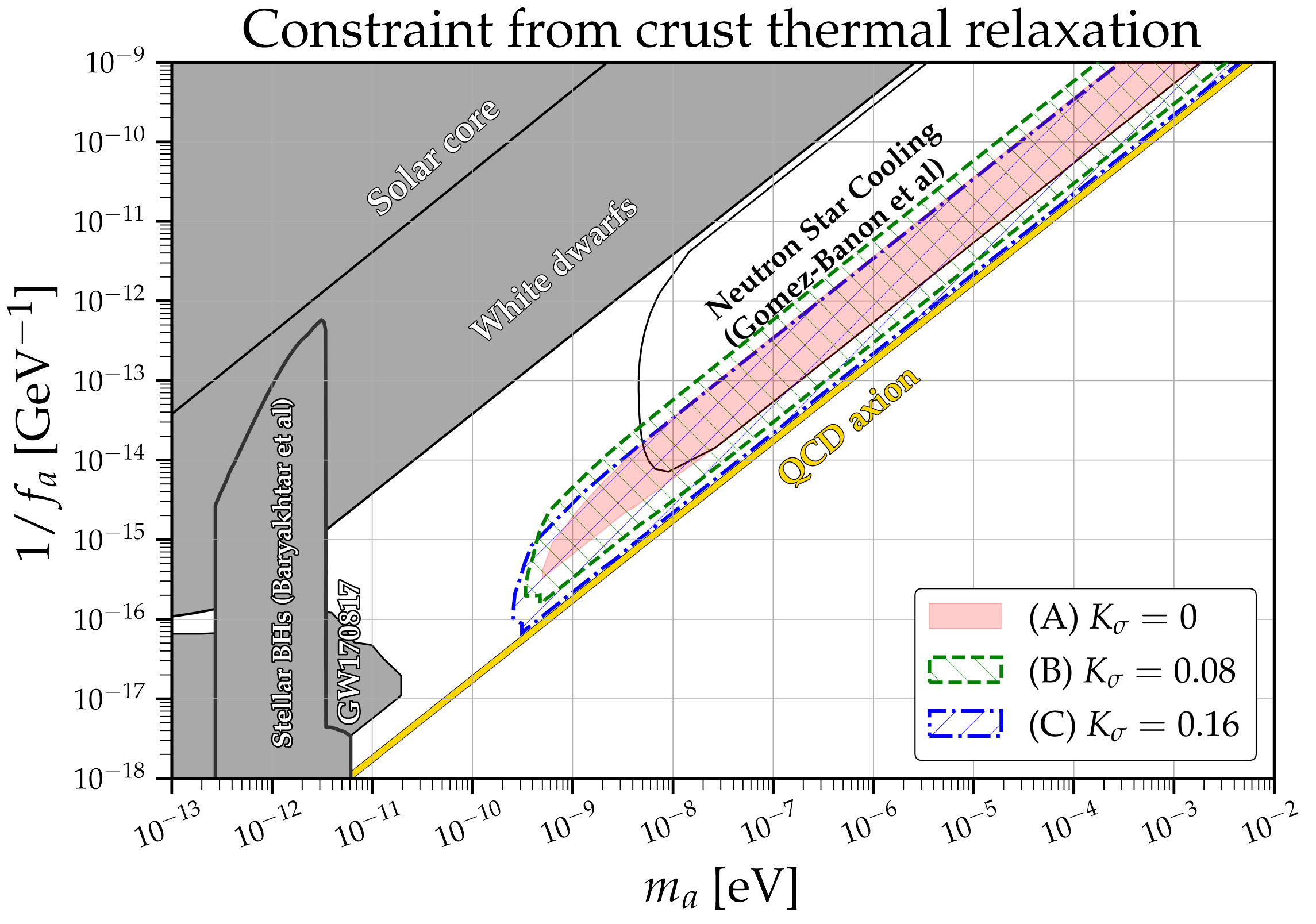}
    \caption{Constraint from crust thermal relaxation assuming a total crust thickness of $\Delta r < 0.2$~km is disfavored by observation of thermal relaxation timescales. Three scenarios are shown, assuming no change to nuclear interactions (A), which we take as the conservative bound, and moderate (B) or large (C) increases of attractive nuclear interactions in $\theta=\pi$ matter (see Sec.~\ref{sec:nuclear matter} for details). The grey shaded regions are excluded by prior studies of the Sun and white dwarf stability~\cite{Hook:2017psm,Balkin:2022qer}, black hole superradiance~\cite{Baryakhtar:2020gao,Hoof:2024quk} and neutron star inspiral gravitational wave emission~\cite{Zhang:2021mks}. Recent constraints from isolated neutron star cooling~\cite{Gomez-Banon:2024oux} are shown as a contour. Plots of past constraints are generated with code from Ref.~\cite{ciaran_o_hare_2020_3932430}.}
\label{fig:constraint relaxation}
\end{center}
\end{figure}

Figure~\ref{fig:crust_thickness_eps_ma} shows the total crust thickness as a function of $m_a$ and $\varepsilon$ for a $1.4\, M_\odot$ neutron star for our three scenarios in the RMFT model. The white band covers regions of the parameter space where the entire normal crust is present (i.e., the phase transition happens at a pressure greater than the crust-core boundary in a normal star) and the hatched region shows where the in-medium axion mass is low enough in the domain wall region that the constant chemical potential approximation fails. The solid black line divides the region where the domain wall is at the surface of the star and the region where it is inside the star. Based on the conservative estimate that the total crust thickness must be at least 0.2 km, our constraint is shown for our three scenarios in Fig.~\ref{fig:constraint relaxation}. Table~\ref{tab:crust constraint} shows the constraint on $\varepsilon$ in our three scenarios for $m_a \gtrsim 10^{-8} \, {\rm eV}$ where the domain wall becomes insignificant and the constraint is dependent on $\varepsilon$ only. 

Scenario B gives the strongest constraint for small $\varepsilon$ from crust thermal relaxation because the pressure at the crust-core phase transition that determines at what $\varepsilon$ the crust appears is non-monotonic as a function of $K_\sigma$ (cf. Fig.~\ref{fig:rmf phase diagram}). As $K_\sigma$ is increased from zero, the pressure for the crust-core phase transition decreases because the neutron matter surrounding nuclei in the inner crust becomes more attractive. At the same time, pressure in the outer crust that is dominated by electrons is increasing as nuclei become more neutron-rich. The pressure at the crust-core phase transition will decrease as $K_\sigma$ is increased, giving a stronger bound on small $\varepsilon$, until the crust-core phase transition occurs at the same pressure as the transition from inner to outer crust (neutron drip pressure). The pressure at the crust-core phase transition will grow if $K_\sigma$ is increased further, giving a weaker bound on $\varepsilon$, since there are no more dripped neutrons in the crust and the pressure due to electrons increases with $K_\sigma$.

\begin{table}[tb]
\renewcommand{\arraystretch}{1.2}
    \setlength{\tabcolsep}{15pt}
    \caption{Constraint on $\varepsilon$ from crust thermal relaxation for $m_a \gtrsim 10^{-8} \, {\rm eV}$.}
    \centering
        \begin{ruledtabular}
    \begin{tabular}{lccc}
    % \hline
        & (A) $K_\sigma = 0$ & (B) $K_\sigma = 0.08$ & 
        (C) $K_\sigma = 0.16$ \\
        \hline 
        Min. excluded $\varepsilon$ & $2.2 \times 10^{-3}$ & $8.7 \times 10^{-4}$ & $2.5 \times 10^{-3}$\\
         % \hline
        Max. excluded $\varepsilon $& $0.10$ & $0.32$  & $0.69$  \\
         % \hline 
    \end{tabular}
        \end{ruledtabular}
    \label{tab:crust constraint}
\end{table}

\subsection{Constraint from isolated neutron star cooling}
\label{ssec:isolatedcooling}
The cooling of isolated neutron stars has been used to constrain light QCD axions in Ref.~\cite{Gomez-Banon:2024oux}. Their constraint hinges on the envelope being thinner in a star with an axion-condensed crust than in a normal neutron star, providing less thermal insulation between the core and exterior of the star and shortening the cooling timescale of the core. Taking the boundary of the envelope to be at an energy density of $10^{10}$ g/cm$^3$ when the energy density of the axion is not included, we find that for the range of $\varepsilon$ we consider, the entire envelope is in the domain wall region. 

\begin{figure}[tb] %[ht]
\begin{center}
\includegraphics[width=\textwidth]{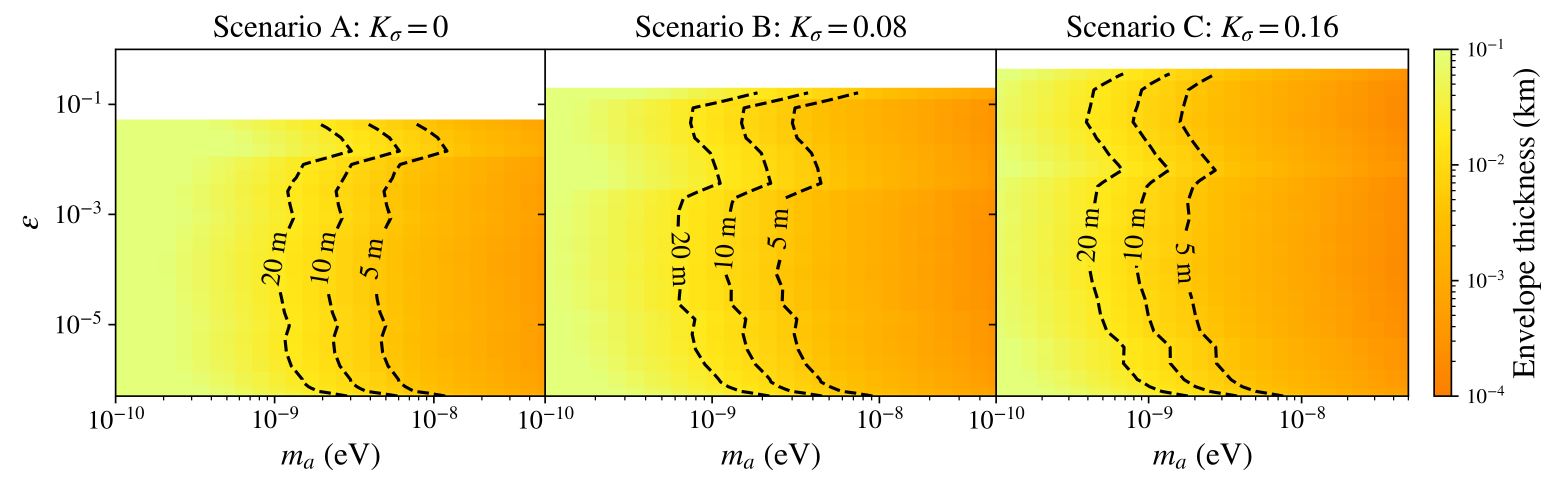}
\caption{Envelope thickness of a $1.4 \, M_\odot$ neutron star as a function of $m_a$ and $\varepsilon$ in the RMFT model for the three  $K_\sigma$ scenarios. In the white band, the entire normal envelope is present.}
\label{fig:envelope_thickness_eps_ma}
\end{center}
\end{figure}

\begin{figure}[tb] %[ht]
\begin{center}
    \includegraphics[width=0.7\textwidth]{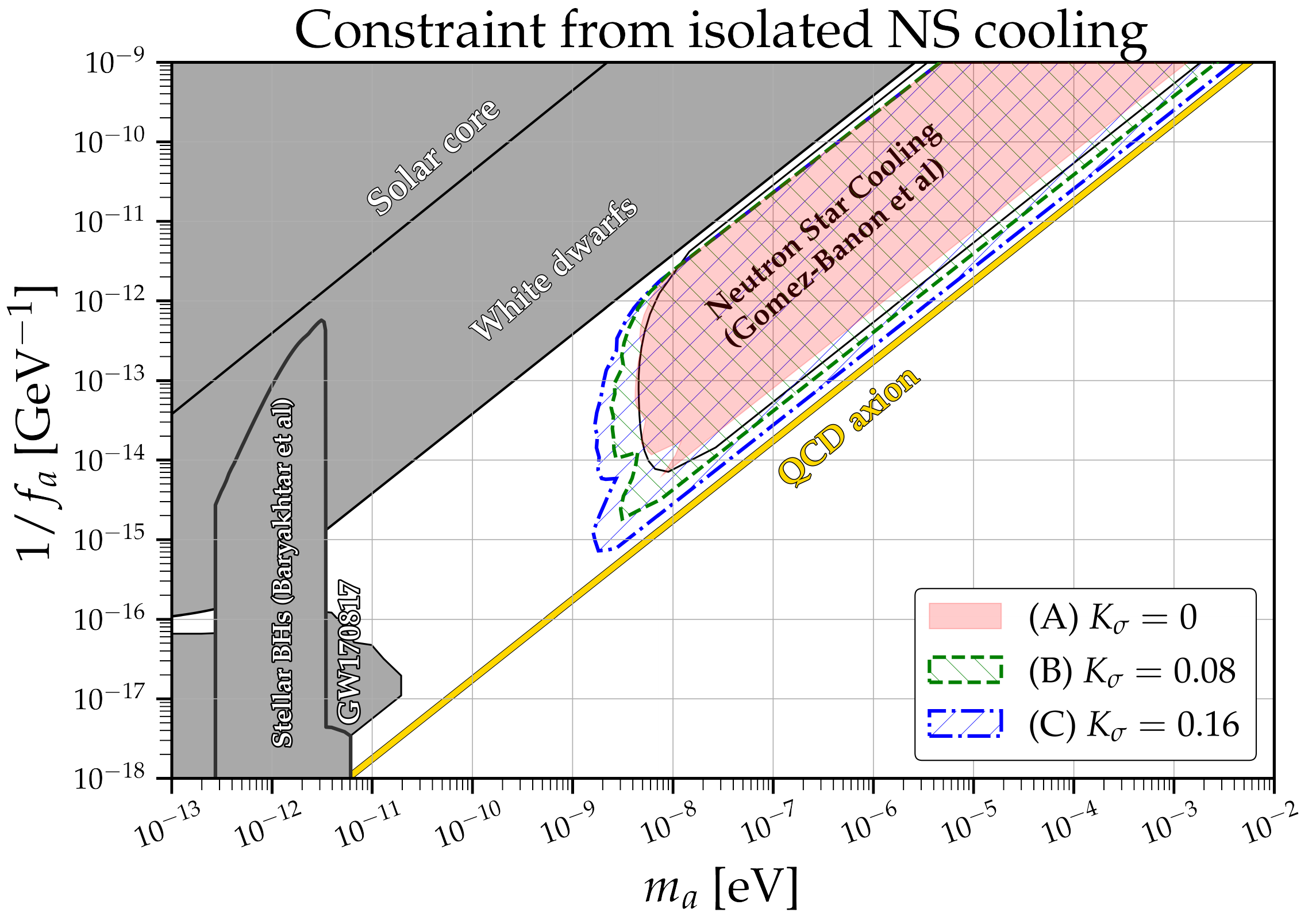}
    \caption{Constraint from isolated neutron star cooling based on the lack of an envelope in our three scenarios. We use the envelope thickness of $5$~m as a conservative minimum consistent with neutron star cooling. Our scenario (A) is in good agreement with the neutron star cooling bound of Ref.~\cite{Gomez-Banon:2024oux}. Parameter space as in Fig.~\ref{fig:constraint relaxation}.}
\label{fig:constraint env}
\end{center}
\end{figure}

Figure~\ref{fig:envelope_thickness_eps_ma} shows the thickness of the envelope as a function of $m_a$ and $\varepsilon$. In the white band, the phase transition to axion condensed matter occurs inside the star and the entire normal envelope is present. The feature around $\varepsilon \simeq 10^{-2}$ in all three panels is a result of a jump in the EOS just as the crust-core phase transition is at the bottom edge of the domain wall. Ref.~\cite{Gomez-Banon:2024oux} found that when the envelope is at least a few times thinner than its normal $\sim 100$ m, the star cools too quickly and this region of parameter space is excluded. Taking a conservative estimate that the envelope must be at least 5 m thick to explain neutron star cooling data, we show the constraint for our three scenarios in Fig.~\ref{fig:constraint env}. Notably, this indicates that the constraint in Ref.~\cite{Gomez-Banon:2024oux} extends to lower axion masses and larger $\varepsilon$ than previously considered when the modification to the nuclear force is taken into account. This occurs because the chemical potential in the crust is significantly lower for larger $K_\sigma$ due to nuclei being more strongly bound. At a chemical potential much lower than that of normal nuclei, the baryon density goes to zero much more rapidly as $\theta$ decreases in the domain wall, resulting in a thinner envelope. The envelope constraint applies for all $\varepsilon$ we consider where the domain wall is at the surface of the star (given by the first row of Table~\ref{tab:crust ranges}). At larger $\varepsilon$, the domain wall is inside the star and the entire normal envelope is present. Table~\ref{tab:env constraint} shows the constraint on $\varepsilon$ in our three scenarios for $m_a \gtrsim 10^{-8} \, {\rm eV}$ where the domain wall becomes insignificant and the constraint is dependent on $\varepsilon$ only.

\begin{table}[tb]
\renewcommand{\arraystretch}{1.2}
    \setlength{\tabcolsep}{15pt}
    \caption{Constraint on $\varepsilon$ from isolated neutron star cooling for $m_a \gtrsim 10^{-8} \, {\rm eV}$. }
    \centering
        \begin{ruledtabular}
    \begin{tabular}{lccc}
    % \hline
        & (A) $K_\sigma = 0$ & (B) $K_\sigma = 0.08$ & 
        (C) $K_\sigma = 0.16$ \\
        \hline 
        Min. excluded $\varepsilon$ & $5.2 \times 10^{-7}$ & $5.8 \times 10^{-7}$ & $5.6 \times 10^{-7}$\\
         % \hline
        Max. excluded $\varepsilon $& $0.068$ & $0.27$  & $0.61$  \\
         % \hline 
    \end{tabular}
        \end{ruledtabular}
    \label{tab:env constraint}
\end{table}

We mention in passing that a detailed calculation of constraints from neutron star cooling should include the modification to the Direct Urca threshold from axion condensation.  When the central density of a neutron star exceeds the Direct Urca threshold (set by the density at which $k_{Fn} < k_{Fp} + k_{Fe}$, equivalently $Y_p \geq 0.11$ in matter containing neutrons, protons, and electrons), rapid cooling becomes possible.  The isovector contribution to the nucleon masses from finite $\theta$ will result in a larger proton fraction at $\theta = \pi$, lowering the Direct Urca threshold. Additionally, a more attractive nuclear force with a lighter effective nucleon mass will result in a larger symmetry energy and a lower Direct Urca threshold. In the RMFT model, the symmetry energy is given by
\begin{equation}
    a_{\rm sym} = \frac{k_F^2}{6 \sqrt{k_F^2 + m^{*2}}} + \frac{n_B}{8} \left( \frac{g_\rho}{m_\rho} \right)^2\,,
\end{equation}
where $k_F$ is the Fermi momentum of SNM and $m^*$ is the effective nucleon mass, decreased from its vacuum value by the effect of the $\sigma$ meson. Finite $K_\sigma$ results in a smaller effective nucleon mass $m^*$, increasing the symmetry energy and lowering the Direct Urca threshold. The exact value of the Direct Urca threshold in nuclear matter is not known and the effects of the axion at the relevant baryon densities is uncertain. Table~\ref{table:du thresholds} shows the baryon density at the Direct Urca threshold for the IUFSU$^*$ RMFT parameter set we have used throughout. These values should be considered illustrative of possible impacts of axion condensation on neutron star cooling and not a concrete prediction.

\begin{table}[tb] %[ht]
\renewcommand{\arraystretch}{1.2}
    \setlength{\tabcolsep}{15pt}
\caption{Baryon density at the Direct Urca threshold ($k_{Fn} < k_{Fp} + k_{Fe}$) in the RMFT model.}
\centering
    \begin{ruledtabular}
\begin{tabular}{ccccc}
    % \hline
    & No axion & (A) $K_\sigma = 0$ & (B) $K_\sigma = 0.08$ & (C) $K_\sigma = 0.16$ \\
    \hline
    $n_{DU}$ (fm$^{-3}$) & 0.61 & 0.54 & 0.48 & 0.42 \\
    % \hline
\end{tabular}
    \end{ruledtabular}
\label{table:du thresholds}
\end{table}

\subsection{Constraint from pulsar glitches}
\label{ssec:glitches}
Pulsars are spinning neutron stars that emit regular pulses of EM radiation at twice their spin frequency. Pulsar timing studies since the 1970s have cataloged about 3000 sources that exhibit remarkable stability. Precise measurements of the spin period and its steady rate of change have provided useful insights about neutron star populations and their magnetic fields. Some pulsars, however, have been observed to glitch, a phenomenon in which the rotation frequency rapidly increases on timescales of less than a minute. For a concise review of pulsar timing and glitches see \cite{Manchester:2017ykr}. 

In crust-based models of neutron star glitches, the phenomenon is thought to result from a sudden transfer of angular momentum from an interior superfluid that stores angular momentum in quantized vortices to the crust. As the star gradually spins down, the superfluid cannot adjust smoothly because its vortices are pinned—or otherwise hindered—by a lattice of neutron-rich nuclei \cite{andersonitoh_glitches1975, Ruderman_glitches1976, Pines_superfluidityinNS1985, Link_2022}. Over time, this creates a buildup of angular momentum stress between the superfluid and the rest of the star. Eventually, this tension is released in a sudden, dramatic, and still poorly understood event that unpins many vortices at once, rapidly spinning up the star.

The spin-up is proportional to the ratio of the moment of inertia of the superfluid component, denoted by $I_s$, to the moment of inertia of the star, denoted by $I$.         
%This is typically understood as resulting from a rapid transfer of angular momentum from the core to the neutron superfluid in the crust of a normal neutron star \cite{andersonitoh_glitches1975, Ruderman_glitches1976, Pines_superfluidityinNS1985}. 
%After including the effects of coupling the solid and superfluid components of the crust, this provides a lower limit on the fractional moment of inertia in the crust. 
The Vela pulsar places the most stringent constraint on glitch models, requiring that the ratio of superfluid moment of inertia to the total moment of inertia lies between 1.6\% and 7\% \cite{PRL_crustnotenough}. While there remains ongoing debate about whether nuclear physics models can account for values at the upper end of this range \cite{PRL_crustalentrainment_chamel,PRL_crustnotenough,PRC_crustmaybenough}—and speculation about additional angular momentum reservoirs being necessary to fully explain Vela's glitches \cite{PRD_glitches_2016}—the conventional model, involving a neutron superfluid in the inner crust, remains both natural and viable.

The glitches of the Vela pulsar are challenging to describe in terms of other available models for glitch behavior.  Starquakes are a possible alternate explanation for pulsar glitches \cite{PRL_starquakes, BaymPines1971} and recent observations suggest that the Vela pulsar may experience starquakes \cite{Bransgrove_2020}. However, to explain the high frequency of glitches in the Vela pulsar with starquakes requires either an extreme equation of state in tension with observations of NS masses and radii \cite{Negi_starquakes_eos} or results in the Vela pulsar having an unrealistically small mass \cite{Negi_starquakes_smallmass} with the connection between starquakes and glitches still typically requiring the presence of a neutron superfluid~\cite{Quakesneedneut}. While the possibility of a phase transition deep in the neutron star is potentially relevant for glitches, without any concrete evidence for such a phase transition, this explanation remains speculative. 

In the following, we constrain $\varepsilon$ within the framework of the standard glitch model, which requires a substantial superfluid component in the inner crust with $I_s/I>1.6\%$. In our Scenarios B and C, when axions condense in the crust, the crust-core phase transition occurs at a baryon density below the neutron drip, and there is no region of coexistence of solid and superfluid matter in the crust. Consequently, glitches cannot be explained in the presence of an axion condensate at the surface of the star when the nuclear force is more attractive. Thus, within the standard glitch models, $\varepsilon < 0.27$ is ruled out for Scenario B and $\varepsilon < 0.61$ is ruled out for Scenario C provided that $f_a$ is not so large that the surface of star has small $\theta$ due to the large domain wall. Even in the domain wall, the increasing nucleon mass brings matter below neutron drip before the interactions become sufficiently repulsive to force neutrons out of nuclei. This result is in part a consequence of the fact that in the ansatz we use in the RMF model, the nuclear force becomes monotonically more attractive as $\theta$ increases. If nuclear matter with moderate isospin asymmetry were only more attractive near $\theta = \pi$ and rapidly became repulsive for intermediate $\theta$, it is possible that enough neutrons could be present in the domain wall to explain glitches when $m_a \ll 10^{-8} \, \mathrm{eV}$. This is an issue that warrants further exploration in light of the prediction from our MBPT calculation that neutron matter will be more repulsive at intermediate $\theta$ and the work of past authors \cite{Cohen:1991nk, Kaiser:2007nv, Plohl:2007dp, Kruger:2013iza} that showed that reducing the pion mass made nuclear interactions in neutron matter more repulsive in the vicinity of the physical pion mass (but not at significantly reduced pion mass, as in the axion condensed phase).

We estimate this critical $f_a$ by treating the neutron star as a uniform sphere of saturation density matter. The approximation of constant density is conservative since most of the star is at a higher density than the reference density used but is also not particularly impactful since within this approximation $f_a^2 \propto n_B^s$ and $n_B^s$ changes by only a factor of at most $3-5$ within the core of a $1.4 \, M_\odot$ star. Within this constant density approximation, the maximum $f_a$ can be found by solving an inhomogeneous Poisson's equation.
\begin{equation}
    \nabla^2 \theta = - \frac{1}{r_0^2} \frac{\sin \theta}{f(\theta)}
\end{equation}
where $r_0$ is given by
\begin{equation}
    r_0^{-2} = \frac{\sigma_N \nsat (1 + \delta^{\rm (loc)})}{f_a^2} - m_a^2 \,.
\end{equation}
In practice, $m_a$ has a very small effect on this constraint for $\varepsilon$ that are not very close to the QCD axion line, which this part of the constraint does not touch. In the regime of $m_a \ll 1 / R_{NS} \simeq 10^{-10} \, \mathrm{eV}$, this can be solved by matching the surface boundary condition to the lowest energy configuration of a massless field, $\theta(r) = \theta_0 R_{NS} / r$, where $\theta_0$ is the critical $\theta$ at which dripped neutrons can appear at positive pressure. The bound on $f_a$ we find is below the value necessary for the system to be stable against condensation due to the large extent of the axion field. At the $\varepsilon$ relevant for this part of the constraint, $\beta$-equilibrated matter at or above saturation density is unstable to axion condensation and the system can adiabatically evolve to its ground state. If a more detailed calculation of the nuclear forces shows that $\beta$-equilibrated matter becomes significantly more repulsive for intermediate $\theta$ before becoming attractive at large $\theta$, the exact limits of this bound will need to be revisited but the findings will be qualitatively unchanged.

In Scenario A where dripped neutrons can appear in an axion condensed crust or in any scenario when the critical pressure occurs in the outer crust of a normal star, if the domain wall is large a significant number of neutrons can be present in the domain wall region. For this case, we limit $m_a > 10^{-8} \, \mathrm{eV}$ so that the domain wall has a thickness of $\mathcal{O}(10 \, \mathrm{m})$, more than an order of magnitude smaller than is normally seen. Since the neutron superfluid mechanism for glitches as currently understood requires nearly all of the superfluid neutrons to participate in a normal star, a reduction this severe excludes this possibility.  This applies for Scenario A for $0.0083 < \varepsilon < 0.17$, for Scenario B for $0.27 < 0.42$, and for Scenario C for $0.61 < \varepsilon < 0.79$. 

\begin{figure}[ht]
\begin{center}
    \includegraphics[width=0.7\textwidth]{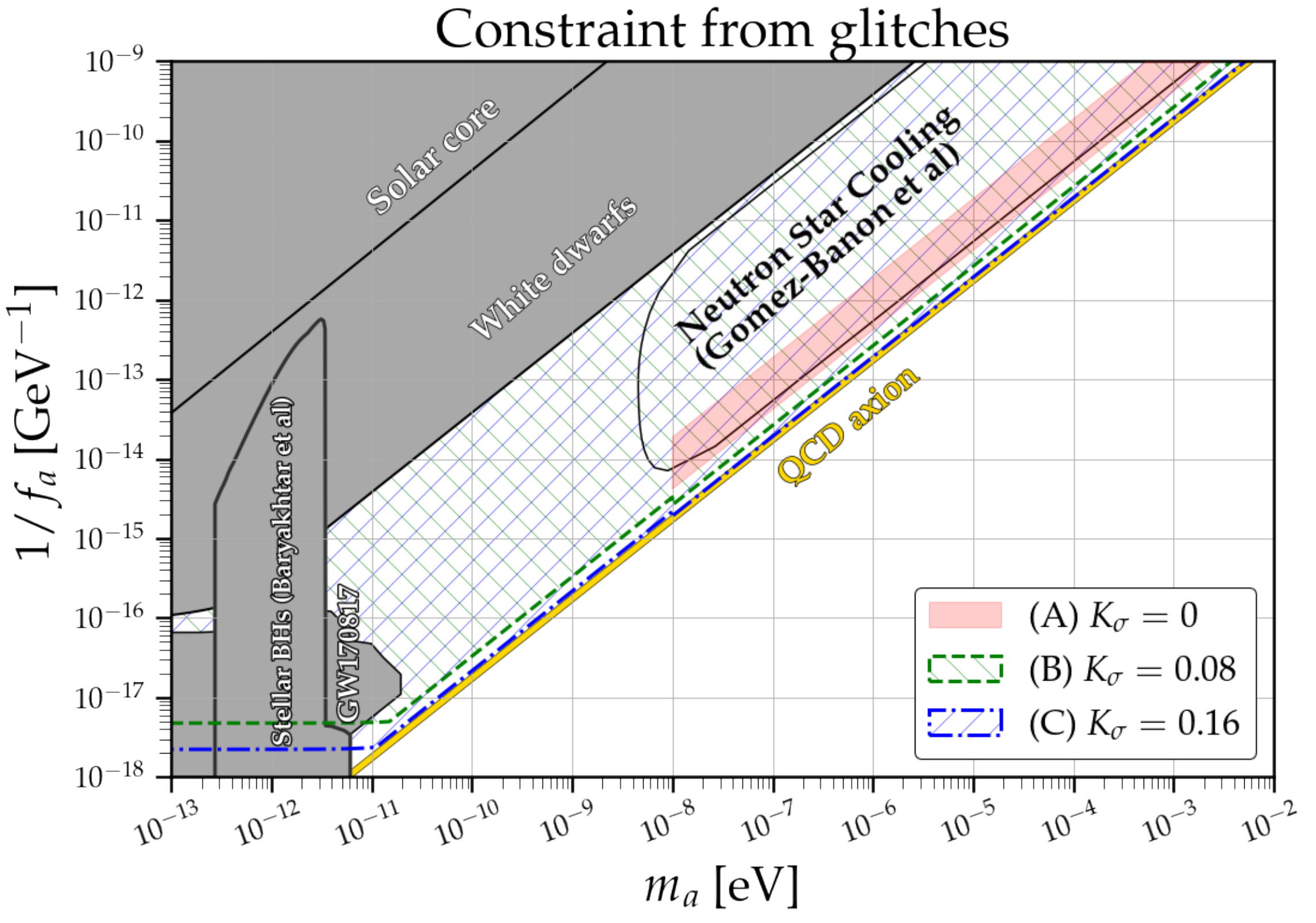}
    \caption{Constraint from observation of crust-based pulsar glitches in our three scenarios requiring the quantity of neutrons be reduced by less than an order of magnitude. Parameter space as in Fig.~\ref{fig:constraint relaxation}.}
\label{fig:constraint glitches}
\end{center}
\end{figure}

Lacking a complete and detailed picture of the crust glitch mechanism, we refrain from making a concrete numerical estimate of exactly how many neutrons are necessary to be in tension with observations and rely on the fact that near the boundaries of our constraint the quantity of neutrons changes by many orders of magnitude for relatively small changes in axion parameters. Scenario B and C are particularly notable for the fact that the neutron dripped phase is completely absent for much of the parameter space. The constraint that the quantity of neutrons is reduced by less than an order of magnitude is shown for our three scenarios in Fig.~\ref{fig:constraint glitches}. Table~\ref{tab:moi constraint} shows the constraint on $\varepsilon$ in our three scenarios for $m_a \gtrsim 10^{-8} \, {\rm eV}$, where, as before, the domain wall does not play an important role and the constraint only depends on $\varepsilon$. Note that in Fig.~\ref{fig:constraint glitches}, there is a small notch in the constraint curves for Scenarios B and C at $m_a = 10^{-8} \, \mathrm{eV}$. While this is a small effect on the typical log scale for constraint plots, it is a relevant difference when assessing how close constraints come to the QCD axion line at large and small axion masses. It should be noted that only this region of larger axion masses can be confidently constrained within this scheme when using the \ZN model, which will be discussed in greater detail in the next section.

%\begin{figure}[tb]
%\begin{center}
%\includegraphics[width=\textwidth]{moi_frac_eps_ma.pdf}
%\caption{Fractional crust moment of inertia fraction of a $1.4 \, M_\odot$ star as a function of $m_a$ and $\varepsilon$ in the RMFT model for three values of $K_\sigma$. Note the slightly different vertical scale; for smaller $\varepsilon$, the crust moment of inertia fraction stays nearly constant. In the hatched region, the domain wall approximation fails. In the white band, the entire normal crust is present.}
%\label{fig:moi_eps_ma}
%\end{center}
%\end{figure}

%Figure~\ref{fig:moi_eps_ma} shows the fractional of the moment of inertia of a $1.4\, M_\odot$ star as a function of $m_a$ and $\varepsilon$. For scenarios B and C, the moment of inertia in the crust is strongly suppressed when there is an axion-condensed crust (cf. Fig.~\ref{fig:crust properties}). Again, the white band shows where the whole normal crust is present and there is no constraint, the hatched region shows where the domain wall approximation fails, and the solid black line divides the region where the domain wall is at the surface of the star and inside the star. The constraint that the fractional moment of inertia exceeds 0.1\% is shown for our three scenarios in Fig.~\ref{fig:constraint moi}. Table~\ref{tab:moi constraint} shows the constraint on $\varepsilon$ in our three scenarios for $m_a \gtrsim 10^{-8} \, {\rm eV}$, where, as before, the domain wall does not play an important role and the constraint only depends on $\varepsilon$.

\begin{table}[tb]
\renewcommand{\arraystretch}{1.2}
    \setlength{\tabcolsep}{15pt}
    \caption{Constraint on $\varepsilon$ from observation of crust-based pulsar glitches for $m_a \gtrsim 10^{-8} \, {\rm eV}$.}
    \centering
        \begin{ruledtabular}
    \begin{tabular}{lccc}
    % \hline
        & (A) $K_\sigma = 0$ & (B) $K_\sigma = 0.08$ & 
        (C) $K_\sigma = 0.16$ \\
        \hline 
        Min. excluded $\varepsilon$ & $8.3 \times 10^{-3}$ & $0$ & $0$\\
         % \hline
        Max. excluded $\varepsilon $& $0.17$ & $0.42$  & $0.79$  \\
         % \hline 
    \end{tabular}
        \end{ruledtabular}
    \label{tab:moi constraint}
\end{table}

\subsection{\ZN model}
\label{ssec:ZNmodel}

The discrete \ZN symmetry proposed in  Ref.~\cite{Hook:2018jle} provides a concrete realization of the exceptionally light QCD axion. This model predicts a suppression factor $\varepsilon$ of the axion mass for odd integer $\mathcal{N}$, given by
\begin{equation}
    \varepsilon^{1/2} \equiv \frac{m_a}{m_a^{\rm (QCD)}} \approx \sqrt{ \frac{1}{\sqrt{\pi}} \mathcal{N}^{3/2} z^{\mathcal{N} - 1} (1 + z) \sqrt{1-z^2}} \approx \frac{1}{2^{{\mathcal{N}}/2 - \log \mathcal{N}}},
\end{equation}
where $z \equiv m_u / m_d$ and the large $\mathcal{N}$ limit has been taken~\cite{DiLuzio:2021pxd}.  For small $\mathcal{N}$, the \ZN model results in a modest change to the axion mass: for $\mathcal{N}=3$, $m_a/m_a^{\rm (QCD)} = 0.94$. The first significant decrease comes at $\mathcal{N} = 5$ giving $m_a / m_a^{\rm (QCD)} = 0.65$ and $\mathcal{N} = 11$ corresponds to approximately an order of magnitude suppression.

In the \ZN model for the exceptionally light QCD axion, the axion potential has a period of $2\pi /\mathcal{N}$ and the negative axion pressure is suppressed by an additional factor of $\mathcal{N}^2$ \cite{Hook:2018jle}. 
In the simple case where the domain wall connects $\theta = 0$ and $\theta = \pi$ smoothly and the axion mass is sufficiently large that the domain wall pressure is unimportant, the results are identical to our existing calculation but with a modified axion potential provided $\mathcal{N}$ is small enough that normal matter at saturation density is unstable to the nucleation of the $\theta = \pi$ phase. Thus, in the range $m_a \gtrsim 10^{-8}\, {\rm eV}$, the constraint on the \ZN model parameter space can be inferred directly from the constraint on $\varepsilon$ we have derived. The lower limit on $\mathcal{N}$ is determined from the difference in vacuum energy between the maximum at $\theta = \pi$ and minimum at $\theta = 0$ of the potential given in Ref.~\cite{DiLuzio:2021pxd} since the dependence on the relationship between $m_a$ and $f_a$ and the change to the period of the axion potential is minimal in this region. This lower limit on $\mathcal{N}$ for the various effects and scenarios is given in Table~\ref{tab:zn constraint}. For $m_a \gtrsim 10^{-8}\,{\rm eV}$, the combined constraint is the same for all three scenarios, $\mathcal{N} \geq 19$. Combined with the upper limit from white dwarf of $\mathcal{N} \leq 31$~\cite{Balkin:2022qer}, this leaves a gap of $19\leq \mathcal{N} \leq 31$. On the other hand, the upper limit on $\mathcal{N}$ is also determined by the conditions for the appearance of the domain wall in Tab.~\ref{tab:crust ranges}. 
Although $\mathcal{N} = 3$ and $5$ appear to be constrained in Scenario A and $\mathcal{N} = 3$ appears to be constrained in Scenario B (see Table~\ref{tab:zn constraint}), for those choices of $\mathcal{N}$ and $K_\sigma$, the axion mass remains positive near $\theta = 0$ in the \ZN model. Therefore, even though the ground state of matter is axion condensed, since matter at saturation is only metastable in these cases, it is conceivable that the axion mass squared is always positive near $\theta = 0$ anywhere in the neutron star, and these $\mathcal{N}$ cannot be conclusively excluded. However, it is possible that if axion condensation is favored at a sufficiently large density where the axion mass squared turns negative near $\theta = 0$, these $\mathcal{N}$ values could also be excluded in Scenarios A \& B. 

This might not be a complete story, if we were to only solve for the possible axion field profile inside a neutron star, while neglecting how neutron stars formed. For large $\mathcal{N}$, when only modifications to the nucleon masses are included, it was pointed out in Ref.~\cite{Balkin:2022qer} that the ground state of matter below the critical density is no longer $\theta = 0$  but $\theta = (\mathcal{N} -1 ) \pi / \mathcal{N}$, resulting in a vanishing axion potential but lighter nucleon masses . If this ground state is found, the full crust and envelope could be present at $\theta = (\mathcal{N} -1 ) \pi / \mathcal{N}$ and the constraint from neutron star cooling would not apply. %It is possible that a domain call connecting $\theta = 0$ and $\theta = \pi$ is metastable on long timescales.
In Scenario A, the constraints from crust thermal relaxation and glitches also likely do not apply in this case since a normal crust would be present. When nuclear interactions, and their dependence on $\theta$ are taken into account, the conclusion of Ref.~\cite{Balkin:2022qer} can be modified. In Scenarios B and C, the \ZN model may still be excluded since the lack of a dripped neutron phase precludes crust-based glitches. Additionally, in Scenarios B and C, the dependence of the free energy as a function of $\theta$ is modified and the minimum $\mathcal{N}$ where the $\theta = \pi$ state is stable (instead of $\theta = (\mathcal{N} -1 ) \pi / \mathcal{N}$) at zero pressure decreases. In our Scenario C, for $\mathcal{N} = 17\, \& \,19$, the ground state remains at $\theta = \pi$ at zero pressure, and is disfavored as a result. 

A better understanding of the nuclear physics at $\theta = \pi$ is necessary to provide a constraint on the \ZN model that is independent of our knowledge about how neutron stars are created, and how the axion condensed phase could form following a supernova. On the other hand, it is unclear how a profile with $\theta = \pi$ in the core of a neutron star, and $\theta = (\mathcal{N} - 1) \pi / \mathcal{N}$ in most of inner and outer crust 
could form in the event of neutron star formation in, for example, a core collapse supernova. In an event of a supernova, as the density of the core grows past the critical density in Eq.~\ref{eq:crit density},  
a region of $\theta = \pi$ emerges in the core, with a domain wall interpolating between the original $\theta = 0$ region and the newly formed $\theta = \pi$ region. This domain wall, in particular the part that interpolates between $\theta = (\mathcal{N} - 1) \pi / \mathcal{N}$ and $\theta = 0$, will need to expand significantly such that most of the inner and outer crust is at $\theta = (\mathcal{N} - 1) \pi / \mathcal{N}$ (see Fig.~8 of~\cite{Balkin:2022qer}). During this process, as described in~\cite{Hook:2019pbh}, the domain wall gains energy that is $\mathcal{O}(\sigma_N/m_n)$, roughly $5\%$ of the total mass energy of the progenitor star, or at least the total mass energy of the inner and outer crust of the proto-neutron star. This energy, as outlined in~\cite{Hook:2019pbh}, will be released slowly, heating up the medium surrounding the neutron star, leading to persistent bright X-ray/optical emissions from the supernova remnant that is not observed. Whereas it is possible that the axion field profile form nearly at rest and are not subsequently accelerated outward, it is hard to imagine how the field profile could gain this much energy through interactions with the baryons, while at the same time, experience exact the right amount of friction to land exactly at the boundary of the outer crust in every single supernova we observed~\cite{Hook:2019pbh,Balkin:2021zfd}. As a result, we do not consider this scenario further.

For axion masses below $10^{-8}\,{\rm eV}$, the distribution and properties of  nuclear matter in the axion domain wall region are significantly affected by the change of the period of the axion potential in the \ZN model, which requires further study. The parameter space in the gaps, as well as the lower axion mass region, might be probed by other considerations of the formation of neutron stars during a supernova explosion~\cite{Hook:2019pbh}. We leave a more detailed study of the \ZN model to future work.

\begin{table}[tb]
\renewcommand{\arraystretch}{1.2}
    \setlength{\tabcolsep}{15pt}
    \caption{Minimum allowed $\mathcal{N}$ in the \ZN model for $m_a \gtrsim 10^{-8} \, {\rm eV}$ when the domain wall connects $\theta = 0$ and $\theta = \pi$. In Scenario A, the envelope constraint also permits $\mathcal{N} = 3$.  $\mathcal{N}$ must be an odd integer for the axion to solve the strong-CP problem in these models. Note that these bounds assume the inner region is at $\theta =\pi$, as is preferred in a dynamical formation of the neutron star \cite{Hook:2019pbh}, rather than the critical density minimum at $\theta = (\mathcal{N} - 1) \pi / \mathcal{N}$. See text for further discussion.}
    \centering
        \begin{ruledtabular}
    \begin{tabular}{lccc}
    % \hline
        Constraint & (A) $K_\sigma = 0$ & (B) $K_\sigma = 0.08$ & (C) $K_\sigma = 0.16$ \\
        \hline 
        Crust thickness & $7$ & $9$ & $7$\\
         % \hline
        Envelope & $19$ & $19$  & $19$  \\
        % \hline 
        Glitches& $7$ & $5$  & $5$  \\
         % \hline
    \end{tabular}
        \end{ruledtabular}
    \label{tab:zn constraint}
\end{table}

\section{Conclusions}
\label{sec:conclusion}

\begin{figure}[tb]
\begin{center}
    \includegraphics[width=0.49\textwidth]{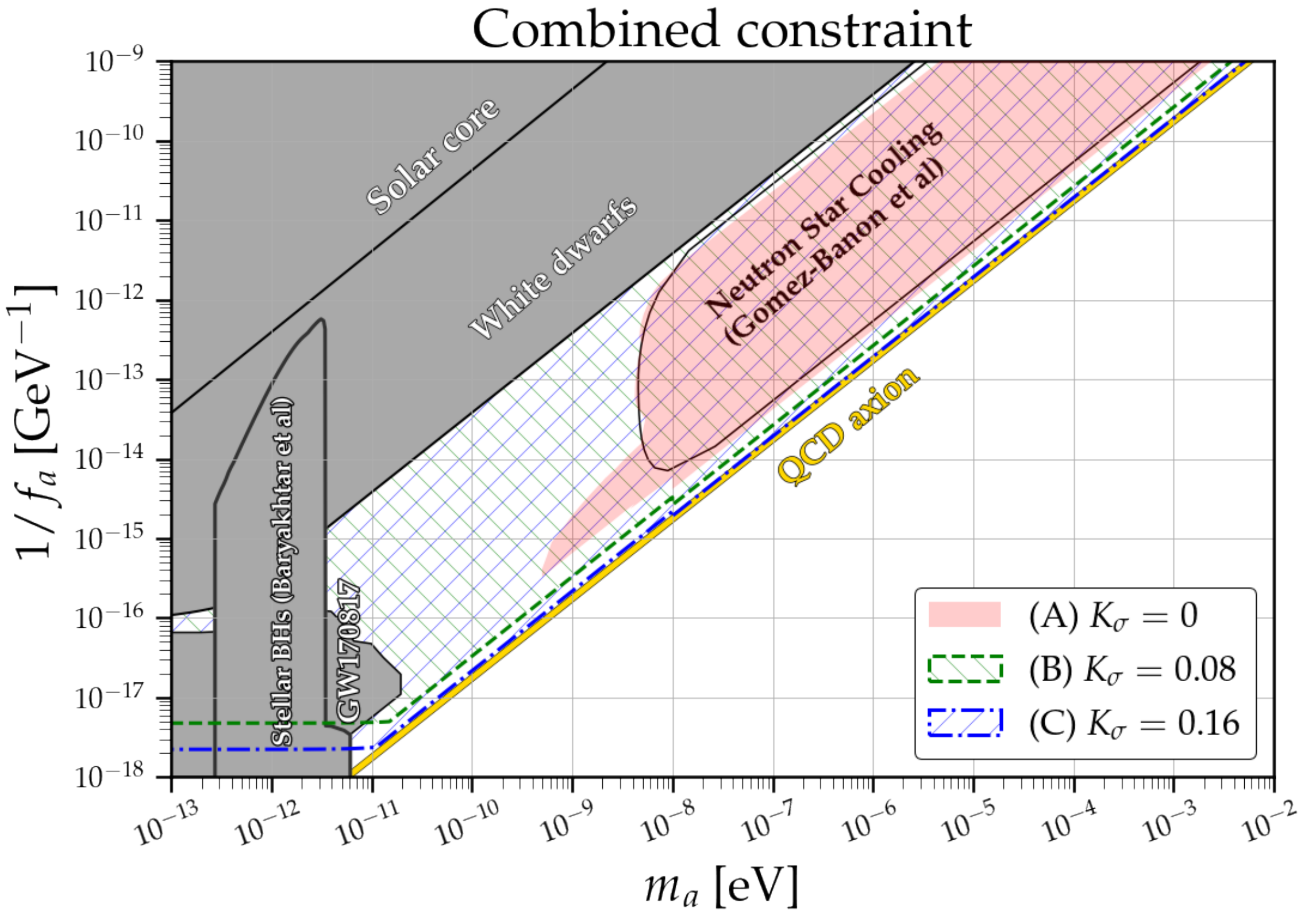}
    \hfill
    \includegraphics[width=0.49\textwidth]{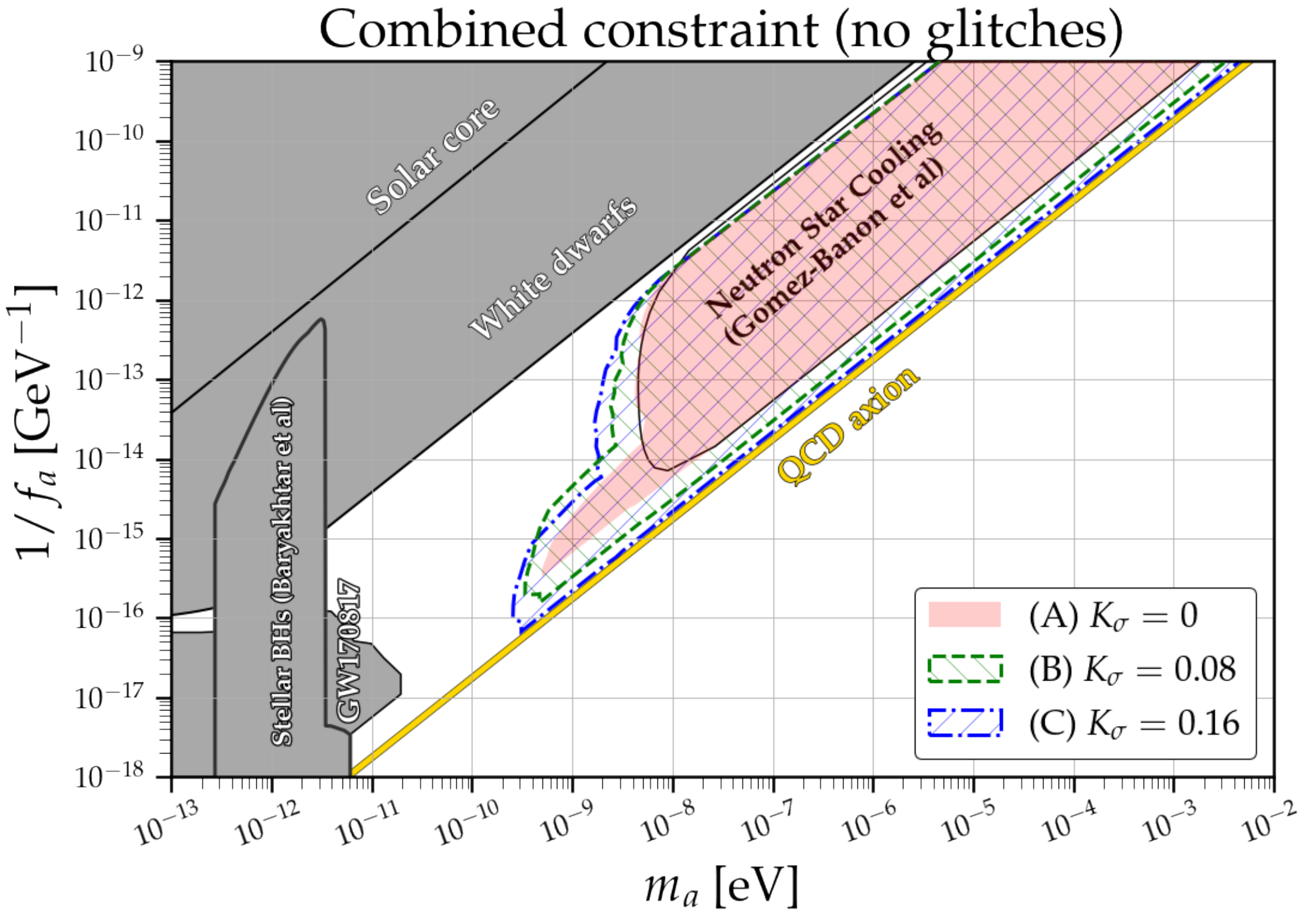}
    
    \caption{Combined constraint from crust thermal relaxation (Sec.~\ref{ssec:thermalrelaxation}), neutron star cooling (Sec.~\ref{ssec:isolatedcooling}), and pulsar glitches (Sec.~\ref{ssec:glitches}) (left) and without glitches (right) in our three scenarios accounting for modification of nuclear interactions in $\theta=\pi$ matter. In Scenario A, neutron star cooling is the dominant signature for $\varepsilon\ll 1$ while crust thermal relaxation and pulsar glitches set the strongest constraints for $\varepsilon\sim 1$. For Scenarios B and C, the glitch constraint covers the entire excluded parameter space. We consider the pink shaded region (Scenario A, no modification to nuclear interactions) to be  the conservative benchmark constraint. Increasingly attractive nuclear interactions in axion condensed matter (Scenarios B and C) would cause the modification of crust structure to occur closer to the QCD axion line.  Parameter space as in Fig.~\ref{fig:constraint relaxation}. }
\label{fig:constraint combined}
\end{center}
\end{figure}

In this paper, we present a comprehensive study of the neutron star equation of state in the presence of exceptionally light QCD axions. In dense objects, the axion potential receives corrections proportional to the nucleon density due to axion couplings to QCD. For a wide range of $\varepsilon$ (see Eq.~\eqref{eq:defineep}), a neutron star is both dense enough such that the minimum of the axion potential shifts to $\theta = \pi$, and large enough such that it is energetically favorable for a $\theta = \pi$ region to emerge at the center of the star and an axion domain wall profile between the $\theta = \pi$ inside and $\theta = 0$ outside to develop.
In the $\theta = \pi$ region, there are three main effects, namely, a negative axionic pressure of order $-m_a^2 f_a^2$, lighter nucleon masses, and a decrease of the pion mass from $138 \,{\rm MeV}$ to about $80 \,{\rm MeV}$. The latter can lead to the modification of nuclear interactions. 

We study these changes with Chiral Effective Field Theory and Relativistic Mean Field Theory. The results of ChiEFT are somewhat inconclusive at $\theta = \pi$ and a pion mass of about $80 \, {\rm MeV}$, with the convergence of many-body perturbation theory becoming worse as the pion mass is decreased. Nevertheless, an initial assessment indicates that the nuclear force may become significantly more attractive at $\theta = \pi$ leading to self-bound neutron matter and axion condensation at $n_B \simeq n_{\rm sat}$ even for $\varepsilon \simeq 1$. We model this reduction of the pion mass in RMFT with a decrease of the $\sigma$ mass ($K_{\sigma}>0$, see definition given by Eq.~\eqref{eq:defineKsigma}, Table~\ref{tab:Ksigma}, and related discussion), resulting in an increase of the strength of the attractive part of the nuclear interaction. Using RMFT, we calculate modifications to the equation of state of nuclear matter at different densities at $\theta = \pi$. Combining these modifications to the nucleon masses and nuclear interaction strength, the negative axion pressure, axion domain wall profile, and gravity, we compute the neutron star equation of state and structure at different $m_a$ and $\varepsilon$ shown in Figs.~\ref{fig:rmf phase diagram}, \ref{fig:no ax and norm crust}, and \ref{fig:axcrusts}. Qualitatively, the negative axion pressure, the lighter nucleon masses, and the more attractive nuclear force at $\theta = \pi$ lead to neutron stars with a reduced or missing inner crust and a much reduced outer crust that is partially or entirely at a negative pressure in the axion domain wall for $\varepsilon \lesssim 0.1$. Our preliminary investigations of the $m_\pi-$dependence of the nuclear EOS in \ChiEFT suggest the intriguing possibility of axion condensation for $\varepsilon \simeq 1$ at $n_B \lesssim 2 \nsat$ and will be explored separately in future work. 
Our future work will also more closely examine the $m_\pi$-dependence of the underlying nuclear interactions, which is generally not well understood, especially those terms that are typically assumed to just renormalize the low energy constants.

This finding has observational consequences that are significantly constrained by isolated neutron star cooling (also recently considered in Ref.~\cite{Gomez-Banon:2024oux}), crust thermal relaxation, and pulsar glitches. The combination of these constraints is shown in Fig.~\ref{fig:constraint combined} with and without pulsar glitches as the pulsar glitch mechanism is the least understood of these three observational constraints and most amenable to alternative explanations, difficult to fully justify as those may be. These all depend primarily on the equation of state at $n_B \lesssim n_{\rm sat}$ where our treatment of the interaction between axions and nucleons is robust. As the most conservative bound, Scenario A in the figure shall be considered as the benchmark constraint, while Scenarios B and C shall be considered as a generous estimate of the uncertainty and possibly stronger constraints to be drawn once the $\theta$-dependence of QCD has been explored in greater detail. The possibility that much of the parameter space can be excluded by a lack of dripped neutrons in the crust invalidating the crust-based glitch mechanism for pulsars is an intriguing prediction which warrants further exploration. In particular, the pion mass dependence of the low energy constants in ChiEFT must be determined from Lattice QCD and the convergence of nuclear many-body perturbation theory at decreased pion mass must be addressed. Additionally, in the domain wall region, parity-violating nuclear interactions may become relevant and should be included in a detailed treatment in ChiEFT. More details are available in App.~\ref{app:pv_dw}.

We also suggest two intriguing observational consequences that require further observational and theoretical investigation, also previously noted in Ref.~\cite{Balkin:2023xtr} (see also Ref.~\cite{Gao:2021fyk} for similar effects of scalar fields with Yukawa couplings to SM fermions). As shown in Fig.~\ref{fig:mr}, the mass-radius relation of neutron stars with a $\theta = \pi$ core is indicative of rapid stiffening of the equation of state at intermediate density. Additionally, for some $\varepsilon$ there are missing crust layers, resulting in very small radii for low-mass neutron stars, similar to strange quark stars. With data from the NICER x-ray observations as well as neutron star merger measurements with LVK and future gravitational wave observatories, a precisely known mass-radius relation would strongly constrain the presence or absence of a $\theta = \pi$ core. A second intriguing possibility is self-bound objects at $\theta = \pi$ which we call $\pi$-balls, with densities close to nuclear density and stabilized by the attractive axion potential rather than gravity. These objects can have different masses and sizes, ranging from planet to white dwarf masses. Further studies are required to understand their formation in violent events such as supernovae or neutron star mergers, their growth in the interstellar and intergalactic medium, as well as the resulting observational signatures.

\section*{Acknowledgments}

\noindent  
We thank Stefan Stelzl for helpful discussions and correspondence, as well as Anson Hook and Ken McElvain for helpful discussions. The U.S. Department of Energy supported the work of M. K. and S. R. under Grant No. DE-FG02-00ER41132. S. R. also thanks the members of the N3AS Physics Frontier Center, funded by the NSF Grant No. PHY-2020275 for useful conversations. M.B. is supported by the U.S. Department of Energy Office of Science under Award Number DE-SC0024375 and the Department of Physics and College of Arts and Science at the University of Washington. M. B.  is also grateful for the hospitality of Perimeter Institute, where part of this work was carried out. Research at Perimeter Institute is supported in part by the Government of Canada through the Department of Innovation, Science and Economic Development and by the Province of Ontario through the Ministry of Colleges and Universities. This work was also supported by a grant from the Simons Foundation (1034867, Dittrich).
C. D. acknowledges support from the National Science Foundation under award PHY~2339043.
This material is based upon work supported by the U.S. Department of Energy, Office of Science, Office of Nuclear Physics, under the FRIB Theory Alliance award DE-SC0013617. 

\section*{Data availability statement}
The data that support the findings of this article are openly available \cite{axionns-public}.

\appendix{}
\section{Additional nuclear interactions in the axion domain wall}
\label{app:pv_dw}

The energy density and pressure of the axion domain wall has a somewhat weak effect on the EoS of typical neutron stars. However, inside the axion domain wall, the properties of nuclear matter can be affected by the following additional terms~\cite{Ubaldi:2008nf},
\begin{equation}
\begin{split}
     \mathcal{L}_{\rm DW} &\supset i(c_- +4 c_4) \frac{m_u m_d\sin\theta}{\left[m_u^2+ m_d^2 +2 m_u m_d \cos\theta\right]^{1/2}}\bar{\mathcal{N}}\gamma_5 \mathcal{N} \\
     &\quad + c_+ \frac{m_u m_d\sin\theta}{\left[m_u^2+ m_d^2 +2 m_u m_d \cos\theta\right]^{1/2}} \bar{\mathcal{N}} \frac{\pi^a\tau^a}{f_{\pi}} \mathcal{N},
\end{split}
\end{equation}
where $\mathcal{N}$ contains both nucleon fields and $\tau_3$ is the third Pauli matrix in isospin space. This parity-violating mass term and pion-nucleon interaction only turn on in regions of finite $\theta$ between $0$ and $\pi$. The LECs $c_-$, $c_+$, and $c_4$ are expected to contribute at high order in the chiral expansion since the interaction is isospin breaking, the Lagrangian has a quark mass insertion, and in the case of the mass correction  the parity violation in the nucleon fields is relativistic. If the LECs are all order unity, the interaction between nucleons and pions is smaller than the leading nucleon pion interaction strength by $m_q/f_{\pi} \simeq 10^{-1}$ and the parity-violating mass is a correction of order $m_q / m_N \simeq 10^{-3}$. We do not expect these interactions to matter for sufficiently large axion masses. However, they might lead to interesting signatures for small axion masses (thick domain wall), especially in the \ZN model.

\section{RMFT corrections to the domain wall calculation}
\label{app:deltaloc}
In general, $\delta^{\rm (loc)}$ is given by:
\begin{equation}
    \delta^{\rm (loc)} = \frac{1}{\sigma_N n_B^{(s)}} \frac{\partial \Omega}{\partial m_\sigma} \frac{\dd m_\sigma}{\dd \theta}
\end{equation}
where $n_B^{(s)}$ is the total scalar density of baryons. In the outer crust and core, the expression in terms of the mean-field value of the $\sigma$ field is simple.
\begin{equation}
\begin{split}
    \delta^{\rm (loc)}  &= \frac{d_\sigma}{\sigma_N  [2 + 2 d_\sigma f(\theta)]} \frac{\chi}{n_B^{(s)}} \left( \frac{m_\sigma}{g_\sigma} \right)^2 \\
    &\times \left[ (g_\sigma \langle \sigma \rangle)^2  + \frac{\kappa_3}{3m_N^{\rm phys}} (g_\sigma \langle \sigma \rangle)^3 + \frac{\kappa_4}{12(m_N^{\rm phys})^2} (g_\sigma \langle \sigma \rangle)^4 \right]
\end{split}
\end{equation}
where $g_\sigma$, $\kappa_3$, and $\kappa_4$ are parameters of the RMFT model, $\chi$ is the volume fraction of nuclei (or one in the core), and $g_\sigma \langle \sigma \rangle = m_N (\theta) - m^*_N (\theta)$ for $m^*_N (\theta)$ the effective mass of the nucleon with mean field effects included. In the inner crust, this expression becomes more complicated.
\begin{equation}
    \begin{split}
        \delta^{\rm (loc)} &= \frac{d_\sigma}{\sigma_N  [2 + 2 d_\sigma f(\theta)]} \frac{1}{n_B^{(s)}} \left( \frac{m_\sigma}{g_\sigma} \right)^2 \\
        &\times \left\{ \chi \left[ (g_\sigma \langle \sigma \rangle_{\rm nuc})^2  + \frac{\kappa_3}{3m_N^{\rm phys}} (g_\sigma \langle \sigma \rangle_{\rm nuc})^3 + \frac{\kappa_4}{12(m_N^{\rm phys})^2} (g_\sigma \langle \sigma \rangle_{\rm nuc})^4 \right] \right. \\
        &+ \left. (1 - \chi) \left[ (g_\sigma \langle \sigma \rangle_{\rm neut})^2  + \frac{\kappa_3}{3m_N^{\rm phys}} (g_\sigma \langle \sigma \rangle_{\rm neut})^3 + \frac{\kappa_4}{12(m_N^{\rm phys})^2} (g_\sigma \langle \sigma \rangle_{\rm neut})^4 \right] \right\}
    \end{split}
\end{equation}
where $\langle \sigma \rangle_{\rm nuc}$ is the mean field value of the sigma field in nuclei and $\langle \sigma \rangle_{\rm neut}$ is the mean field value of the sigma field in the dripped neutron phase. Since $n_B^{(s)}$ is dominated by the dripped neutron phase except in the vicinity of the neutron drip phase transition, but the $\sigma$ field has a small mean field value in the dripped neutron phase, this suppresses $\delta^{\rm (loc)}$ in the inner crust.

\bibliographystyle{apsrev4-1}
\bibliography{axion}
\end{document}